# The Exosphere as a Boundary: Origin and Evolution of Airless Bodies in the inner Solar System and beyond including Planets with Silicate Atmospheres

Helmut Lammer[1], Manuel Scherf[1], Yuichi Ito[2,3], Alessandro Mura[4], Audrey Vorburger[5], Eike Guenther[6], Peter Wurz[5], Nikolai V. Erkaev[7,8,9], P. Odert[10]

[1]Space Research Institute, Austrian Academy of Sciences, Schmiedlstr. 6, 8042 Graz, Austria

[2]Dept of Physics and Astronomy, Faculty of Maths and Physical Sciences, University College London, Gower Street, WC1E 6BT, London, UK

[3]National Astronomical Observatory of Japan, Osawa 2-21-1, Mitaka, Tokyo 181-8588, Japan

[4]Istituto de Fisica dello Spazio Interplanetario-CNR, Rome, Italy

[5]Physikalisches Institut, University of Bern, Bern Switzerland

[6]Thüringer Landessternwarte Tautenburg - Karl-Schwarzschild-Observatorium, Sternwarte 5, D-07778 Tautenburg, Germany

[7]Institute of Computational Modelling, Siberian Branch of the Russian Academy of Sciences, 660036 Krasnoyarsk, Russian Federation

[8]The Applied Mechanics Department, Siberian Federal University, 660074 Krasnoyarsk, Russian Federation

[9]Institute of Laser Physics, Siberian Branch of the Russian Academy of Sciences, 630090, Novosibirsk, Russian Federation

[10]Institute of Physics/IGAM, University of Graz, Graz, Austria

**Abstract:**

In this review we discuss all the relevant solar/stellar radiation and plasma parameters and processes that act together in the formation and modification of atmospheres and exospheres that consist of surface-related minerals. Magma ocean degassed silicate atmospheres or thin gaseous envelopes from planetary building blocks, airless bodies in the inner Solar System, and close-in magmatic rocky exoplanets such as CoRot-7b, HD219134b and 55 Cnc e are addressed. The depletion and fractionation of elements from planetary embryos, which act as the building blocks for proto-planets are also discussed. In this context the formation processes of the Moon and Mercury are briefly reviewed. The Lunar surface modification since its origin by micrometeoroids, plasma sputtering, plasma impingement as well as chemical surface alteration and the search of particles from the early Earth's atmosphere that were collected by the Moon on its surface are also discussed. Finally, we address important questions on what can be learned from the study of Mercury's environment and its solar wind interaction by MESSENGER and BepiColombo in comparison with the expected observations at exo-Mercurys by future space-observatories such as the JWST or ARIEL and ground-based telescopes and instruments like SPHERE and ESPRESSO on the VLT, and vice versa.

# 1. Introduction

The exosphere is a thin gaseous atmospheric layer that surrounds a planet or natural satellite where atoms and/or molecules are gravitationally bound, but where the gas density is so low that the particles are collisionless. In ordinary planets with dense atmospheres, the exosphere is the uppermost atmospheric layer, where the gas density thins out and becomes one with outer space (e.g., Chamberlain, 1963). The lower boundary of an exosphere is called exobase or critical level, which corresponds to the altitude where barometric conditions no longer apply, and the atmosphere temperature becomes nearly a constant. At this altitude, the mean free path of upward traveling atmospheric species is equal to the scale height (e.g., Chamberlain, 1963; Bauer and Lammer, 2004). Under these conditions, the pressure scale height is equal to the density scale height of the main constituent. Since the dimensionless Knudsen number *Kn* is defined as the ratio of the mean free path length to a representative physical length scale, it means that the exobase lies in the atmospheric level where $Kn \approx 1$.

However, there are several planetary bodies in inner planetary systems that have no dense atmospheres with the Earth-centric structures like troposheres, stratospheres, mesospheres, thermospheres below their exospheres. Rocky bodies from planetesimals to magmatic planetary embryos, asteroids, natural satellites like the Moon and low gravity close-in planets like Mercury or even some "rocky" exoplanets that orbit at close-in distances around their host stars possess only an exosphere or a gravitationally bounded atmosphere/exosphere environment that consist of outgassed and surface released (moderately) volatile elements.

At planetary bodies where an exosphere acts as a boundary between the surface layer and the surrounding environment, the exobase corresponds to the surface. Such an exosphere is a thin gaseous envelope where atoms and molecules are released from the surface by various processes such as thermal release, photon- or electron-stimulated desorption, particle sputtering and micrometeorite impact vaporization (e.g., Domingue et al. 2014; Wurz et al. 2021, this issue).

Atoms and molecules that are emitted from the surface are ejected on ballistic trajectories until they collide again with the surface, where they can alter the chemistry of the surface material, modifying optical surface properties, etc. together with external plasma (e.g., Killen et al. 2007; Wurz et al. 2021, this issue). In case the particles are released into the exosphere with energies that are larger than the escape energy, they are lost from the body since they move through a collisionless environment. Smaller planetary bodies (i.e., planetesimals, planetary embryos, asteroids, small moons, etc.), where the surface material experiences escape velocity are not considered to have exospheres.

Recent studies related to the outgassing of noble gases (Ar, Ne, Kr, Xe) and the before mentioned moderately volatile elements and their losses from planetesimals and growing magmatic low mass planetary embryos indicate that evaporation processes should have depleted the planetary building blocks from their initial abundance in non-chondritic and chondritic rocky materials (Hin et al., 2017; Young et al., 2019; Sossi et al., 2019; Benedikt et al., 2020; Lammer et al., 2020a; Lammer et al. 2020b). These findings further indicate that accretional vapor loss from magmatic planetary building blocks shapes planetary compositions. A steady-state rock vapor or an atmosphere/exosphere envelope that consists of the body's minerals forms above magma oceans within minutes to hours and results from a balance between rates of magma evaporation and atmospheric escape (Young et al. 2019; Benedikt et al. 2020).

A main evidence for evaporation-related losses during planet formation is heavy isotope enrichment in several rock-forming elements relative to chondrites that are found in various differentiated bodies in the Solar System (Young et al., 2019; Sossi et al., 2019; Benedikt et al., 2020). It was also shown by Young et al. (2019) and Benedikt et al. (2020) that magmatic planetary embryos with masses that are lower than that of the Moon, the gravity is too weak for the build-up of a dense silicate atmosphere. Because the low gravity and hot surface temperatures act together, all outgassed elements will escape immediately to space and the planetary building block will be depleted in noble gases and moderately volatile elements (Lammer et al., 2020a).

One can expect that terrestrial planets that formed close to their star, or in case of Mercury to the Sun, might have accreted significantly volatile depleted material after the gas disk dissipated and during the so-called giant impact phase. Lower mass planets that formed further out and migrated inward to close-orbital distances would have lost their primordial $H_2$-He-dominated atmospheres due to EUV-driven hydrodynamic escape (Owen and Wu 2017; van Eylen et al. 2018; Armstrong et al. 2019). That there is a sub-Neptune-desert or a photoevaporation valley in close-orbital distances is also confirmed by exoplanet observations with the Kepler space telescope (McDonald et al. 2019). This sub-Neptune desert or so-called Fulton gap (Fulton et al. 2017; Fulton and Petigura 2018) is an observed scarity of planets with radii between $\approx 1.4 - 2$ Earth radii ($R_{Earth}$). It is expected that thermal escape of sub-Neptunes in close orbital distances would lead to a population of "hot" rocky cores with smaller radii at small separations from their parent stars, and planets with thick hydrogen- and helium-dominated envelopes with larger radii at larger distances. The bimodality in the distribution was confirmed with higher-precision data in the California-Kepler Survey in 2017 (Fulton et al. 2017), which was shown to match the predictions of the mass-loss hypothesis.

If these bodies also accreted volatile-rich materials after these periods and after they have grown to masses too high for the delivered volatiles to efficiently escape, they may result in planets with a high metal to silicate ratio, while the crust remains volatile-rich such as expected for Mercury (e.g, Peplowski et al. 2012; Nittler et al. 2018). At more massive higher metal/silicate ratio exoplanets at close-in orbits around their host stars, dayside surface temperatures above 1500 K can originate so that magma oceans or magma lakes remain over the planet's lifetime (Schaefer and Fegley 2009; Valencia et al. 2010; Ito et al. 2015; Miguel et al. 2019; Venot et al. 2020). In such cases, rock vapor atmospheres can originate above the hot surface and stellar wind plasma interactions with the mineralogical atmosphere/exosphere environment will occur (Mura et al. 2011; Guenther et al. 2011; Vidotto et al. 2018).

During the planet formation process large planetesimals, planetary embryos and the growing protoplanets develop magma oceans due to heating of the decay of radioactive elements, particularly due to the short-lived $^{26}$Al, $^{60}$Fe (Lichtenberg et al. 2016; O'Neill et al. 2020), gravitational energy released upon accretion (Albarédé and Blichert-Toft 2007, Elkins-Tanton 2012) and collisions (e.g., Morbidelli et al. 2012; Brasser 2013; Johansen et al. 2015; 2021; Lammer et al. 2021). Analysis of some rocks on the Moon, Mars, and Vesta indicate such an early widespread silicate melting and fractional crystallization afterwards. The crystallization ages of these rocks agree with the age range of primary planetary formation until $\leq 4.4$ Gyr (Elkins-Tanton 2012; and references therein), indicating that accretionary and radiogenic heat produces mantle melting. The lifetime of a magma ocean depends on a number of parameters such as:

- the size of the planetary body;

- the amount of accreted radioactive elements;

- the temperature of the accreting material;

- the time between collisions with large planetary embryos;

- the existence of a conductive boundary layer;

- and the existence of an atmosphere.

The before mentioned magmatic bodies can be separated in two categories where transient magma oceans are present, first: magmatic planetesimals and planetary embryos that belong to the building blocks of planets with too low masses for their gravity to bind outgassed constituents (Hin et al. 2017; Young et al. 2019; Benedikt et al. 2020). On such bodies, outgassed silicates and moderately volatile elements escape to space or form only tiny atmospheres near the surface, which are in balance between outgassing and escape rates but are lost when the outgassing process decreases, and the magma ocean solidifies.

The second kind of bodies at which transient magma oceans are present are the early planets after they finished their accretion. When the magma ocean solidifies, depending on the oxidation stage of the magmatic mantle, either predominantly $H_2O$, and $CO_2$ for oxidized, and $H_2$, and $CO$ for reduced conditions, respectively, will be outgassed and steam atmospheres build up (Lebrun et al. 2013; Salvador et al. 2017; Nikolaeou et al. 2019; Bower et al. 2019; Herbort et al. 2020). The further evolution of these dense atmospheres depends on the stellar XUV flux evolution, orbit location, water inventory, and the volcanic activity of the respective planet.

Until the photometrical discovery of the first detected "rocky" exoplanet Corot-7b by the French-led CoRoT space telescope in February 2009 (Legér et al. 2009), only transient magma oceans as discussed above were know. With Corot-7b's radius of $1.58 \pm 0.1$ $R_{Earth}$ (Legér et al. 2009) and a mass of $7.42 \pm 1.21$ Earth masses ($M_{Earth}$; Hatzes et al. 2011) the planet's bulk density lies close to the density-radius relationship of Mercury. Due to its close orbital distance $d$ of $0.0172 \pm 0.00029$ AU the planet has an equilibrium temperature of $\leq 1800$ K at its dayside which is hot enough to expect a magma ocean or magma ponds. Since this discovery, more of such close-in "rocky" exoplanets with expected permanent magma oceans/ponds on their dayside were found. One can also expect that this type of planets will outgas volatile and moderately volatile elements from such magmatic hot regions so that detectable silicate atmospheres may build up until the elemental reservoir in the magma ocean depletes. If the gravity of such budies is high enough, however, to keep atmospheres that consist of the planet's silicates, they experience extreme stellar wind plasma interactions that shape their exospheres to cometary-like structures (see Section 5.3 for a detailed discussion). This can be observed in the future by large ground and space-based telescopes.

In this review, we are focusing on the origin and evolution of airless bodies in the inner Solar System and on rocky exoplanets in close-in orbits. We, therefore, do not discuss planets with magma ocean outgassed primary atmospheres.

In Section 2 we discuss the latest knowledge on the radiation and plasma environment of young stars and the Sun. This is important if one is interested in radiation and particle related release processes of minerals from the surfaces of planetary building blocks as well as the historical exposure of the Hermean and Lunar surfaces. In Section 3 we discuss the depletion and fractionation of rock-forming elements from planetesimals to magmatic planetary embryos. Section 4 investigates the origin of the Moon and its exosphere evolution including fingerprints from the Earth's ancient atmosphere, in Section 5 we address the characteristics of Mercury and its formation hypotheses, and compare the planet with more massive close-in rocky exoplanets where the stars luminosity form magma oceans and related silicate atmospheres with extended exospheres. Before we conclude the review, we discuss in Section 6 the possibilities for observations of silicate-like atmospheres/exospheres from close-in hot higher metal/silicate ratio type exoplanets.

## 2. Radiation and plasma environment of young stars and the Sun
### 2.1. X-ray and EUV evolution

To understand the evolution of airless bodies in the inner Solar System and rocky close-in exoplanets, it is important to reconstruct the evolution of the solar and stellar radiation and plasma environments over time. The X-ray and EUV flux evolution of young stars, together often subsumed as XUV ($\leq 91.2$ nm), is particularly important since short wavelength radiation drives loss processes, not only on planetary bodies with extended atmospheres but also within the exospheres of airless bodies (Wurz et al. 2021, this issue). The XUV flux from the young Sun, for instance, leads to photoionization of particles in the exospheres of the Moon and Mercury. A higher number of particles gets ionized for higher XUV fluxes, which reduces the return flux onto the surface of these bodies (see Section 4.3.3), thereby increasing escape. The radiation from the young Sun also heats up the thermospheres of magma ocean degassed atmospheres (see Section 3) which leads to atmospheric expansion and strong thermal escape rates (e.g., Benedikt et al. 2020). Since stellar radiation scales with $1/d^2$, close-in rocky exoplanets experience a far more extreme radiation environment than any Solar System objects. CoRoT-7b, for instance, at an orbit of 0.0172 AU is irradiated by ~3400 times higher XUV fluxes than present-day Earth at 1 AU. Since such high radiation significantly affects exospheres of such bodies, it is crucial to understand the early radiation environments of young stars and the Sun.

The XUV flux evolution of a star is dependent on its initial rotation rate with faster rotating stars showing higher initial fluxes than moderate or slow rotators (e.g., Tu et al. 2015; Johnstone et al., 2015a, 2015b; Johnstone et al. 2020). For solar-like stars, all rotators, however, show an initial saturation phase with XUV fluxes being as high as 400 to 500 times the present-day solar value that can last for about 5 to 150 million years (Myr), depending on whether the star was a slow or a fast rotator or something in between (Tu et al. 2015). The torque from the stellar mass loss, however, slows down the initial rotation rate until the different rotators converge towards one single track which, for solar-like stars, happens after about 1 billion years (Gyr) (Johnstone et al. 2015b). The convergence of the different tracks happens later for lower-mass stars, but the difference between the various rotational tracks gets less and less pronounced for decreasing stellar masses (Johnstone et al. 2020). Moreover, Johnstone et al. (2020) found that the total emitted XUV flux from M- or K-type stars is generally lower than for G- or even F-type stars. On the other hand, any body that receives the same amount of bolometric luminosity as around a G star would, therefore, experience much higher and longer lasting XUV flux exposure corresponding to its orbit.

Figure 1 shows the evolution of the X-ray surface flux for slow, moderate, and fast rotators with masses of 1, 0.75, 0.5, and 0.25 solar masses ($M_{Sun}$) as an example scaled to the corresponding habitable zones of these stars according to Johnstone et al. (2020). Here, the different rotators are defined as the 5[th] (slow), 50[th] (moderate), and 95[th] (fast) percentiles of the rotational distribution of the investigated stellar sample of Johnstone et al. (2020). To show the difference of the short-wavelength radiation received by a body for the same bolometric luminosity at different host stars, it is common to illustrate this effect within the habitable zone. As mentioned above, close-in airless bodies will likewise receive an even higher radiation than depicted within the exemplary Figure 1.

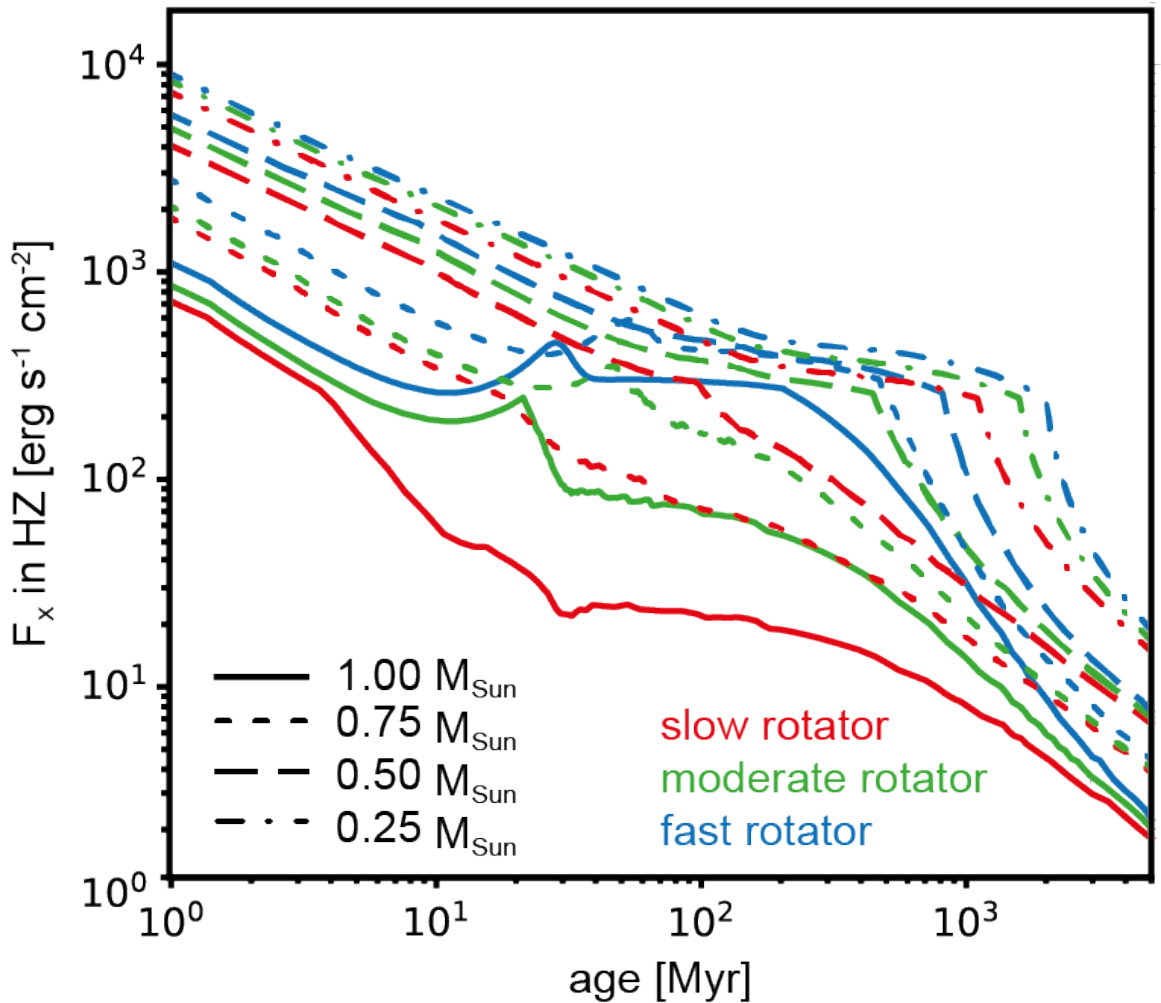

*Figure 1. Evolution of the X-ray surface flux for slow, moderate, and fast rotators scaled to the respective habitable zones of stars with masses of 0.25, 0.5, 0.75, and 1.0 $M_{Sun}$ according to Johnstone et al. (2020).*

This also illustrates that the determination of the early evolution of the Sun, and hence also of the planets, is not straight forward since it is not possible to infer its initial rotation rate from its present-day value. However, there are several different studies suggesting that the Sun likely was a slow, or at most, a slow to moderate rotating young G-type star (e.g., Saxena et al. 2019; Lammer et al. 2020a,b; Johnstone et al. 2021). An $N_2$-dominated atmosphere would not have been stable under the strong XUV flux of a fast rotator during the Earth's Archean eon (Johnstone et al. 2021), while the present-day noble gas ratios of Ar, Ne, and Ar/Ne in the atmospheres of Earth and Venus can only be reproduced in case that the young Sun was a slow, or a slow to moderate rotator. Likewise, the present-day moderately volatile surface composition of the Moon can also be reproduced through sputtering if the Sun was a slow rotator (Saxena et al. 2019).

Besides the XUV flux evolution, one also must take into account the related evolution of solar and stellar flares since their increased burst of radiation can significantly affect escape processes at the airless bodies in the Solar System and beyond (see Wurz et al. 2021, this issue). Again, fast rotators happen to flare more often than moderate or slow rotators with more massive stars, having more energetic flares than lower-mass stars, and flare rate generally decreasing for older stars (e.g. Davenport et al. 2019; Johnstone et al. 2020). While at present-day, the strongest flares observed for the Sun are in the range of ~$10^{32}$ erg (e.g., Emslie et al., 2005, 2012), this value can reach up to $10^{37}$ erg for solar-like stars and ~$10^{33}$ erg for low-mass stars in the range of $0.1 - 0.2$ $M_{Sun}$ (e.g., Wu et al. 2015; Yang and Liu 2019). Also, here, it has to be pointed out that even though the flare energies are lower for lower-mass stars, their input into their respective HZs is likely higher than in the case of more massive stars (Johnstone et al. 2020).

The far-ultraviolet (FUV, 91.2-200 nm) and the ultraviolet flux (UV, 200-400 nm) are also increasing towards the past, but less significant (e.g., Ribas et al. 2005; Claire et al. 2012). UV related processes like photo-stimulated desorption (PSD; Wurz et al., 2021, this issue) were, therefore, likely also more efficient than at present-day.

**2.2. Bolometric luminosity evolution**

The bolometric luminosity $L_{bol}$ of a star is the integrated flux over all wavelength ranges and is predominantly shaped by the optical. On the contrary to the XUV evolution, for a G star the bolometric luminosity generally increases over time after reaching the zero-age main sequence (ZAMS; Figure 2). After the arrival of the Sun at ZAMS, $L_{bol}$ was about 30% lower than at present-day (Baraffe et al. 2015; Gough 1981; e.g., Newman and Rood 1977). The reason for the subsequent increase in $L_{bol}$ is due to nuclear fusion of hydrogen to helium in the core of the Sun (e.g., Feulner 2012; Gough 1981). While He is accumulating, the molecular weight of the core is also increasing, which leads to a contraction of the core and a therewith connected increase in heat to keep the star stable, with the latter resulting in a higher luminosity output.

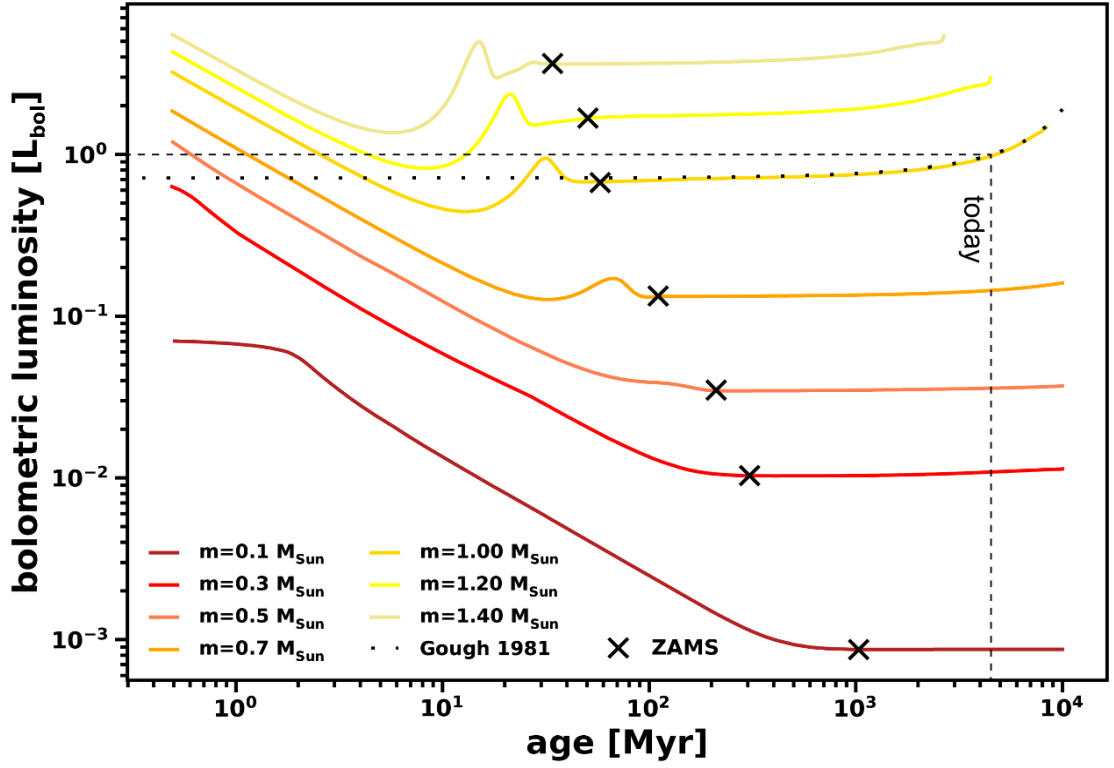

*Figure 2. The evolution of $L_{bol}$ over time for different stellar masses according to the stellar evolution model of Baraffe et al. (2015). The black crosses indicate the respective ages at which these stellar masses reach the zero age main sequence (ZAMS).*

For the Sun, $L_{bol}(t)$ can be estimated as a function of time *t* through the following approximation by Gough (1981), i.e.,

$$L_{bol}(t) = \left[1 + \frac{2}{5}\left(1 - \frac{t}{t_\odot}\right)\right]^{-1} L_{bol,\odot},$$

where $L_{\text{bol},\odot} = 3.85 \times 10^{26}$ W is the present-day bolometric luminosity, and $t_\odot = 4.57$ Gyr is the age of the Sun. This equation correlates very well with the evolution of the Sun's bolometric luminosity except for the first ~0.1 Gyr. This can be seen in Figure 3, which shows the evolution of $L_{\text{bol}}$ for different solar masses as calculated with the stellar evolution model of Baraffe et al. (2015). One can also see the Sun's settling onto the main sequence which is accompanied by radial shrinking and the conversion of gravitational energy into heat. When the proton-proton nuclear reaction chain sets in (e.g., Bethe 1939), $L_{\text{bol}}$ suddenly increases before settling again 30% below the present-day value. Compared to this steady increase of ~1% per 100 Myr, fluc-tuations over a whole solar cycle are rather low, being in the range of 0.1% (Solanki et al. 2013).

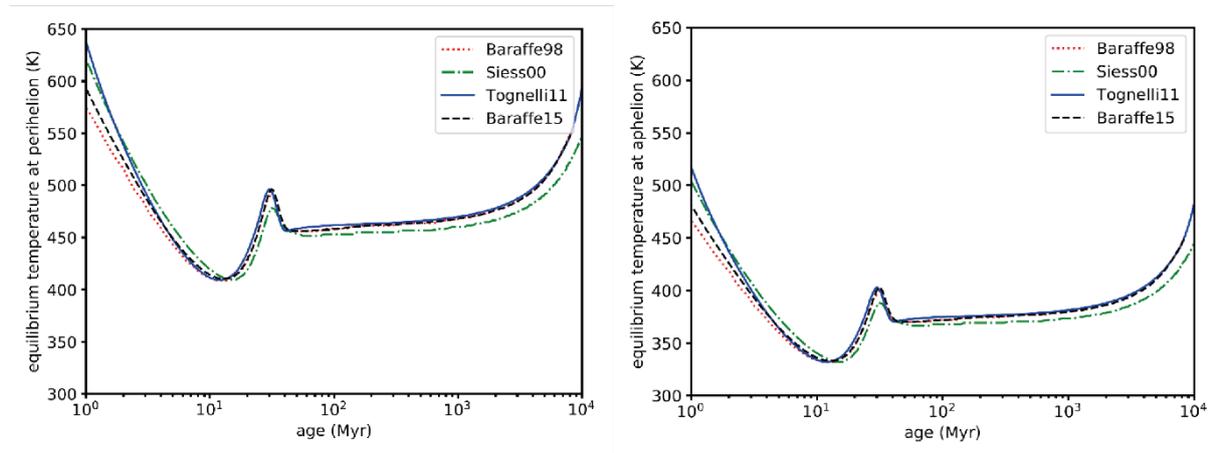

*Figure 3. The evolution of the equilibrium temperature for Mercury at perihelion (0.307 AU; left), and aphelion (0.466 AU; right) for different bolometric luminosity evolution models by Baraffe et al. (1998), Siess et al. 2000, Tognelli et al. 2011, and Baraffe et al. (2015).*

While solar-like and more massive stars show an increase in bolometric luminosity already after the first few 10s of Myr, low mass stars need much longer to settle onto the main sequence. They show a significant decrease of $L_{\text{bol}}$ for the first few 100 Myr (e.g., Baraffe et al. 2015; Spada et al. 2013); as can be seen in Figure 2, late M stars with a mass of 0.1 $M_\odot$ decrease in $L_{\text{bol}}$ by about two orders of magnitude within the first ~500 Myr. Such a strong decrease over a relatively long timeframe does not only affect the potential early habitability of terrestrial planets orbiting such stars but might also affect the early evolution of airless bodies' exospheres significantly.

The bolometric luminosity can be used for an estimate of the equilibrium temperature $T_{\text{eq}}$ of Mercury and other more or less airless rocky close-in exoplanets. As an example, Figure 3 shows estimates of the evolution of $T_{\text{eq}}$ for Mercury's orbital distance at aphelion and perihelion over the lifetime of the Solar System by using various stellar evolutionary tracks for a solar-mass star. The models and selected parameters are: [M/H] = 0, Y = 0.282, α = 1.9 (Baraffe et al., 1998); Y = 0.28, Z = 0.02 (Siess et al., 2000); Y = 0.288, Z = 0.02, α = 1.68 (Tognelli et al., 2011); [M/H] = 0, Y = 0.28, α = 1.6 (Baraffe et al., 2015). Here, [M/H] is the metallicity, Y the He content, Z the metal content, and α the mixing length parameter. Particularly for the first ~30 Myr, $T_{eq}$ is significantly varying which is important if one wants to simulate magma ocean degassed atmospheres at Mercury, but also at close-in exoplanets of other stars. Due to these hotter $T_{eq}$-phases and hence environments, magma oceans of planetary embryos solidify slower so that the outgassing and related depletion of rock-forming elements takes longer as well. Moreover, thermal release of surface-elements is more efficient on airless bodies in inner systems. Depending on the gravity of close-in planets or planetary embryos, higher luminosities can also enhance the thermal escape of primordial atmospheric gas so that bodies in close-in orbital distances loose their gaseouse envelopes via boil-off (Owen and Yu, 2016; Lammer et al., 2016; 2018).

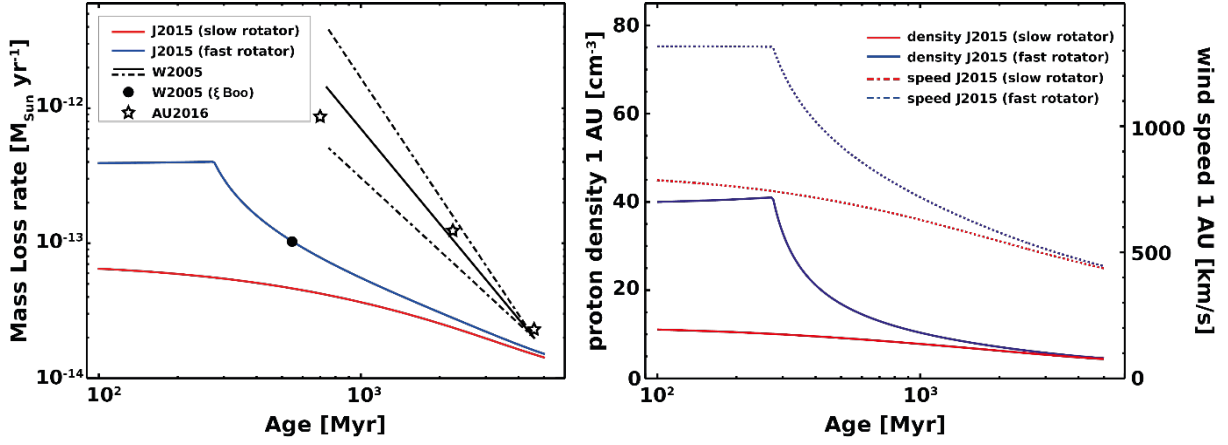

*Figure 4. Left: The mass loss evolution of solar like stars. While observations of stellar astrospheres (black solid and dashed lines as mean and upper and lower limits, respectively) by Wood et al. (2005) indicate a significant increase in solar mass loss for younger stars, a contradicting observation of the binary star ξ Boo (black circle) by Wood et al. (2005) suggest clearly lower loss rates for stars younger than ~700 Myr (however, further observations might be necessary to support smaller loss rates for young ages). The three displayed stars show simulations by Airapetian and Usmanov (2016) which lie well within the mass loss evolution estimate by Wood et al. (2015). The red and blue lines show the simulated evolution of the mass loss for a slow and fast rotator by Johnstone et al. (2015a,b). Right: The evolution of proton density (solid lines) and wind speed (dotted lines) of the slow wind for slow and fast rotating solar-like stars in the Model A by Johnstone et al. (2015a,b). Left figure adopted from Scherf and Lammer (2021)*

**2.3. Plasma environment in time**

As for the XUV flux evolution, solar and stellar mass loss is dependent on the rotational evolution of the respective star. Even though the solar wind did not remove a significant amount of mass from the Sun, it removes angular momentum through magnetic field stresses thereby leading to a rotational spin down and, thus, consequently, to a decrease in mass loss and wind over time (e.g., Kraft 1967, Skumanich 1972, Wood 2004, Johnstone et al. 2015a,b, Vidotto 2021). That the solar mass loss and the therewith connected solar wind might have been higher in the past is also supported by observational studies of stellar astrospheres (e.g., Wood et al. 2002, 2005) for solar-like stars older than ~700 Myr. For younger stars, however, these studies might even indicate a significantly lower mass loss, as can be seen in Figure 4a (ξ Boo compared). Wood et al. (2005) only extrapolated the mass loss evolution of the Sun (black solid and dash-dotted lines) back to 700 Myr since the young binary star ξ Boo (with spectral types G8V and K4V; black dot in Figure 4a) was found to have a significantly lower mass loss rate.

More recent simulations by Airapetian and Usmanov (2016) with a three-dimensional magnetohydrodynamic Alfvén wave driven solar wind model, retrieved mass loss rates for solar-like stars that are in the same range as found by Wood et al. (2005), as can also be seen in Figure 4a. Further observations of the mass loss rates of solar-like stars, particularly of those of a young age, might be needed to retrieve a clearer picture on the evolution of the solar and stellar plasma environments. Another important factor when considering the plasma environment and its impact on the evolution of airless bodies' exospheres are Coronal Mass Ejections (CMEs) which often show a significant increase in ambient particle density and velocities up to >2000 km/s (e.g., Gopalswamy 2004; Chen 2011; Webb and Howard 2012). While a specific planetary body in the present Solar System is only hit by about ~6 – 16% of all CMEs, this increases to more than 31% for solar-like stars with an age of about 700 Myr (Kay et al. 2019). As Kay et al. (2019) found for these ages via studying the solar twin $k^1$ Ceti, CMEs are more frequently focused onto the Ecliptic plain due to the coronal magnetic field reflecting it towards the astrospheric current sheet. In addition, CMEs might have been significantly more frequent in the past than at present-day (e.g., Odert et al. 2017), and likely even stronger (e.g., Airapetian et al. 2016). Extreme space weather events might, therefore, have played a crucial role in the evolution of airless bodies.

# 3. Outgassing from transient magma oceans and depletion of elements from low-mass embryos

Planets accrete mass by numerous collisions between small objects, which accumulate to planetesimals, and planetary embryos. During these collisions transient magma oceans originate in many planetary bodies in the early solar and extrasolar systems, determining the initial conditions for diverse evolutionary paths of terrestrial planets (Deng et al., 2020). Within the first ≈3 Myr after the origin of the Solar System, planetesimals and larger planetary embryos develop magmatic pools, oceans, and some can perhaps completely melt by the heating of short-lived radioactive elements (e.g., Urey, 1955; Fish et al., 1960; Elkins-Tanton, 2012; Lichtenberg et al., 2016, 2018, 2019) such as $^{26}$Al, and $^{60}$Fe, frequent collisions (e.g., Safronov and Zvjagina, 1969; Wetherill, 1980; Tonks and Melosh, 1993; Schlichting et al. 2015), and gravitational energy (Albarède and Blichert-Toft, 2007). It was found by Lichtenberg et al. (2016, 2018) and Neumann et al. (2020) that the heating by the before mentioned short-lived radioisotopes, followed by internal differentiation and fast volatile outgassing determined to a large extent the thermal history and interior structure of these planetary building blocks and, hence, their final composition during the earliest stages of planetary formation. Neumann et al. (2020) studied the energy balance in small bodies that are heated by decay of radioactive elements and compaction-driven water-rock separation in a dust-water/ice-empty pores mixture. Addionally, these authors considered also second-order processes, such as accretional heating, hydrothermal circulation, and ocean or ice convection and found that precursors of bodies like Ceres in the inner Solar System could have been wet and/or dry.

Collisional erosion (O'Neill and Palme, 2008; Carter et al., 2015; Bonsor et al., 2015; Boujibar et al., 2015; Carter et al. 2015; Allibert et al. 2021) fractionates elements such as Si, Fe and Mg (i.e., Fe/Mg, Si/Fe) according to their incompatibility with mantle minerals during melting, while losses of outgassed elements preferentially remove volatiles. Furthermore, it was shown by Sossi et al. (2019) that moderately volatile species can be fractionated from each other through their loss from large planetesimals or planetary embryos in dependence of their equilibrium pressure.

Depending on the body's oxygen fugacity (e.g., Sossi et al., 2019), its temperature, bulk composition and solidification path of the magma ocean (Elkins-Tanton, 2008, 2012), thermodynamic studies indicate that moderately volatile rock-forming elements such as Na, K, Mg, Ca, Si, etc. are outgassed from the magmatic surface (Schaefer and Fegley, 2007; Fegley et al., 2016; Odert et al. 2018; Young et al., 2019; Sossi et al., 2019). Certainly, losses of these outgassed species caused by thermal escape and collisional erosion modified the bulk composition not only of the planetary building blocks but also the composition of the terrestrial planets where they have been incorporated during accretion (Lammer et al. 2020a). The elevated Mn/Na ratio of smaller rocky bodies relative to chondrites most likely reflects the oxygenation of the magma ocean stage, because Na is more volatile under oxidized conditions than Mn (e.g., O'Neill and Palme, 2008; Siebert et al., 2018). Furthermore, Earth's Si/Mg ratio indicated that proto-Earth most likely evolved through escape from the accreting building blocks (Fegley et al., 2016).

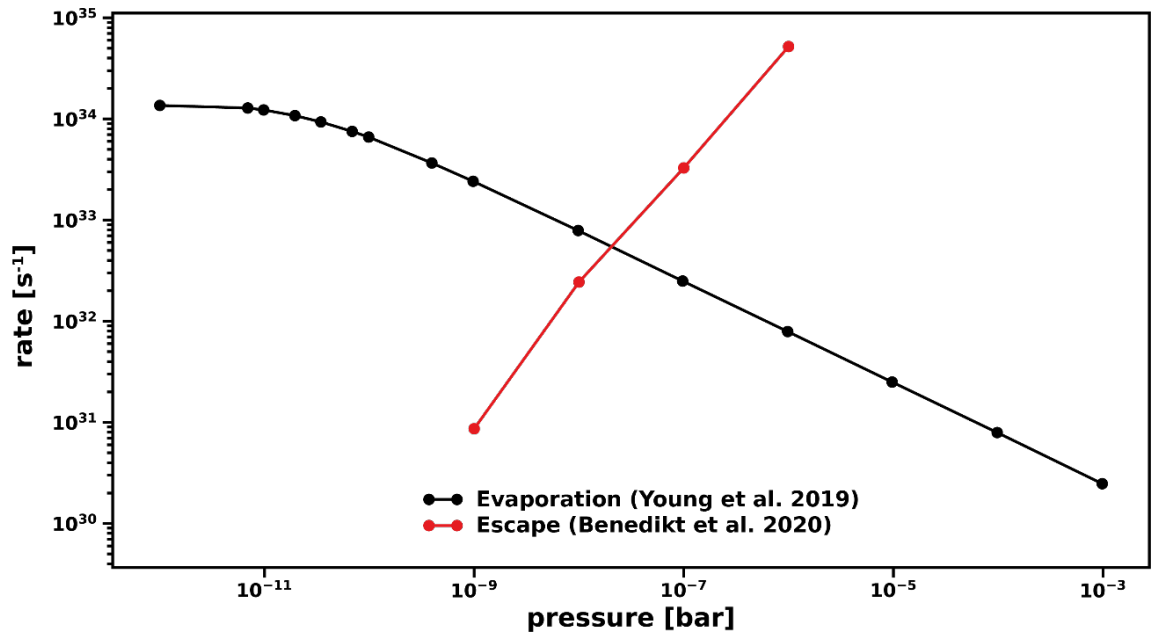

*Figure 5. Outgassing of a silicate atmosphere with a mean weight of 34 unified atomic mass units (amu) from the magma ocean of a planetary embryo with m = 0.1 moon masses ($M_{Moon}$) and a surface temperature of $T_{surf}$ = 1500 K as calculated by Young et al. (2019), and the corresponding escape flux as simulated by Benedikt et al. (2020). Figure adopted from Benedikt et al. (2020)*

The fast escape of the outgassed radioactive isotope $^{40}$K, from magmatic planetesimals or larger building blocks may further alter their composition and structure and hence the accreting protoplanets. The amount of the $^{40}$K isotope is very important since its radioactive decay contributes to the thermal evolution of the interiors of young telluric planetary bodies (i.e., planetary embryos, terrestrial planets, satellites, see O`Neill et al., 2020), and processes such as the development of long-lived magnetic dynamos (e.g., Turcotte and Schubert, 2002; Murthy et al., 2003; Nimmo et al., 2004; Nimmo and Kleine, 2015).

Hin et al. (2017), Young et al. (2019) and, more recently, Benedikt et al. (2020) studied the losses of magma-related outgassed rock-forming elements from large planetesimals and planetary embryos. Benedikt et al. (2020) showed that for Moon-mass and smaller bodies with magma layers, no dense silicate atmosphere or even an outgassed steam atmosphere can build up. It was found by these studies that, if these bodies have magma oceans and a sufficiently high surface temperature (i.e., 1500 – 3000 K), then escape should be immediate. Because of the high surface temperature and the low gravity of these bodies, the so-called unitless escape or Jeans parameter $\lambda = (GM_{pl} m_i/kT_{surf}r_{pl})$, which compares the gravitational energy with the thermal energy, of the outgassed rock-forming elements indicate an immediate hydrodynamic loss (Young et al., 2019; Benedikt et al., 2020), so that a dense rock-vapor atmosphere cannot build up. $G$ is the gravitational constant, $M_{pl}$ and $m_i$ the masses of the protoplanetary body and a particular element, $k$ is the Boltzmann constant; $T_{surf}$ is the temperature at the surface of the body with radius $r_{pl}$.

The thermal escape regime changes over a narrow range of $\lambda$ where the escape is purely hydrodynamic for values that are ≤ 2 to 3, whereas for $\lambda ≥ 6$ it is not (Volkov et al., 2011; Erkaev et al., 2015). Benedikt et al. (2020) found that for a Moon-mass-embryo that is surrounded by a 2500 K magma ocean all elements such as Na, Mg, Si, K, Ar, Ne, O, $H_2O$ would have a $\lambda$ that is < 6, i.e., they are immediately lost to space. Heavier elements and molecules such as Fe, FeO or SiO, however, have escape parameters that are slightly above the critical value. For these elements, one can expect that they will also experience high thermal loss rates; they, furthermore, will be ionized and picked-up by the solar wind.

Young et al. (2009) and Benedikt et al. (2020) modeled the outgassing and loss of a rock-vapor atmosphere with an average mass of $m_{av} \approx 34$ g/mol from a planetary embryo with the mass of $\approx 0.01 M_{Moon}$ by using a melt composition of an Enstatite chondrite without Fe. Figure 5 shows the outgassing rates as a function of pressure for an escaping silicate atmosphere of such a body. One can see that the equilibrium between outgassing and escape is reached at a pressure of $\approx 2 \times 10^{-8}$ bar and a hydrodynamic escape rate of $\approx 5.5 \times 10^{32}$ s$^{-1}$. It should also be noted that for such high escape rates such bodies can also lose noble gases like Ar and Ne, but as it was shown by Benedikt et al. (2020) and in Lammer et al. (2020b), the initial fractionation of $^{36}Ar/^{38}Ar$ and $^{20}Ne/^{22}Ne$ does not change significantly. From the results of these studies one can expect that the building blocks of proto-Mercury, Venus, and Earth, etc., were highly depleted in volatile and moderately volatile elements. This would indicate that growing protoplanets accreted significantly volatile depleted material, which is also in agreement with Sossi et al.

(2019), Lammer et al. (2020a, 2021) and Herbort et al. (2020) who showed that Earth's volatiles are the result of the accretion of smaller building blocks which experienced various levels of volatile losses.

Elemental data that have been collected from Venus (e.g., Morgan and Anders 1980; Basilevsky 1997), Earth (e.g., Lyubetskaya and Korenaga 2007; Arevalo et al. 2009), and Mars (e.g., Taylor 2013; Yoshizaki and McDonough 2020) are consistent with the before mentioned hypothesis. However, it would be expected that Mercury is less volatile compared to Venus, Earth and Mars due to its closer orbit around the Sun (e.g., Cameron et al., 1988) and if a giant impact (see Section 5.2), as suggested by some researchers, was involved (e.g., Smith 1979, Benz et al., 1988), the volatile depletion would be even enhanced (see Section 5.1.2). For example, Earth's Moon that originated by a giant impact–high temperature event (see Section 4.1) resulted in the volatile poorest body analyzed so far (e.g., Hartmann and Davis, 1975). However, from the MESSENGER X-ray and $\gamma$-ray spectrometer data and Earth-based observations of the planet's Na and K exosphere, it is now known that this is not the case and Mercury's crust/mantle is volatile-rich (e.g., Peplowski et al. 2011). As discussed in more details in Sect. 5.1.2, if one compares Mercury's K/Th, K/U or Cl/K surface ratios with the before mentioned terrestrial planets, one finds that these ratios are close to that of Mars but slightly higher (Peplowski et al. 2012; Evans et al. 2015; Nittler et al. 2018). Compared to Mercury, Venus' and Earth's K/Th ratios are $\approx 2.5 - 3.5$ times lower, respectively (Nittler et al. 2018).

Mercury is the most reduced terrestrial planet. From experiments it is known that at low oxygen fugacities, elements that are typically considered lithophile can become more siderophile (e.g., Chabot and Drake, 1999; Bouhifd et al., 2007; Mills et al., 2007; McCubbin et al., 2012). Because of this it is possible that the surface elements, as measured by MESSENGER's instruments, were modified due to Mercury's strongly reduced oxygen fugacity, which could have affected these elements through metal/silicate partitioning during the planet's core formation, as suggested by McCubbin et al. (2012). On the other hand, a later accretion of undepleted chondritic material, at the same time when Earth obtained its volatiles, may be another explanation for the puzzling innermost planet in the Solar System.

To summarize, the studies reviewed within this section indicate that the building blocks of terrestrial planets are depleted in volatile and moderately volatile elements as soon as they formed magma oceans or magma pools. These findings agree with Marty (2012) and Lammer et al. (2020b) who found that Earth accreted ≈ 0.95 $M_{Earth}$ from extremely depleted building blocks.

## 4. Origin and exosphere evolution over the Moon's history

### 4.1. Moon forming hypotheses

The Moon (see Table 1 for a parameter list) formed most likely between 50 to 200 Myr after the origin of the Solar System, with earlier ages being more likely as suggested by several different isotopic dating systems (see, e.g., Lock et al. 2020, for an extensive discussion). The Moon forming event itself created an Earth-Moon system with a few significant characteristics that have to be explained by any Lunar formation hypothesis. Besides the Pluto-Charon system, the Earth-Moon size ratio is the lowest in the whole Solar System, with the Moon having roughly 27% of the Earth's radius. On the other hand, the Lunar metal core only comprises ~1 – 2% of its total mass while the Earth's core makes up about 30% of the whole planet. Besides this difference in composition, the Moon is also significantly more depleted in volatile and moderately volatile elements, while its major elements and isotopic abundances show a striking similarity to the composition of the Earth (e.g., Canup et al. 2021). Other characteristics worth noting are the high total angular momentum of the Earth-Moon system with an initially rapidly spinning Earth, and evidence that the Moon formed hot and held a deep magma ocean.

*Table 1. Lunar parameters (from D. Williams, NASA planetary fact sheet).*

| Mass ($10^{24}$kg) | 0.073 | Rotation Period (hours) | 656 | Orbital Period (days) | 88 | Mean Temperature (C) | -20 |
|---|---|---|---|---|---|---|---|
| Diameter (km) | 3475 | Length of Day (hours) | 709 | Orbital Velocity (km/s) | 47 | Surface Pressure (bars) | 0 |
| Density (kg/m$^3$) | 3340 | Dist. from Sun ($10^6$ km) | 57.9 | Orbital Inclination (°) | 5.1 | Global Magnetic Field | No |
| Gravity (m/s$^2$) | 1.6 | Perihelion ($10^6$ km) | 46 | Orbital Eccentricity | 0.055 | Bond albedo | 0.11 |
| Escape Velocity (km/s) | 2.4 | Aphelion ($10^6$ km) | 69.8 | Obliquity to Orbit (°) | 6.7 | Visual geom. albedo | 0.12 |

Several different theories on the Lunar origin were published over the last decades that try to solve these characteristics. Early explanations included co-accretion, capture of a planetary embryo as well as disintegrating capture in which a planetesimal passed through the Roche-lobe of the Earth and re-accreted later, and fission from a rapidly spinning Earth (see, e.g., Wood 1986, for a discussion of these earlier theories). However, these were not able to account for most of the specific characteristics listed above. The co-accretion hypothesis, for instance, was not able to explain the high angular momentum of the system and the small Lunar iron core, while the capture from a different orbital position seems to be difficult to reconcile with the similar isotopic composition between the Earth and the Moon. Another model, however, that emerged in the 1970s was the giant impact model (e.g., Hartmann and Davis 1975; Cameron and Ward 1976) which seemed to be able to account for the small Lunar iron core, the high angular momentum, the similar oxygen isotopic composition, and the deep magma ocean of the Moon (Wood 1986). While older impact models (e.g., Benz et al. 1986, 1987, 1989; Cameron and Benz 1991) were able to reproduce the iron-poor core of the Moon, it was later found that only quite a narrow range of grazing impact scenarios with a Mars-mass embryo colliding approximately with mutual escape velocity was able to additionally reproduce the angular momentum of the system, as well as the Lunar size (Canup and Asphaug 2001; Canup 2004a,b, 2008).

The canonical (Figure 6, left panel) and other giant impact scenarios, however, predicted that the Moon should have primarily been formed by material from the impactor by typically 70 – 80% (Canup et al. 2021), thereby leading to a distinct Lunar isotopic composition, even though different studies showed a significant isotopical similarity between the Earth and the

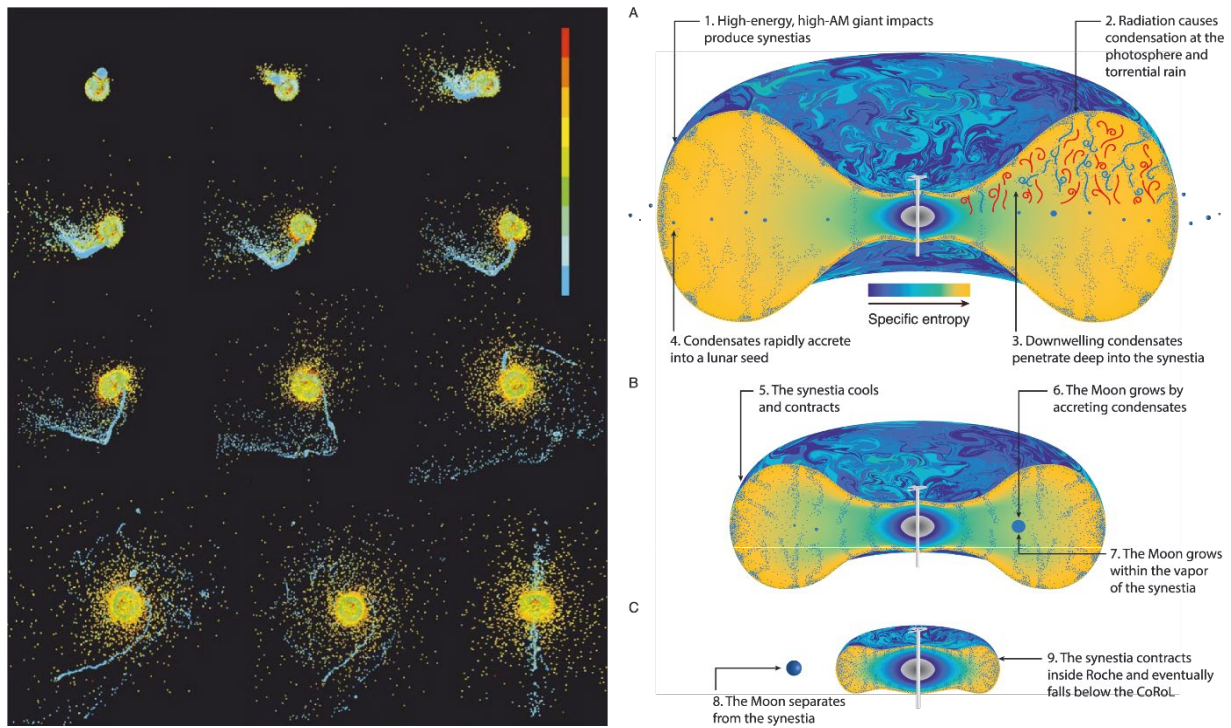

*Figure 6. The canonical giant Moon forming impact (left) by Canup and Asphaug (2001) vs the new concept of a synestia (right) by Lock and Stewart (2017); see text for further information. The left figure covers the first 23 hours from the impact, the color scale illustrates the thermal state of the matter with blue and dark green being condensed matter. AM in the right figure stands for angular momentum, CoRoL for corotation limit. Left figure from Canup and Asphaug (2001); right figure from Lock et al. (2018).*

Moon (e.g., Zhang et al. 2012; Kruijer et al. 2015; Touboul et al. 2015). If the impacting embryo was similar to Mars with regard to its oxygen isotopic composition, less than 5% of this body should have been re-accreted to the Moon while >95% should have been delivered from the Earth's mantle (e.g., Canup 2012; Canup et al. 2021). However, it is yet unclear how the composition of the impactor could have looked like. While some argue that it might have been comparable to Mars (e.g., Pahlevan and Stevenson 2007), other studies point towards an Earth-like composition (e.g, Dauphas et al. 2014; Mastrobuono-Battisti et al. 2015; Dauphas 2017; see also Canup et al. 2021 for a review of this topic). In case that the impactor differed isotopically from the Earth, equilibration (e.g., Pahlevan and Stevenson 2007; Lock et al. 2018) must have taken place, a process through which the isotopic signature of the impactor gets erased by vapor mixing between the post-impact planet and the debris disk.

In more recent years, it was shown that the canonical impact scenarios can be extended by a broader range of impact parameter space that can explain most of the above Earth-Moon system characteristics. A hit-and-run collision, for instance, with an increased impact velocity and steeper impact angle was proposed by Reufer et al. (2012). These simulations result in disks that are composed only by about 40 – 60% from the impactor but angular momentums that are about 30 – 40% too high, meaning that some mechanisms for losing angular momentum has to act afterwards. Multiple giant impacts (Rufu et al. 2017) and an impact onto a magma ocean dominated proto-Earth (Hosono et al. 2019) have also been shown to potentially be able to create the Moon. In addition, it was demonstrated that the angular momentum could have also been significantly altered after the Moon forming impact through lunar tidal evolution, offering more energetic events than previously thought (e.g., Canup 2012; Cuk and Stewart 2012; Wisdom and Tian 2012; Cuk et al. 2016; Tian et al. 2017). Kokubo and Genda (2010) even found that low velocity impacts, as in the canonical Moon forming scenario, are relatively rare, so that high energetic events might be favorable.

Such high energy events might commonly form so-called synestias, as was found by Lock and Stewart (2017) and Lock et al. (2018). Synestias (Figure 6, right panel) are partially vaporized and rapidly rotating large biconcave disk-shaped objects with an angular momentum that exceeds the corotation limit. Lock and Stewart (2017) and Lock et al. (2018) proposed that equilibration within the high entropy regions of the synestia occurs due to turbulent mixing, which could ultimately lead to a Moon with Earth-like composition. However, the synestia is differentiated into iron and silicate layers and about 75% of the silicate mass is comprised by a low-entropy inner layer followed by a 25% high-entropy outer layer. While the outer layer is mixed well, there is only little mixing between the different layers (Lock et al. 2018). The Moon itself forms in the outer region of the synestia, where droplets condense onto a Lunar seed that was accreted from large debris of the impact. Lock et al. (2018) proposed that radial transport of silicate rain droplets due to gas drag might be able to mix and homogenize the different regions, but only up to about 50% of the emerging body. Intra-impact mixing will, thus, still be needed, in case that the different bodies have significant diverging isotopic compositions (Lock et al. 2020). However, synestias generally allow a greater mixing than the canonical impact theory (see, e.g., Fig. 9 in Lock et al. 2020, where synestia simulations are compared with canonical impact scenarios for various impactors with different isotopic compositions). While synestia simulations (Lock et al. 2018; Lock and Stewart 2019) can successfully mix bodies with a difference in $^{17}$O of up to 0.3‰, canonical impacts (Canpu 2004, 2008) were only successful for $\Delta^{17}$O ~ 0.01‰ (Lock et al. 2020).

It finally has to be noted that, as suggested by Lock and Stewart (2017), previously investigated high energy impacts (e.g., Canup 2012; Cuk and Stewart 2012) are also likely to produce synestias, but this was not yet known at earlier times. This is also exemplified by a similarly successful mixing of planetary bodies with up to $\Delta^{17}O \sim 0.3‰$ within the simulations of Cuk and Stewart (2012) than for synestias (Lock et al. 2020). Lock and Stewart (2017) further suggest that almost all planets transition through a synestia at least once during their accretion. However, as stated above, the impactor might nevertheless have been isotopically similar to proto-Earth to explain the similarities between the Earth and the Moon. Such a similarity was recently also suggested by Nielsen et al. (2021) who found that the vanadium isotopic composition of the Moon is offset from the bulk silicate Earth's value by $0.18 \pm 0.04$ parts per thousand towards the chondritic value. These authors propose that this isotopic fractionation resulted from terrestrial core formation prior to the giant impact which further suggests that no post-giant impact equilibration through a synestia or other alternative impact geometries could have taken place. According to Nielsen et al. (2021), this result also implies evidence for the canonical giant impact scenario and for a common isotopic reservoir in the inner Solar System out of which the impactor and proto-Earth must have accreted. Table 2 summarizes the various Moon-forming hypotheses and their physical plausibility, discussed before.

*Table 2. Summary of Moon-forming hypotheses and their physical plausibility.*

|  | Earth-Moon size ratio | Small metal core | Depletion of volatile elements | Isotopic similarity to Earth | Angular momentum of Earth-Moon system | hot formation of the Moon | Physical plausibiliy |
|---|---|---|---|---|---|---|---|
| Co-accretion[1] | maybe | no | maybe | maybe | no | maybe | maybe |
| Capture of planetary embryo[1] | maybe | no | maybe | likely no | maybe | maybe | likely no |
| Disintegrative capture[1] | maybe | maybe | maybe | no | maybe | maybe | likely no |
| Fission from rapidly rotating Earth[1] | no | yes | maybe | maybe | no | yes | no |
| Canonical impact scenario[2,3,4,5,6] | yes | yes | yes | likely no | only narrow of scenarios | yes | yes |
| Hit-and-run collision[7] | yes | yes | yes | likely no | likely no | yes | yes |
| Multiple giant impacts[8] | yes | yes | yes | likely yes | yes | yes | yes |
| Impact onto magma ocean dominated proto-Earth[9] | yes | yes | yes | likely yes | yes | yes | yes |

| High energetic impact events, synestias[10,11,12,13,14,15] | yes | yes | yes | likely yes | yes | yes | yes |

[1]Wood (1986), [2]Hartmann and Davis (1975), [3]Cameron and Ward (1976), [4]Benz et al. (1986) [5]Cameron and Benz (1991), [6]Canup et al. (2021), [7]Reufer et al. (2012), [8]Rufu et al. (2017), [9]Hosono et al. (2019), [10]Canup (2012), [11]Cuk et al. (2016), [12]Kokubo and Genda (2010), [13]Lock and Stewart (2017), [14]Lock et al. (2018), [15]Nielsen et al. (2021).

One can see in Table 2, that, by our current understanding, multiple giant impacts, an impact onto a magma ocean proto-Earth or high energetic impact events might best reproduce the formation of the Moon. After the Moon might have formed by one of these catastrophes (or a potential combination of these) and after the magma ocean solidified, its surface was exposed to frequent impacts, a decreasing XUV flux over time and the solar wind over the whole history of the Solar System. Different processes, as described within Section 4.3, weathered the upper surface layers of the Moon which also modified its composition over time.

### 4.2. The Lunar water inventory

From various Apollo missions, pyroclastic glass beads indicate that their water content, which can be traced back to different eruptive events, is ≤ 100 ppm and in many cases approach the detection limits of the Secondary-Ion Mass Spectrometry (SIMS) and Fourier-Transform-Infrared spectroscopy (FTIR) detection limits (Hauri 2017). Saal et al. (2008) estimated that lunar magma has lost ≥ 90% of their pre-eruptive $H_2O$-budget via degassing.

This is in agreement with Hauri et al. (2011), who measured up to 1,200 ppm $H_2O$ in melt inclusions contained within olivine crystals from Apollo 17 orange glass samples and showed that this lunar magma contained F, Cl, and S in abundances similar as discovered in Earth's mid-ocean ridge basalts. The findings of magmatic $H_2O$ in volcanic lunar samples indicate that the origin of the Moon and its evolution must have had processes that allow for the accretion and retention of most volatiles that were present in the Solar System (Saal et al. 2008; Hauri et al. 2011, 2017; Füri et al. 2014; Chen et al. 2015).

Additional studies regarding water in lunar apatite that found analogous abundances to Earth's apatite, indicate further evidence that the Moon's interior contains significant amounts of magmatic water (e.g., Boyce et al. 2010; McCubbin et al. 2010; Anand et al. 2014; Barnes et al. 2014). From these findings, it can be expected that lunar magma likely contained much more water and other volatiles prior to eruption than we currently measure in the degassed glassy melt droplets (e.g., Hauri et al. 2017).

It is expected from Elkins-Tanton et al. (2011) that the lunar magma ocean had a depth of ≈ 500 – 1000 km and crystallized within ≈ 10 Myr if a stable crust existed throughout its crystallization phase. According to Solomon and Longhi (1977) and Meyer et al. (2010), the timescale for the solidification of the lunar magma ocean could be as long as 100 – 200 Myr if the loss of heat was greatly reduced. The difference of a factor of 10, demonstrates that the timescale for the lunar magma ocean solidification is highly sensitive to the details of heat loss through its surface. As discussed in Sect. 3, the mass of the Moon is too low to retain an atmosphere (Hauri et al. 2017, Benedikt et al. 2020), so that the Moon's magmatic surface will release volatiles into the vacuum of space. According to Elkins-Tanton and Grove (2011), the degassing efficiency could have been mitigated or modified by the presence of a surface crust. However, this magma ocean phase is most likely the time period during which the Moon could have gained its $H_2O$ between the formation event and full solidification from impacting hydrous meteorites (Elkins-Tanton et al. 2011; Hauri et al. 2015, 2017).

Besides the detection of water in Apollo samples, hydroxyl and/or $H_2O$ bearing minerals were recently also detected through combined observations of the Indian space mission Chandrayaan-1, and the flybys of NASA's Deep Impact and Cassini missions at the surface of the Moon within permanently shaded regions (PSR) at the poles (Clark 2009, Pieters et al. 2009, Sunshine et al. 2009; see also Figure 7a). This water may be produced due to solar wind interaction with the surface (Tucker et al. 2019, Jones et al. 2020; see also Figure 7b) and is subsequently trapped within PSR resulting in an abundance of 10 to 1000 ppm and locally even higher (Clark 2009). This was confirmed by Li et al (2018), who found direct evidence for surface-exposed water ice in the polar regions of the Moon with the Moon Mineralogy Mapper instrument, again on Chandrayaan-1. Observations with the NASA/DLR Stratospheric Observatory for Infrared Astronomy (SOFIA) detected molecular water to be present at the Lunar surface with an abundance of 100 to 400 ppm (Honniball et al. 2020). These authors propose that distribution of water might be the result of local geology and may be restricted to a small latitude range. Hayne et al. (2021) further investigated the surface distribution of potential cold traps and found a total area of ~40,000 km$^2$ out of which 60% is located in the south with the majority at latitudes >80°. This opens up a wider distribution of water ice at the Moon which might be accessible and an important resource for future human exploration (e.g., Hayne et al. 2021).

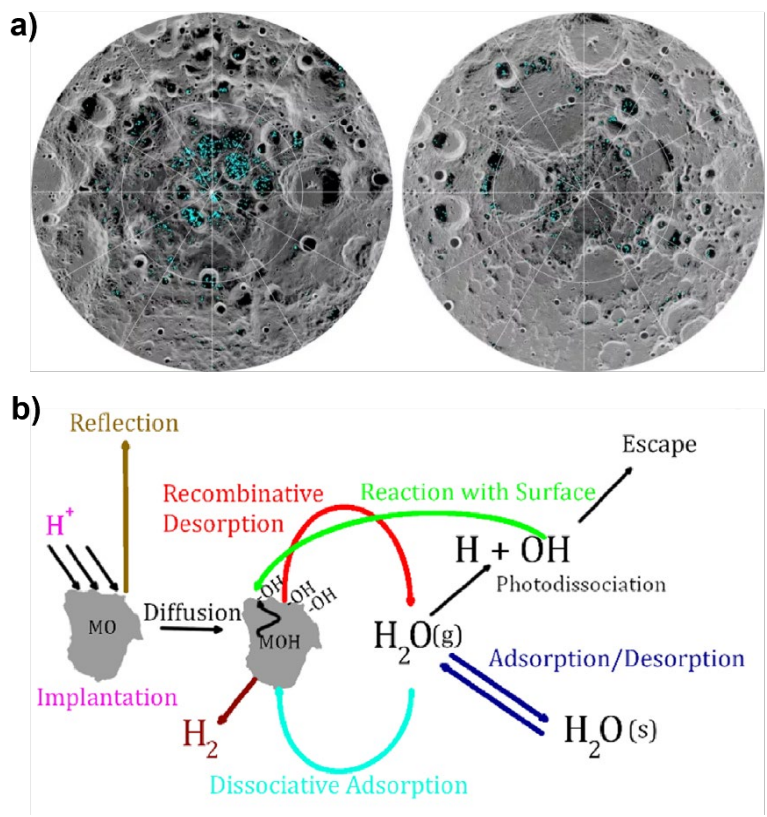

*Figure 7. a) Distribution of surface ice (cyan) within permanently shaded regions at the Moon's south (left) and north pole (right) as detected by NASA's Moon Mineralogy Mapper instrument (Image credit: NASA). b) Illustration of the kinetic scheme that describes the $H^+$ induced $H_2O$ cycle that is related to recombinative desorption, dissociative adsorption, adsorption, photodissociation, kinetic escape OH/H radical reaction, photo-stimulated desorption and desorption (after Jones et al., 2020).*

## 4.3. Large impacts, micrometeorites, and sputtering effects on the Lunar surface composition over time

The lunar surface is covered with large and small craters, which originated by impacts. The size of these craters ranging from microscopic sizes up to several hundreds of kilometers in diameters. The larger the impactor, the larger is the excavation depth and the volume of the melt. For instance, Pieters et al. (1993) suggest that the unusual surface composition at the lunar South Pole Aitken (SPA) basin is likely due to the exposure of mantle or lower crustal material. In addition, the ejecta of large impactors may also cover large areas on the Lunar surface. As an example, and contrary to Pieters et al. (1993), it was also suggested that the anomalous chemical composition of SPA could also be due to the accumulation of impact ejecta from the nearside Serenitatis basin, as suggested by (Wieczorek and Zuber 2001). The SPA basin with its diameter of 2900 km has a depth between 6.2 and 8.2 km and it is one of the largest known impact craters in the Solar System, and the largest, oldest, and deepest basin recognized on the Moon (Petro and Pieters 2004). Airless bodies such as the Moon lack generally erosional processes, with the possible exception of volcanism (Spudis 2015), and as a result, impact debris accumulate at the object's surface.

According to Wieczorek and Zuber (2001) the ejecta of large impacts like the one that formed the South Pole Aitken basin could cover a large fraction of the lunar surface or even the whole body. Addionally, more frequent but smaller impactors such as micrometeorites, stir and mix the upper surface layer until today. It has been estimated from the analysis of Lunar Reconnaissance Orbiter (LRO) satellite data that the top centimeter of the lunar surface is overturned every $\geq 80\,000$ years (Speyerer et al. 2016). However, one should keep in mind that the thickness of the ejecta of the early large impactors such as the one that formed the South Pole Aitken basin may exceed the thickness of the layer where micrometeorites affect.

Micro-meteoroid bombardment, for example, has shattered, fragmented, churned, and homogenised the Lunar surface (creating a loose layer of fine-grained regolith in the process), while solar wind plasma induces radiolysis, the chemical alteration of the topmost 1–3 atomic surface layers (Behrisch et al., 2007). In addition, both interaction processes result in particle release, through micro-meteoroid impact vaporization on one hand and through sputtering on the other hand. During their ballistic flight, these particles become part of the Moon's exosphere before they either escape or return to the surface. Since some species have a higher probability to return to the Lunar surface than others, these processes result in a chemical alteration (fractionation) of the Lunar surface over time. In this section we will review how the micro-meteoroid flux has changed in the past ~4 Gyr, how much this process and sputtering have contributed to the Lunar exosphere, and to what degree they have chemically altered the Lunar surface as a function of time.

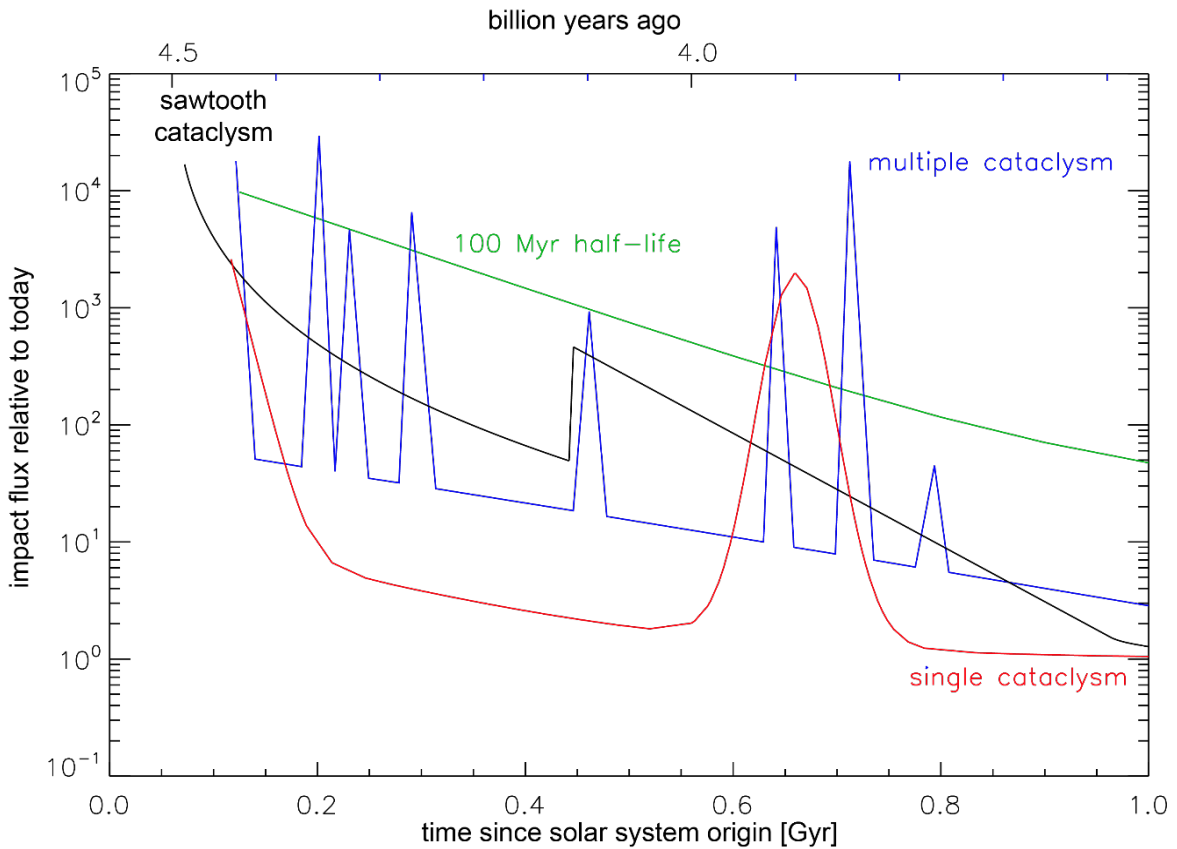

*Figure 8. Different models for the Lunar impact history for the first billion years (see text for further information).*

Micro-meteoroid impacts were much more frequent in the early days than they are today, though it is difficult to tell by how much exactly. Unlike large meteoroids that leave impact craters withstanding time, micro-meteoroids unfortunately leave no lasting observational evidence. The early meteoroid impact rates thus have to be estimated from the better-known flux of larger, crater and impact basin producing impactors. Several Lunar bombardment models have been proposed to explain the Lunar crater chronology. These models all fit the derived Lunar cratering chronologies well, but differ notably in shape, especially during the first Gyr, where, according to some models, the late heavy bombardment was said to have taken place.

Four models are mostly discussed by the scientific community: the 'smooth decline' model with half-lives of 50 Myr to 100 Myr (Hiesinger et al., 2012; Neukum et al., 2001; Wilhelms et al., 1987; Zahnle et al., 2007), the 'single cataclysm' model with a late heavy bombardment around 3.9 Ga ago (Ryder, 2002, 2003), the 'multiple cataclysm' models with several spikes in the first Gyr (Tera et al., 1974), and the sawtooth cataclysm (Morbidelli et al., 2012). Figure 8 shows representative samples of all of these models for the first Gyr. Whereas the 'single cataclysm' model also known as Lunar cataclysm or late heavy bombardment model was most commonly favoured in the past (Ryder, 2002, 2003), a paradigm-shift has led to the scientific community nowadays favouring the 'steady decrease' model (see e.g., Fernandes et al., 2013, Fritz et al., 2014; Hopkins and Mojzsis, 2015, Boehnke and Harrison 2016; Mojzsis et al. 2019). According to the 'steady decrease' model, the micro-meteoroid impact flux was roughly in the order of four magnitudes higher at the beginning of the Solar System than at present-day.

Today, micro-meteoroids and solar-wind ions chemically alter the Lunar surface to about equal measures (see, e.g., Killen et al. 2012). This was not the case at the beginning of the Moon's life, though. Monte-Carlo modelling reveals that, in fact, the contribution to the exosphere due to micro-meteoroid bombardment was more than 1'000 times more substantial than the contribution by sputtering in the beginning (Vorburger et al. 2020).

In addition, as mentioned above, different return rates for different species lead to a chemical fractionation of the Lunar surface with time. Figure 9 shows for the 12 most abundant species and uran the fraction of particles that return to the Lunar surface as a function of time. Whereas at the beginning the return rates were rather low (due to the high solar UV flux resulting in high ionisation losses), the return rates today are persistently higher, and range from ~10% (aluminium) to almost 100% (uranium). Out of the 13 species, uranium continuously exhibits the highest return rate (due to its high mass and low ionization rate), whereas aluminum continuously exhibits the lowest return rate (due to its very high ionization rate). Over time, this leads to an enrichment of uranium and a depletion of aluminum in the Lunar surface material. Another species of interest poses the moderately volatile element potassium, because the Moon's surface material is found to be anomalously low in K/U when compared to the terrestrial planets (see e.g., Taylor and Jakes 1974, Anderson 2005, Lucey 2006, Peplowski 2011, Taylor 2014).

As one can see from Figure 9, the return rate for potassium ranges from less than 1% to ~80% of the return rate of uranium. The persistently lower return rate of potassium when compared to uranium has thus led to a decrease in the Lunar K/U ratio with time and might explain the low K/U ratio in the Lunar surface material observed today.

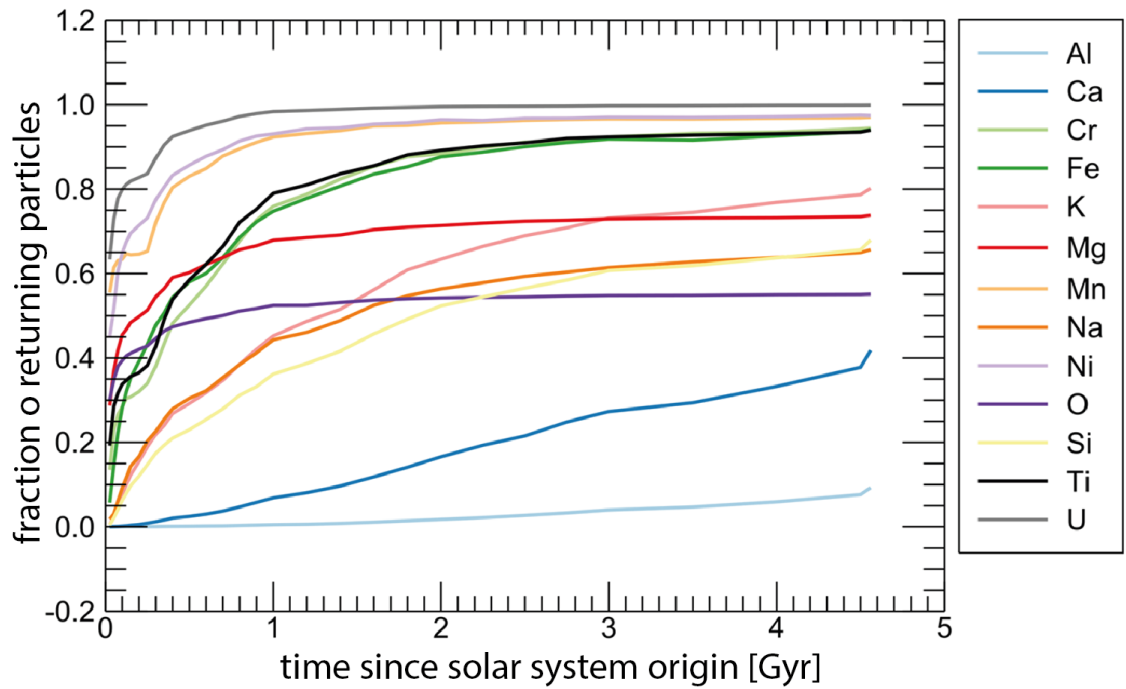

*Figure 9. The 12 most abundant species and uran as the fraction of particles that return to the Lunar surface as a function of time.*

Due to the species' different and varying return rates, the chemical composition of the Lunar surface as we observe it today does thus not reflect the Lunar surface composition right after the Moon's formation. These changes thus have to be factored in when trying to deduce the original Lunar surface composition from today's measurements and observations.

Saxena et al. (2019) modeled the effect of paleo space weather onto the Lunar surface and tried to reproduce the present-day potassium and sodium abundances in the regolith via sputtering of frequent CME impacts and argue that the present abundances can best be reproduced in case that the early Sun was a slow rotator (see also Section 2.1). Although space weather as modelled by Saxena et al. (2019), however, will modify the chemical composition of the upper surface layers, it may not be the most important process. Apollo drill cores of the Lunar surface indicate non-monotonic variations in the composition of the upper meter of regolith. This indicates that the Lunar surface has been overlain many times by impact **ejecta**. As pointed out above, one should note that the modification of the upper layers over evolutionary time scales is very complex. Interplanetary dust, cometary dust and larger meteorites permanently bombard the surfaces of airless bodies like the Moon or Mercury. This processes not only add external material to the upper regolith layers, but churns and vitrifies it. Some of these debris will escape but some will also be retained. Therefore, the results of Saxena et al. (2019) should be taken with care.

**4.4. Fingerprints of early Earth's atmosphere on the Lunar surface**

Early Earth's atmosphere might have been susceptible to strong atmospheric escape, particularly of hydrogen from dissociated $H_2O$ and $CH_4$ (e.g., Zahnle et al. 2019) and nitrogen (e.g., Tian et al. 2008; Lammer et al. 2018, 2019; Gebauer et al. 2020; Johnstone et al. 2021; Sproß et al. 2021) due to the increased XUV flux and plasma environment from the young Sun (see Section 2). It was already suggested some time ago by Marty et al. (2003) and Ozima et al. (2005) that nitrogen from such an "Earth Wind" together with light noble gases could have been implanted onto the Moon. This idea was backed-up by strong variations of N, He, Ne and Ar isotopic extra-Lunar implantations into the Lunar regolith by as much as 30% (Ozima et al. 2005). These authors argued, this cannot be explained due to solar wind implantation alone, since no fractionation process at the Sun or within the solar wind could explain such strong variation.

Ozima et al. (2005) calculated the escape from a terrestrial atmosphere that was directly exposed to the solar wind due to the intrinsic geomagnetic field not yet being present and found loss rates similar to present-day Venus for an atmosphere that was not significantly expanded compared to present-day. These authors further argued that the Earth wind could have only implanted terrestrial ions into the Lunar regolith until Earth did not possess a magnetosphere, since a magnetic field would have prevented strong atmospheric escape; such an effect could have, therefore, also been used as a tracer for the onset of the geomagnetic field. By mixing their escape rates with solar wind implantation, they found scenarios for which the present nitrogen and light noble gas isotopic ratios at the Lunar surface could have been explained. Ozima et al. (2005) further pointed out that this hypothesis could be tested by measurements on the far side of the Moon since the Earth wind could have been only implanted onto the near side.

However, as more recent studies have shown, Earth's atmosphere could have been significantly expanded in the past due to the strong XUV flux from the Sun (Tian et al. 2008; Lammer et al. 2018; Johnstone et al. 2021). Atmospheric escape might have, therefore, also been significant even though an intrinsic geomagnetic field was already present which is also supported by remnant magnetizations in zircons that date back even until 4.1 Ga (e.g., Tarduno et al. 2014, 2015). Polar outflow, for instance, might have even been much more significant in the past than at present-day (e.g., Kislyakova et al. 2020). This is also backed by the "missing xenon paradox" (e.g., Hébrad and Marty 2014; Zahnle et al. 2019), i.e., the isotopic fractiona-tion of heavy xenon isotopes which might best be explained by strong polar outflow of hydrogen ions that dragged away $Xe^+$ ions during the Archean eon (Zahnle et al. 2019).

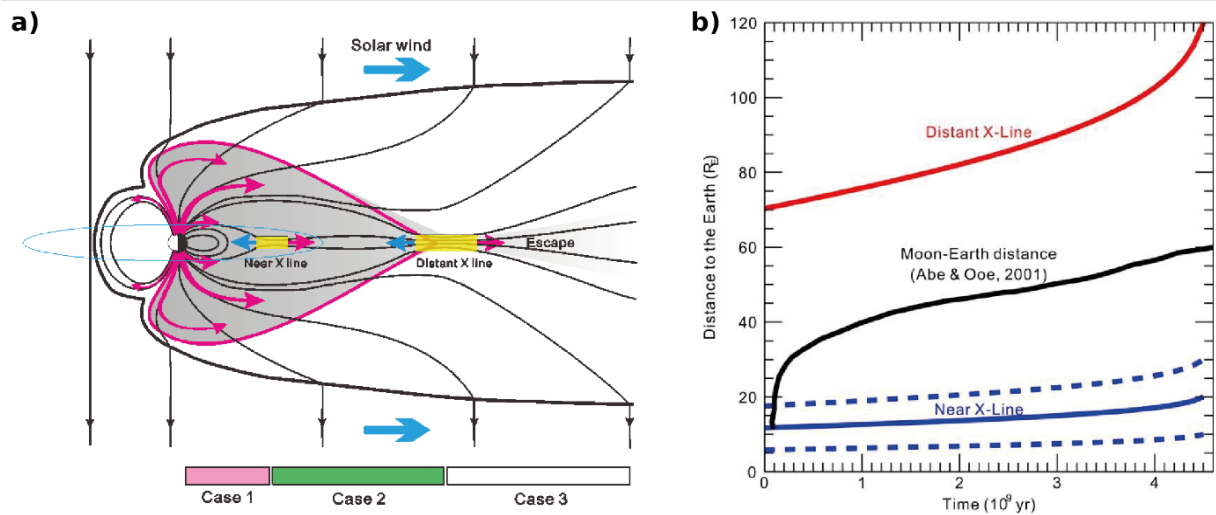

*Figure 10. The transport of Earth's atmospheric ions through the magnetotail onto the Lunar surface according to Wei et al. (2020). Panel a) shows the terrestrial magnetosphere, for Case 1 (pink), the Moon is located within the near X-line, for Case 2 (green), between the distant X-line and the near X-line, and for Case 3 (white) beyond the distant X-lineWei et al. (2020) point out that at present-day mainly Case 2 applies, and Case 1 only for 1% of the time. Panel b) illustrates the distances of the near and distant X-lines and the position of the Moon relative to the Earth over time. Figures from Wei et al. (2020).*

Part of this outflow will be transported along the field lines into the tail of the terrestrial magnetosphere. Some of them will then also be transported back towards the Earth through magnetic reconnection at the distant and near x-lines (see Figure 10), of which the distant one is always beyond the Moon's orbit and the near one depending on the space weather conditions. Since the Moon crosses the Earth's magnetotail, Wei et al. (2020) hypothesized that the outflowing charged particles from the Earth's atmosphere might not only have been implanted into the nearside of the Moon but that also the farside regolith should show a recognizable enrichment of, e.g., $^{15}N$. These authors, in addition, suggest that measuring these implants on both sides provide a way to study the evolution of the Earth's atmosphere and magnetosphere. They are concluding that within the last 3.5 Gyr after the Lunar volcanism and magnetic field shut down, the nearside of the Moon has been impacted by ~400 km/s atmospheric ions for most of this time, as well as the farside by ~100 km/s atmospheric ions that were deflected back from the distant X-line (see Figure 10). In addition, the farside is presently also impacted by about 1% of its time by 400 km/s atmospheric ions originating from the near X-line, even though this might change depending on the Earth's dipole field evolution (see Figure 10b). The farside, however, was not impacted by atmospheric ions at times when the terrestrial dipole field was too weak or at periods of geomagnetic reversals. Before 3.5 Ga, at the time when the Moon potentially yet had its own intrinsic magnetic field, Wei et al. (2020) suggest that atmospheric ions could have been implanted at the Lunar magnetic poles.

Besides nitrogen, oxygen could have also been transported to the Lunar surface. Several different studies (e.g., Hashizume and Chaussidon 2005, 2009; Ireland et al. 2006) found anomalous oxygen isotopic components in the Lunar soil, particularly the provenance of a $^{16}$O-poor component which remained weakly understood. However, already Ozima et al. (2007) proposed that oxygen ejected from the upper atmosphere and transported through the magnetotail onto the Lunar surface might explain this $^{16}$O-poor component. A more recent study by Terada et al. (2017) analyzed measurements of the Japanese Kaguya spacecraft and found that it observed a significant amount of $1 - 10$ keV $O^+$ ions only when the Moon crossed the Earth's plasma sheet. The high energy that would allow penetration depths that would fit to the previous measurements within the Lunar regolith, together with $^{16}$O-poor mass-independent fractionation within the Earth's upper atmosphere (Thiemens 2006) let these authors conclude that this component was indeed implanted into the Lunar surface by the Earth wind with at least $2.6 \times 10^4 \text{ions cm}^{-2}\text{s}^{-1}$. Terada et al. (2017) further concluded, in agreement with Ozima et al. (2005) and Wei et al. (2020), that the Lunar surface might indeed be a window into the last few Gyr for the Earth's atmosphere.

Finally, Wang et al. (2021) suggest that the Earth wind partially contributes to Lunar surface hydration, specifically to the OH/$H_2O$ abundance at the Lunar poles. When the Moon crosses the terrestrial magnetosphere, OH/$H_2O$ production, which is normally triggered by the proton flux from the solar wind, was observed to not decrease, even though the proton flux at the relevant energy range of 1 keV is by two orders of magnitude lower within the magnetotail than in the solar wind. However, Wang et al. (2021) found that other energy ranges (below 325 eV, and above 4 keV) that are more efficient within the magnetotail than outside together with heavy ions from the Earth wind such as $N^+$ and $O^+$ are contributing to the production of OH/$H_2O$, thereby compensating the lower proton flux at 1 keV. $O^+$ ions, for instance, were measured with the Kaguya spacecraft to show an ion flux at the lunar surface when encountering the terrestrial plasma sheet of $\sim 7 \times 10^{-19}$ g cm$^{-2}$s$^{-1}$ (Terada et al. 2017). The present-day micrometeoroite flux, for comparison, is by about 3 orders of magnitude higher, i.e., $\sim 7 \times 10^{-16}$ g cm$^{-2}$s$^{-1}$.

As one can see from this discussion, future Lunar missions that sample the far- and the near side but also spectroscopic missions might be able to reveal not only information about the Lunar surface and exosphere but also about the history of the Earth's magnetosphere and atmosphere.

### 4.5. Noble gas isotopes on the Lunar surface: archive of the Solar wind

No indigenous noble gas component has been unambiguously identified in Lunar rocks (Marty et al., 2003). The composition of noble gases in the Lunar exosphere (largely inferred from studies of gas trapped in Lunar regolith samples) indicates that some of them might be dominated by a solar wind source (e.g. Hodges Jr. and Hoffman, 1975; Wieler et al., 1996), some probably arise from the interior of the Moon or other external sources, like comets. Internal sources are supported by observations of episodic outgassing of radon (Gorenstein et al., 1974a; 1974b; Hodges Jr., 1973), and have been reviewed by Lawson et al. (2005).

Solar wind impinging on the Lunar surface might be a direct contributor of volatile species in the Lunar exosphere. Because the solar wind impinges on the Lunar surface with energies of about 1 keV / nuc H, He and other solar wind species are absorbed in the surface material (the regolith grains, rocks, …), are trapped, and accumulate in the regolith grains. Since the noble gases do not chemical bind within the regolith grains, a fraction of the noble gases is subsequently released via diffusion to the surface to become part of the Lunar exosphere (e.g., Hinton and Taeusch, 1964; Johnson, 1971; Hodges, 1973). Very efficient retention of implanted H and He has been shown long ago, however with prolonged ion irradiation saturation of the implantation occurs (Lord, 1968).

Let us consider the noble gases that are implanted into the Lunar soil first. Assuming that the Lunar soils are saturated with noble gas atoms (Schultz et al., 1978) one obtains an equilibrium between the flux of implanted solar wind noble gas ions $f_{\text{SW,i}}$ and the flux of released noble gas atoms $f_{\text{rel,i}}$ of species $i$ from the soil by diffusion, i.e. $f_{\text{rel,i}} = f_{\text{SW,i}}$. Once the noble gases are released into the exosphere they stay there because they do not chemically bind to the surface, and they become permanent gases of the exosphere. The loss fluxes from the exosphere are Jeans escape $f_{\text{esc}}$ and photo-ionisation $f_{\text{ion}}$. These loss fluxes and loss fractions from the exosphere can be calculated, e.g., with a Monte Carlo code (Wurz et al., 2012). For a stable population of the exosphere with noble gases there has to be an equilibrium between the input by the solar wind and the loss from the exosphere:

$$f_{SW,i} = f_{esc,i} + f_{ion,i}. \qquad (1)$$

Dividing by the released flux $f_{rel,i}$, we get

$$\frac{f_{SW,i}}{f_{rel,i}} = \frac{f_{esc,i}}{f_{rel,i}} + \frac{f_{ion,i}}{f_{rel,i}} = r_{esc,i} + r_{ion,i}, \qquad (2)$$

with $r$ being the loss fraction for the respective process and species. Thus, in equilibrium we get for the apparent flux of noble gases released from the surface

$$f_{rel,i} = \frac{f_{SW,i}}{r_{esc,i} + r_{ion,i}}, \qquad (3)$$

which is higher than the flux by diffusion from the regolith into the exosphere because of the accumulation of noble gases in the exosphere. The density of species $i$ in the exosphere at the surface and the column density are then

$$n_i(0) = f_{rel,i}\sqrt{\frac{8\,k_B T}{\pi\,m_i}} \quad \text{and} \quad N_{C,i} = \int_0^\infty n_i(r)dr. \qquad (4)$$

A possible additional loss process might be cold trapping of some noble gases in the permanently shadowed craters near the poles (e.g. Hodges 1980), which is not included in this calculation. Since the exospheric loss rates are small the noble gases are enriched in the Lunar exosphere until the flux of escaping particles matches the influx by the solar wind. Table shows a calculation of the amount of noble gases in the Lunar exosphere based on a solar wind ion flux of

$$f_{SW_i} = v_{SW} n_{SW_i} \approx 400 \cdot 10^3 \frac{m}{s} \times 8 \cdot 10^6 \text{ m}^{-3} = 3.2 \cdot 10^{12} \text{m}^{-2}\text{s}^{-1}.$$

Table also shows the literature data for the abundance of noble gases in the Lunar exosphere and Figure 11 shows the resulting density profiles for the noble gases resulting from the solar wind implantation into the soil.

For $^4$He the model results in about a factor 5 less than the observed value, but origin of the Lunar $^4$He is mostly from radioactive decay. The contribution of solar wind helium to the Lunar He exosphere is small, with a calculated surface density of $n_0 = 540$ cm$^{-3}$ at the sub-solar point compared to the Apollo measurement of 2'000 cm$^{-3}$ on the dayside (Heiken et al., 1991).

The range for the He exosphere density at the surface given in Table is because it is known already from the Apollo missions that there is a diurnal variation in the densities of $^{40}$Ar and $^{4}$He, and likely by the other volatile species (Stern, 1999; Benna et al., 2015). There is a similar situation for argon as is for helium, the $^{40}$Ar is from radioactive decay of $^{40}$K. The $^{40}$Ar density from the Apollo measurements at the surface is $n_0$ = 40'000 cm$^{-3}$ at sunrise indicating some condensation on the night side (Stern, 1999). From recent LADEE measurements about $n_0$ = 8'000 cm$^{-3}$ at the subsolar meridian (Benna et al., 2015). Table and Figure 11 show the combined $^{36}$Ar and $^{38}$Ar density of solar wind origin as $n_0$ = 374 cm$^{-3}$. The Apollo measurements gave a $^{40}$Ar:$^{36}$Ar ratio of approximately 10:1, implying a $^{36}$Ar surface density of about 800 cm$^{-3}$, which means that the solar wind contribution of argon to the Lunar exospheric argon inventory is at most half. For Ne we get reasonable agreement between the model and the observations, we calculate a surface density of 4900 cm$^{-3}$ that compares favourably with the measurements (4 – 10)·10$^3$ cm$^{-3}$ (Heiken et al., 1991). For Kr and Xe the model predictions are much lower than the existing upper limits from observations.

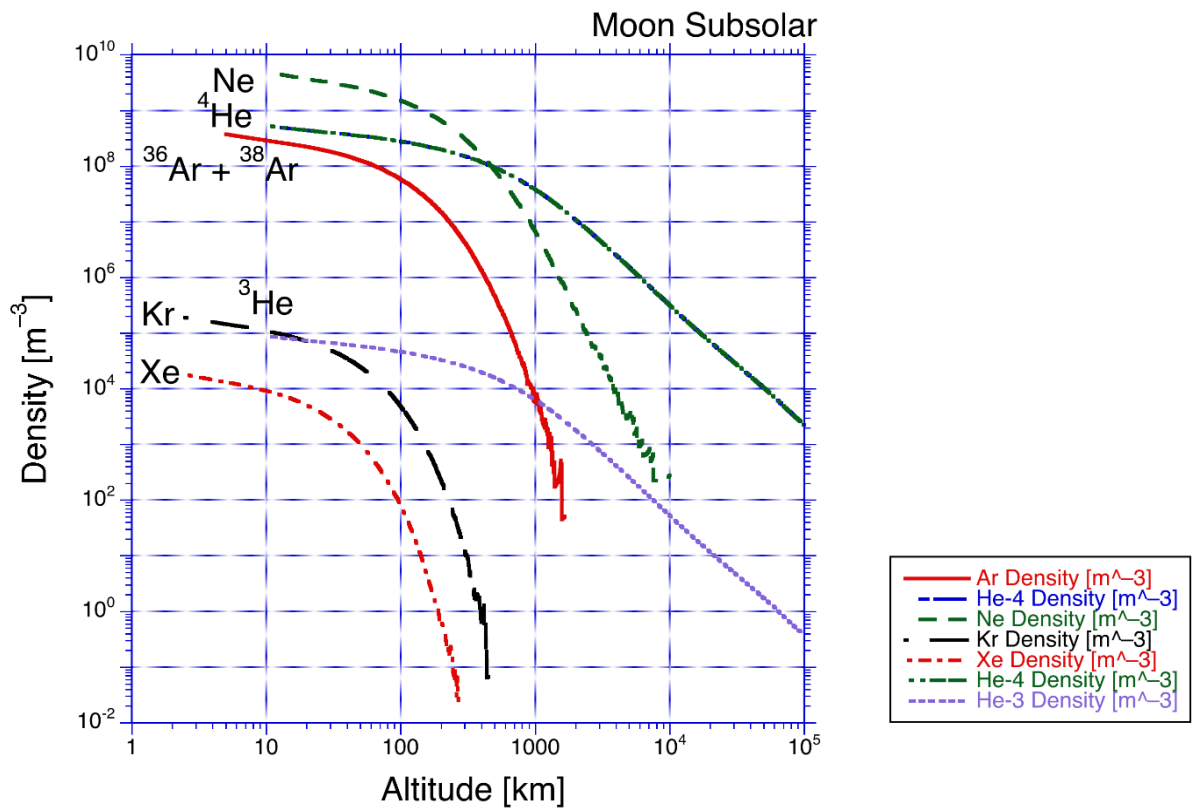

Figure 11. Density profiles for noble gases resulting from solar wind implantation assuming equilibrium between solar wind input and exospheric escape, based on Monte Carlo calculations (Wurz et al., 2007, 2012).

A similar calculation has been made for solar wind protons being implanted in the Lunar soil recently (Wurz et al., 2012), and assuming that the final released species is H$_2$. A density of $n_0 = 2100$ cm$^{-3}$ was predicted from this model, compared to the Apollo measurement of $(2.5 – 9.9) \cdot 10^3$ cm$^{-3}$ on the dayside (Heiken et al., 1991). The recent measurements from the LAMP UV spectrograph on the Lunar Reconnaissance Orbiter gave a value of $n_0 = 1200 \pm 400$ cm$^{-3}$ (Stern et al., 2013).

Table *3*: *Calculation of the amount of noble gases in the Lunar exosphere assuming equilibrium between solar wind input and exospheric escape, based on Monte Carlo calculations (Wurz et al., 2007, 2012).*

| Species | SW flux | Loss fractions | Model results, sub-solar point | Literature values |
|---|---|---|---|---|
| He | $f_{SW}(H) \times 0.03$ | $r_{esc,He} = 0.15$<br>$r_{ion,He} = 1.3 \cdot 10^{-3}$ | $n_0 = 5.44 \cdot 10^2$ cm$^{-3}$<br>$N_C = 1.72 \cdot 10^{10}$ cm$^{-2}$ | $n_0 = (2 – 40) \cdot 10^3$ cm$^{-3}$, day – night (Stern, 1999)<br><br>$N_C = 1 \cdot 10^{11}$ cm$^{-2}$ (Killen and Ip, 1999) |
| $^{36}$Ar, $^{38}$Ar | $f_{SW}(H) \times 3.55 \cdot 10^{-6}$ | $r_{esc,Ar} = 0$<br>$r_{ion,Ar} = 7.8 \cdot 10^{-5}$ | $n_0 = 3.74 \cdot 10^2$ cm$^{-3}$<br>$N_C = 1.90 \cdot 10^9$ cm$^{-2}$ | $n_0 = (0.3 – 30) \cdot 10^2$ cm$^{-3}$, day – night (Stern, 1999) |
| Ne | $f_{SW}(H) \times 7.41 \cdot 10^{-5}$ | $r_{esc,Ne} = 5 \cdot 10^{-6}$<br>$r_{ion,Ne} = 8.8 \cdot 10^{-5}$ | $n_0 = 4.22 \cdot 10^3$ cm$^{-3}$<br>$N_C = 4.54 \cdot 10^{10}$ cm$^{-2}$ | $n_0 = 2 \cdot 10^3$ cm$^{-3}$, $N_C = (4–20) \cdot 10^{10}$ cm$^{-2}$ (Killen and Ip, 1999)<br><br>$n_0 = 3 \cdot 10^3$ cm$^{-3}$, (Das et al., 2016) |
| Kr | $f_{SW}(H) \times 1.91 \cdot 10^{-9}$ | $r_{esc,Kr} = 0$<br>$r_{ion,Kr} = 1.26 \cdot 10^{-4}$ | $n_0 = 1.90 \cdot 10^{-1}$ cm$^{-3}$<br>$N_C = 4.25 \cdot 10^5$ cm$^{-2}$ | $n_0 < 2 \cdot 10^4$ cm$^{-3}$, (Stern, 1999) |
| Xe | $f_{SW}(H) \times 1.86 \cdot 10^{-10}$ | $r_{esc,Xe} = 0$<br>$r_{ion,Xe} = 1.60 \cdot 10^{-4}$ | $n_0 = 1.82 \cdot 10^{-2}$ cm$^{-3}$<br>$N_C = 2.82 \cdot 10^4$ cm$^{-2}$ | $n_0 < 3 \cdot 10^3$ cm$^{-3}$, (Stern, 1999) |

It finally has to be noted that the solar wind might also provide information on the isotopic composition of the solar atmosphere, the bulk Sun and hence the protoplanetary nebula, since it was found that early solar wind was trapped in solar gas rich soils and breccias on the Lunar surface (Anders and Grevesse, 1989; Pepin et al., 1999, Palma et al., 2002). $^{36}Ar/^{38}Ar$ ratio from such samples that did not originate from the lunar mantle but originated from Ar isotopes of the early solar wind show a divergence compared to the modern solar wind ratio, which indicates that the solar wind at some time in the past had a $^{36}Ar/^{38}Ar$ ratio that was above today's values (Becker et al., 1998).

# 5. Mercury vs. "exo-Mercurys"

## 5.1. Mercury: The innermost planet in the Solar System

### 5.1.1. The parameter space of Mercury

Every planet in the Solar System is unique in some respect, making all eight of them precious as archetypes for the studies of exoplanets. Among them, Mercury (see Table 2 for a list of Mercury's parameters) is probably the one with the most uniqueness (Solomon 2003); it is the closest to the Sun, the smallest in size and lightest in mass, but these are only the most obvious peculiarities.

*Table 2. Mercury's parameter list (from D. Williams, NASA planetary fact sheet).*

| Mass ($10^{24}$kg) | 0.33 | Rotation Period (hours) | 1407 | Orbital Period (days) | 88 | Mean Temperature (C) | 167 |
|---|---|---|---|---|---|---|---|
| Diameter (km) | 4879 | Length of Day (hours) | 4222 | Orbital Velocity (km/s) | 47 | Surface Pressure (bars) | 0 |
| Density (kg/m$^3$) | 5427 | Dist. from Sun ($10^6$ km) | 57.9 | Orbital Inclination (°) | 7 | Global Magnetic Field | Yes |
| Gravity (m/s$^2$) | 3.7 | Perihelion ($10^6$ km) | 46 | Orbital Eccentricity | 0.205 | Bond albedo | 0.088 |
| Escape Velocity (km/s) | 4.3 | Aphelion ($10^6$ km) | 69.8 | Obliquity to Orbit (°) | 0.034 | Visual geom. albedo | 0.142 |

In view of possible future direct observations of exoplanets, it is worth noting the way Mercury reflects light. The bond albedo and the geometric albedo are 0.09 and 0.14, respectively, which are very similar to the Moon values, and over certain terrains the albedo is greater than that of similar terrains on the Moon (see, e.g, Mallama 2017). Mercury's reflection depends on its uppermost layer, which is made of regolith, consisting of fragmental material derived from the impact of meteoroids over billions of years that covers more coherent bedrock, formed by processing of the older material (Eggleton 2001). With respect to the Moon, that of Mercury is more mature, with smaller grain sizes and larger proportions of glassy particles (Langevin, 1997), probably because of the differences in the respective size and energy distribution of impactors. Obvious differences in the solar wind, cosmic ray and UV fluxes also play a role when comparing Mercury to the Moon and other objects.

In terms of internal structure, and its importance as a paradigm of planetary evolution, Mercury occupies a very solitary position in the distribution of Solar System objects as well. Two facts are of uttermost importance in this respect: Mercury is, together with the Earth, the only inner planet with an **intrinsic** magnetic field (Ness, et al. 1975), and it also shares with the Earth the highest mean density in the Solar System.

The dipole moment of Mercury is about 330 nT. It is much smaller than the Earth's one, and it has barely the required strength to efficiently interact with the solar wind in a similar way (i.e., resulting in a bow shock, a magnetosheath, and a magnetosphere). Hence the magnetosphere of Mercury is a miniature version of the Earth's one, and, in comparison, Mercury occupies a much larger fraction of it. The observation of such a magnetic field or a magnetosphere is the indirect evidence (although not conclusive) that the planet has an electrically conducting fluid shell surrounding a solid inner core.

Mercury's mean density is almost equal to that of the Earth (5.44 g/cm$^3$ and 5.52 g/cm$^3$, respectively), which is the highest among all planets. However, the Earth's has a much larger internal pressure, which compresses the core and, once this effect is considered, it results that Mercury probably contains a much larger fraction of iron than the Earth and hence of all other planets. Therefore, the ways the masses and volumes are subdivided in core, mantle and crust are quite different. The core, made of liquid iron and other metals accounts for ~42% of Mercury's volume (i.e., >60% of its mass), and only 17% of the Earth's volume (Harder and Schubert 2001; for the inner core to be molted: Margot et al. 2007). The silicates that form the mantle fill a shell of only ≈ 400 – 450 km thickness (Rivoldini and van Hoolst, 2013; Wardinski et al. 2019), while at the Earth this is the most of the volume. Finally, the crust at Mercury, differently from the Earth, has no plate tectonics as the presence of impact craters suggests (Spudis 1998), so that the signs of the impacts on the surface of Mercury can virtually last forever, even though space weathering and surface processes surely play a role (Orsini et al. 2014).

In terms of being the easiest paradigm for close-in exoplanets, we must note that it is not only the closer to its parent star, but also it has the most eccentric (0.205) and inclined (7°) orbit of any other planet in the Solar System. This implies also that the ratio between solar radiation at perihelion and aphelion is the largest (2.3). Even if not the hottest, the surface of Mercury experiences the greatest thermal excursion among all the planets (from ~100 K at night to ~700 K at the subsolar point during part of its orbit). Such extreme thermal gradient should be accounted when considering the net mass loss from the planet (Kang et al. 2009).

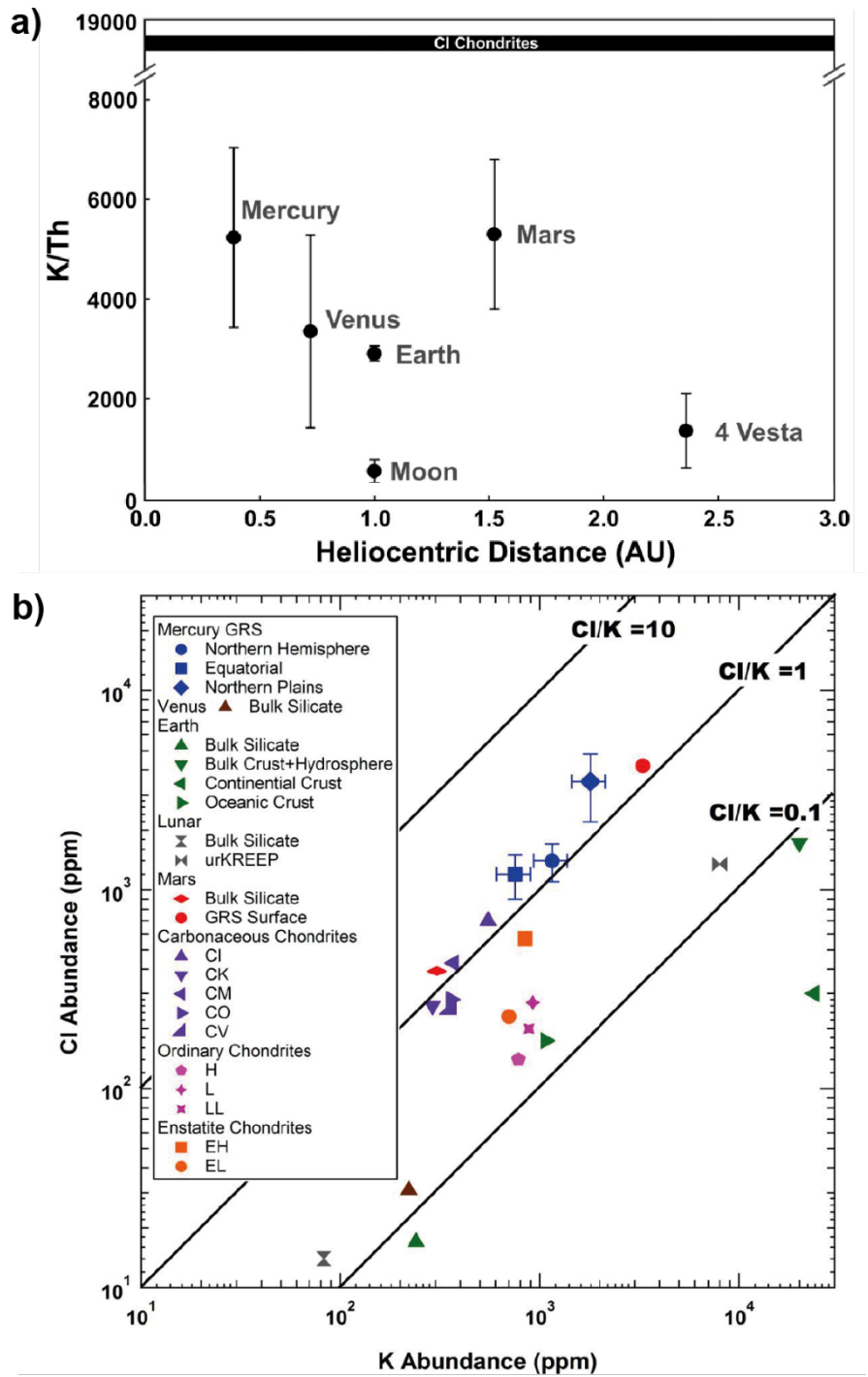

*Figure 12. a) K/Th for different Solar System bodies and for CI Chondrites (figure from McCubbin et al. 2012). b) Cl vs K for different Solar System bodies and chondrites (figure from Evans et al. (2015)*

Close-in extrasolar planets are easily locked to synchronous rotation with their host star because of the strong tidal interaction (Gladman et al. 1996). However, in the Solar System we don't have an example of a tidally-locked planet. Only moons experience complete tidal locking, and the closest, precious example among planets is the 3:2 spin-orbit resonance of Mercury (~59 vs ~88 days, respectively), which has likely been achieved through similar dissipative processes. At perihelion, however, Mercury experience (very slow) retrograde motion for a couple of weeks, so that apparently it behaves almost like a tidally locked planet. Since the obliquity of Mercury is close to 0 it does not experience seasons, but thanks to the 3:2 resonance the same hemisphere always faces the Sun at alternate perihelion passages, producing the so-called hot poles regions at 0° and 180° longitudes (in opposition to the cold poles at 90° and 270°), which may help in understanding the physics of "eyeball" exoplanets, which are tidally locked planets, for which tidal locking induces spatial features in the geography or composition of the planet resembling an eyeball (Angerhausen et al. 2013).

### 5.1.2. Formation hypotheses of Mercury

As already outlined in the previous section, Mercury has some peculiar physical and chemical characteristics that must be considered before constructing a valid formation hypothesis. Besides its high density and metal/silicate ratio, which was already known since Mariner 10 (1974-1975), the more recent MESSENGER mission (2008-2015) revealed and confirmed two additional and slightly surprising characteristics. In contrast to the high metal/silicate ratio that would suggest a strong depletion of volatile and moderately volatile elements if one assumes solar composition, MESSENGER showed that Mercury's surface is indeed quite volatile rich (e.g., Peplowski et al. 2012; Weider et al. 2015; Nittler et al. 2018). The planet has a very high ratio of K/Th (see Figure 12a) and a surface abundance of ~1300 ppm K (Peplowski et al. 2012) which is well above Earth's value but yet below chondritic, while the moderately volatile lithophile elements Si, Ca, Al, and Mg are approximately chondritic (Weider et al. 2015). In addition, the volatile elements Na (2.6-5 wt%) and Cl (1200-2500 ppm) are highly abundant and near-chondritic as well (Peplowski et al. 2015; Evans et al. 2015; see also Figure 12b), while the Cl/K ratio is comparable to Mars (Evans et al. 2015). S, another moderately volatile element, is also highly abundant on its surface with ~4 wt% (Nittler et al. 2011; Evans et al. 2012), a value that is about an order of magnitude higher than at Earth. Even substantial water-ice deposits were discovered within the permanently shadowed craters at the poles (Chabot et al. 2018).

Besides the high amounts of volatiles, MESSENGER also confirmed that Mercury is extremely reduced (McCubbin et al. 2012) and its surface highly depleted in Fe of any form (Murchie et al. 2015). The oxygen fugacity $f(O_2)$ of Mercury's interior was estimated by McCubbin et al. (2012) to be between IW-6.3 and IW-2.6, with the upper limit being relatively unlikely (Zolotov et al. 2013). Here, IW stands for the $f(O_2)$ of the standard equilibrium reaction buffer between iron metal and wüstite. Therefore, IW-6.3 describes an $f(O_2)$ that is $10^{-6.3}$ below IW. Such exceptionally low $f(O_2)$ suggests extremely reduced conditions during the accretion period of the Hermean protoplanet. Earth, for comparison, witnessed significantly less reduced conditions during formation (e.g., Frost et al. 2008); its modern mid-ocean ridge basalts, a tracer for early conditions, show an upper mantle $f(O_2)$ of IW+2 (Cottrell and Kelley 2011). The protocrust of Mars was estimated to have an $f(O_2)$ between IW-1 and IW+1 (e.g., Hirschmann and Withers 2008). Mercury is, therefore, one of the most reduced objects in the Solar System, and only enstatite chondrites and aubrites, as well as some Calcium-Aluminum-rich inclusions (CAIs) potentially show similarly low oxygen fugacities (e.g., Ebel and Stewart 2017). Any formation hypothesis of Mercury should, therefore, be able to explain, i.) the high iron to silicate ratio, ii) the surprisingly high abundance of volatiles, and iii) the extremely reduced nature of the planet. By now, however, none of the theories below either addresses all these characteristics or is even able to explain all of them sufficiently.

One of the oldest theories proposed to explain the high iron/silicate ratio of Mercury is the giant impact hypothesis (Benz et al. 1988). In this proposal, a smaller body impacts with high energy onto a bigger and differentiated proto-Mercury with a similar metal/silicate ratio as the Earth to strip off a substantial part of the silicate mantle while the two iron cores merge into one. Even though part of the stripped-away mantle will be reaccreted back onto the remaining protoplanet, Benz et al. (2007) found that high energy impacts might indeed be able to account for the anomalously high metal to silicate ratio at Mercury, and, therefore, also for its mean density. This is in contrast with a more recent numerical study by Carter et al. (2015) which suggests that reaccreting debris will be problematic and that giant impacts cannot explain such a high variability in the metal/silicate ratio. Furthermore, Stewart et al. (2013, 2016) found that such high energy impact events might even vaporize the entire mantle, and would likely recondense back onto the core, simulations out of which the concept of a synestia evolved (Lock

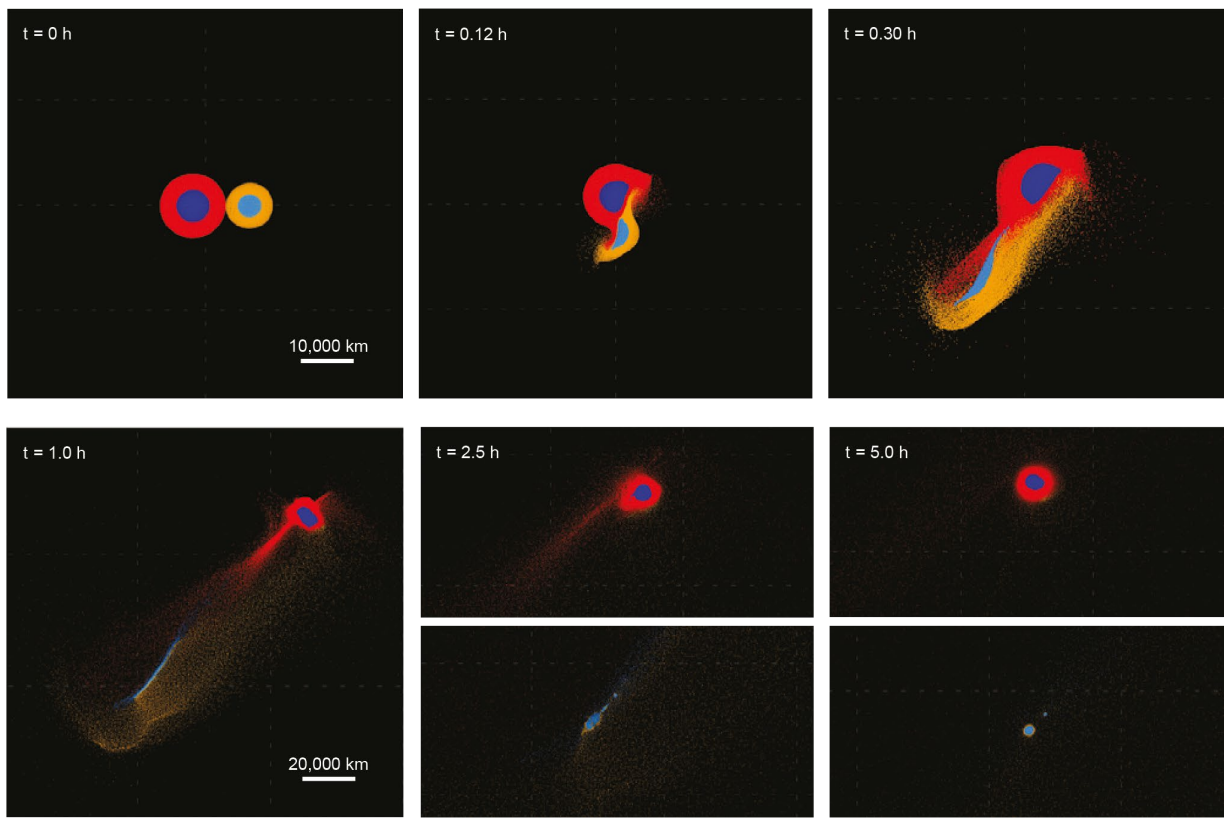

*Figure 13. Forming Mercury ($M_1$=4.52 $M_{Mercury}$) through a single hit and run collision with a more massive object ($M_2$=0.85 $M_{Earth}$) according to Asphaug and Reufer (2014). Tens of hours after the collision, the mantle of $M_2$ is dispersed and a metallic remnant remains. The bigger body might further accrete onto Earth and Venus. Figure from Asphaug and Reufer (2014)*

and Stewart 2017, and Lock et al. 2018), one of the theories for the Lunar Moon forming impact

(see Section 4.1). The giant impact model can, moreover, not account for the volatile rich mantle of Mercury. Whether such an origin might be reconcilable with the low oxidation state of the planet remains by now unknown.

Another theory related to the giant impact theory was developed by Asphaug and Reufer (2014), i.e., the theory of inefficient accretion. Their numerical hydrocode simulations show that Mercury could have been stripped off its mantle by one or several high-speed hit-and-run collisions (with less energy than in the original giant impact model) with a larger target planet (see Figure 13). While proto-Mercury loses most of its mantle, the silicates get then reaccreted onto the larger embryo which might have ended up accreting onto Venus or Earth. Asphaug and Reufer (2014) also illustrated that if Mercury and Mars are relicts out of 20 original planetary embryos that mainly formed the other terrestrial planets, then it is statistically likely that one of the remaining ends up with a stripped-off mantle. While these less energetic hit-and-run collisions might allow the survival of more volatiles than in the giant impact scenario, it is yet neither clear whether this would be sufficient, nor how likely it is that the remaining proto-Mercury will not accrete onto the larger body (e.g., Ebel and Stewart 2017). Helffrich et al. (2019), however, argue that a stripping of Mercury's mantle through impacts might at least be compatible with the high sulfur and low iron composition of its surface.

A more recent study by Chau et al. (2018) simulated different impact scenarios, i.e., the giant impact, hit-and-run, and a multiple collision scenario. They found that a single giant impact as well as the hit-and-run scenario require highly tuned impact parameters to achieve Mercury's mass and Fe/Si ratio, while a multiple-collision impact scenario escapes fine-tuning, and allows a volatile rich Hermean surface due to the relatively low impact energies. However, this scenario is constrained by timing several collisions within a relatively short timeframe, and by the volatile-rich composition of the planet's surface. Chau et al. (2018) finally conclude that it might be possible to form Mercury through collisions, but that it is difficult. The latter is in agreement with a study by Clement et al. (2019) who found that it is highly unlikely with a probability of less than 1% to form Mercury, and the Mercury-Venus dynamical spacing together with the correct terrestrial planets' orbital excitations through a set of collisions. In a subsequent work, however, Clement et al. (2021a) tested another hypothesis based on Volk and Gladman (2015) who suggested that Mercury might have been the lone surviving relic out of a cataclysm of several large planets inside the orbit of Venus. While Volk and Gladman (2015) assumed the cataclysmic planets to have masses similar or higher than Earth, Clement et al. (2021a) performed numerical simulations with a multiplanet system inside of Venus of Mars-sized mass each. They found that perturbations and collisions could have indeed resulted in a higher metal/silicate ratio inner planet. In a follow-up study, Clement et al. (2021b) were even able to recreate Mercury and its orbital spacing with Venus through collisions and a mass-depleted disk of ~0.1 – 0.25 $M_{Earth}$ inside 0.5 AU. However, these authors neither address the potentially volatile rich mantle of Mercury nor its reduced nature.

Finally, O'Neill and Palme (2008) suggested that collisional erosion, i.e., the stripping of mantle and crust material through smaller impacts, could have significantly altered the iron/silicate ratio of planetary bodies. Svetsov (2011) subsequently simulated whether such impacts could have also stripped away most of the mantle of Mercury and found that a proto-Mercury with an iron/silicate ratio comparable to the Earth could have been altered to the present Hermean composition through small impactor with an impact velocity of ~30 km/s that together added up to a mass bigger than present Mercury. However, as already written above, the more recent study by Clement et al. (2019) found that such impacts will likely not significantly alter

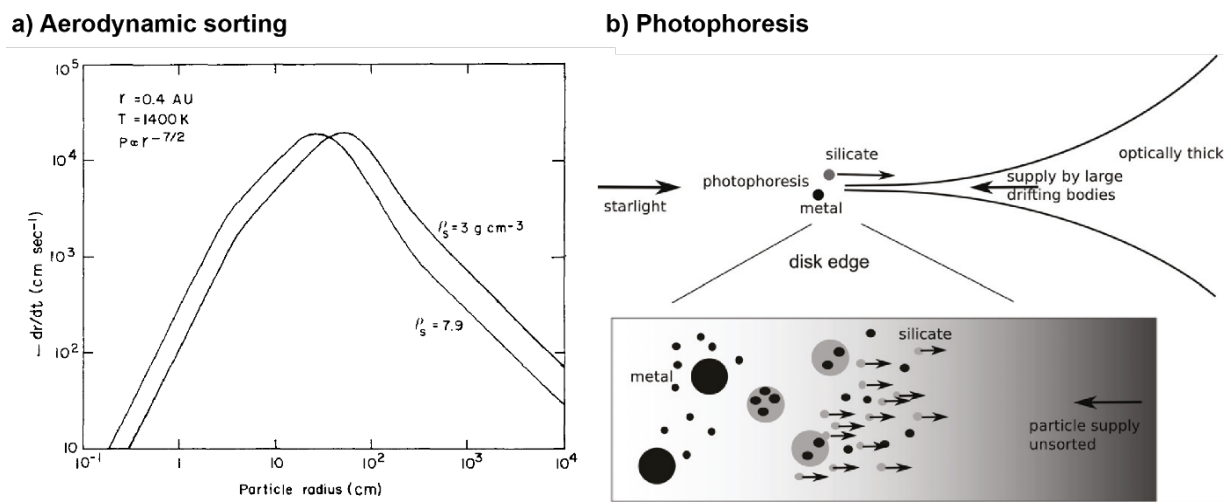

*Figure 14. While aerodynamic sorting (a), as proposed by Weidenschilling (1978), preferentially drags silicate boulders into the Sun (illustrated by the drag-induced radial velocity difference between the less dense "silicate" and denser "iron" in (a) as calculated by Weidenschilling 1978), photophoresis, as suggested by Wurm et al. (2013), pushes silicate grains outwards, away from the Sun. For the latter, the nebula gas already has to be dissipated, while aerodynamic sorting only works during the nebula phase. Both processes are potential contributors to Mercury's anomalous Fe/Si ratio. Left figure from Weidenschilling (1978); right from Wurm et al. (2013)*

the Fe/Si ratios of the proto-planets due to reaccretion.

Besides a collisional origin, there are two further main theories that were developed to explain Mercury's extraordinary characteristics, i.e., the post-accretion evaporation models and metal-silicate fractionation within the solar nebula. The post-accretion evaporation model was first suggested by Cameron (1985) who proposed that very high temperatures within the solar nebula led to the volatilization of the Hermean mantle which was later carried away by the strong solar wind of the young Sun. Fegley and Cameron (1987) calculated that about 70-80% from an initially chondritic mantle should have been evaporated to lead to the present metal/silicate ratio of Mercury. However, this model is neither in agreement with the volatile-rich remaining mantle as measured by MESSENGER, nor with the present knowledge about atmospheric escape, since the initial protoplanet with a mass more than twice as big as the present Mercury, would likely have been too massive to allow for such strong thermal or non-thermal escape processes at Mercury's orbit around the young Sun to evaporate a total amount greater than its present-day mass.

To form Mercury through metal-silicate fractionation within the solar nebula, several different processes were by now suggested. The first one was already put forward by Weidenschilling (1978) who suggested that aerodynamic sorting prior to accretion might result in a metal-rich inner nebula (Figure 14a). While the gas drag decays the orbit of lighter silicate-dominated boulders with sizes > 1 m faster than denser metal-rich bodies, the latter might get enriched within the feeding zone of proto-Mercury, thereby leading to an elevated Fe/Si ratio within the accreting proto-planet. Another fractionation process was explored by Wurm et al. (2013) who suggested that photophoresis (Figure 14b) might contribute to the anomalous metal to silicate ratio of Mercury. Photophoresis could fractionate millimeter-sized high thermal conductivity materials such as iron from lower thermal conductivity solids such as silicates since the latter would be preferentially pushed outwards into the colder and optically thin disk due to the thermal gradient that would build up within the silicate grains. This process would deplete the innermost part of the solar nebula from silicates, as pointed out by Wurm et al. (2013). For this process to work the feeding zone of Mercury should have been optically thin, which might have been the case as observations of cleared inner disks within extrasolar nebulae might suggest (e.g. Espaillat et al. 2014). Both processes – aerodynamic fractionation and photophoresis, however, could have counteracted each other, and, as pointed out by Ebel and Stewart (2017), a model addressing different chemical and physical factors influencing small solids within the inner disk might be needed to better justify the influence of these processes onto the Hermean iron/silicate ratio.

Another fractionation process within the solar nebula was put forward by Ebel and Alexander (2011) who found that the equilibrium condensation in systems enriched in C-enriched and O-depleted dust can form condensates with Fe/Si ratios reaching up to 50% of bulk Mercury at a temperature of 1650 K and a total pressure of 10 Pa, since Si remains in the vapor at such conditions. Ebel and Alexander (2011) further point out that such conditions can explain the formation of enstatite chondrite and aubrite parent body compositions and might also explain Mercury's anomalous composition since the planet could have formed from enstatite chondrite parent planetesimals. That Mercury's feeding zone might have indeed an enstatite-rich environment was also proposed by Pignatale et al. (2016) through simulating the vertical settling and radial drift of dust grains using a thermodynamic equilibrium model. However, origin mechanism proposed by Ebel and Alexander (2011) relies on the assumption that C-rich dust reached the inner Solar System, which is yet unknown (e.g., Peplowski et al. 2016; Vander Kaaden and McCubbin 2016).

A final fractionation process suggested so far was first put forward by Hubbard (2014) who proposed magnetic fractionation. They argue that dust grains rich in metallic iron can attract each other magnetically, and that magnetically induced collision speeds might be high enough to knock-off loosely bound silicates from the grains, thereby enriching and growing metal grains. Hubbard (2014) further argues that the magnetic field requirements for "magnetic erosion" are only fulfilled within the inner disk. This work motivated Kruss and Wurm (2018) to perform experiments on how magnetic fields up to 7 mT influence the aggregation and size of dust clusterings. They found that the cluster size depends on the strength of the magnetic field and the ratio between iron and quartz. If planetesimal formation is sensitive to the largest aggregates, Kruss and Wurm (2018) conclude, then planetesimals will preferentially grow iron-rich in the inner region of protoplanetary disks. In a follow-up study, Kruss and Wurm (2020) extended their experiments by adding pure quartz aggregates to the iron-rich aggregates. They found that their mechanism still works, but a certain fraction of iron-rich material has already to be present to trigger the magnetic enrichment. In case that there are more than 80% nonmagnetic aggregates, the mechanism will halt. A sufficient iron fraction, however, should have been present in the inner disk, as Krumm and Wurm (2020) argue. While these formation hypotheses can address the high iron to silicate ratio of Mercury and might partially also address its volatile rich mantle, none of these models can address the extremely reduced nature of Mercury.

Furthermore, it also has to be pointed out that there might be an additional way to at least reproduce Mercury's high Fe/Si ratio and its volatile rich mantle, which we will call "accretional evaporation", but a comprehensive study on the following idea is yet missing. As discussed in Section 3 it has been shown by Young et al. (2019) and Benedikt et al. (2020), small planetary embryos are significantly affected by the loss of volatile and moderately volatile elements. At early stages of planetary accretion and after these bodies are set free from the dissipating solar nebula, such elements will outgas from the magma ocean of the embryos and immediately be lost to space through hydrodynamic escape (see Section 3.1). This evaporation is strongest for smaller bodies close to the Sun and might lead to a significant loss of the silicate mantle in case that the magma ocean can be protracted over several Myr through frequent impacts, which should have been likely as was at least already shown for Mars (Maindl et al. 2015). When these planetary embryos grow, either through collisions with each other or through accretion of chondritic material, the escape will be significantly diminished after the escape regime changes from hydrodynamic to Jeans escape, a shift that might happen after the embryo reaches several Moon masses (Mercury's core holds about 4 Moon masses). After this change in the escape regime, the proto-planet could proceed accreting volatile rich material, for instance from carbonaceous chondrites, thereby building up a small volatile-rich mantle until accretion stops. However, no one modeled this potential formation theory in detail, and it also remains unclear so far, how such a process might have affected the redox state of Mercury.

*Table 4. Summary of different Mercury-formation hypotheses.*

|  | High metal/silcate ratio | Volatile rich surface | Extremely reduced nature of Mercury |
|---|---|---|---|
| Classical giant impact hypothesis[1,2,3,4,5] | only fine-tuned parameters | no | unknown |
| Inefficient accretion[6,7,8] | only fine-tuned parameters | only fine-tuned parameters | unknown |
| Multiple collision scenario[9,10] | unlikely | unlikely | unknown |
| Cataclysmic relic[11,12,13] | likely yes | unknown | unknown |
| Erosion through smaller impacts[14,15] | maybe | unknown | unknown |
| Post-accretion evaporation model[16,17] | likely no | no | unknown |
| Aerodynamic sorting[7,18] | maybe counteracting photophoresis | no | unknown |
| Photophoresis[7,19] | maybe counteracting aerodynamic sorting | no | unknown |

| Equilibrium condensation[20] | maybe - depending on whether C-rich dust reached inner Solar System | unknown | likely yes |
| Magnetic fractionation[21] | maybe | maybe | unknown |
| Accretional evaporation[22] | likely yes | yes | unknown |

[1]Benz et al. (1988), [2]Benz et al. (2007), [3]Carter et al. (2015), [4]Stewart et al. (2013), [5]Stewart et al. (2016), [6]Asphaug and Reufer (2014), [7]Ebel and Stewart (2017), [8]Helffrich et al. (2019), [9]Chau et al. (2018), [10]Clement et al. (2019), [11]Clement et al. (2021a), [12]Volk and Gladman (2015), [13]Clement et al. (2021b), [14]O'Neill and Palme (2008), [15]Svetsov (2011), [16]Cameron (1985), [17]Cameron (1987), [18]Weidenschilling (1978), [19]Wurm et al. (2013), [20]Pignatale et al. (2016), [21]Hubbard (2014), [22]hypothesis discussed in this review article.

Table 4 summarizes the before discussed formation hypotheses of Mercury. As one can see, there is at present no hypothesis that either sufficiently explains the planet's high metal/silcate ratio, its volatile rich surface and the extremely reduced nature within just one theory, or that is sufficiently well studied to already allow answering all of these three characteristics. It is, furthermore, certainly also likely that several of the mentioned hypotheses and processes acted together in the formation of the Solar System's innermost planet.

Further studies are, therefore, needed to resolve the puzzle of Mercury's formation. It might, however, be likely that several of the above-mentioned processes might have played out together to form the innermost planet of the Solar System.

**5.2. Close-in rocky exoplanets with high metal/silicate ratios**

Exoplanets are unique objects in planetary science, because they have a wide variety of characters regarding masses, sizes, orbital elements, and host star types; some are similar to the Solar System planets. The size of discovered exoplanets ranges from sub-Earth- to super-Jupiter-size, and they orbit around the various type of their host stars such as Sun-like stars and M dwarfs with semimajor axis ranging from one hundredths of Earth's orbit, as shown in Figure 15a.

After the initial discovery phase of extrasolar planets in which gas-giants were discovered, the focus has now shifted to planets with less than ten Earth-masses, corresponding to objects that are smaller than two, or three times the Earth. Was the mere discovery of objects the main achievement of the past years, we are now approaching the realm of studying their nature. It has now become clear that there are at least two different species of low-mass planets. One kind has extended atmospheres that are presumably hydrogen-dominated, others at first glance seem to be bare rocks.

Most of them are close to their host stars and discovered by transit measurement (more exactly, by the Kepler transiting exoplanet survey) because the measurement has a strong bias in favor of close-in planets. Also, their masses distribute over a range from Mars-mass to a few Jupiter-masses (see also Winn and Fabrycky 2015, for a review). In the last decade, the number of detected small exoplanets rapidly have increased, as shown in Figure 15b.

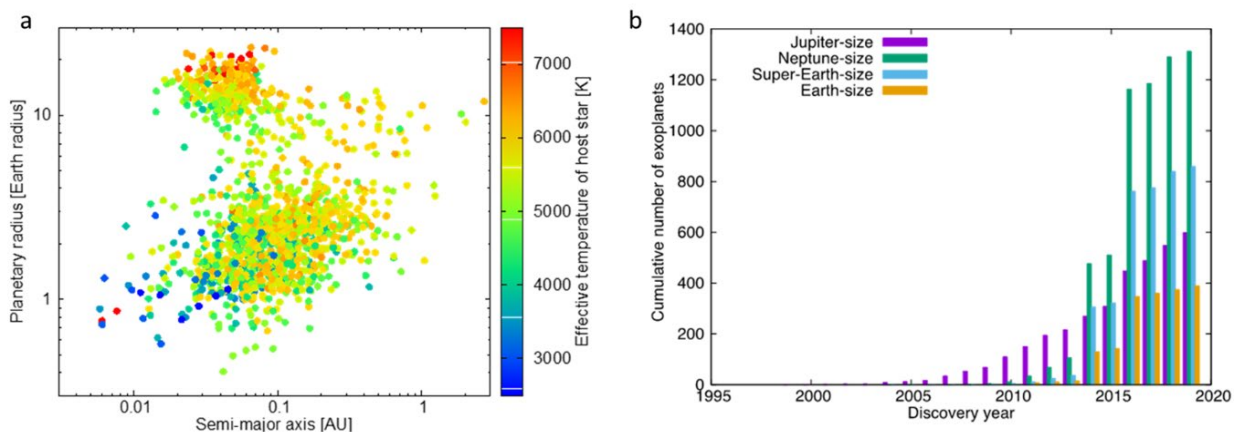

*Figure 15. Number and property of exoplanets detected until 2020. Panel (a) Two-dimensional distribution of discovered exoplanets in sizes and orbits smaller than 3 AU. The colors show the different effective temperatures of their host stars. Panel (b) Cumulative number of discovered exoplanets with different sizes; Jupiter-size (violet, $r_{pl} > 6R_{Earth}$), Neptune-size (green, $6R_{Earth} > r_{pl} > 2R_{Earth}$), super-Earth-size (light blue, $2R_{Earth} > r_{pl} > 1.25R_{Earth}$) and Earth-size (orange, $1.25R_{Earth} > r_{pl}$). The data has been taken from an open exoplanet catalog database (https://exoplanetarchive.ipac.caltech.edu, 27/7/2020).*

Up to now 96.8% of the known planets smaller than 3 $R_{Earth}$ orbit closer than Mercury in our Solar System (http://exoplanet.eu). The measurements of the radii of the planets alone, however, gives us only an incomplete picture; what we need are measurements of the radii and the masses, which would give us the density of the planets. Fridlund et al. (2020) argued that an accuracy of better than 15% for the masses and better than 5% are required to find out what the nature of the planets are (Figure 15a). Then, over 1000 exoplanets whose radii are less than

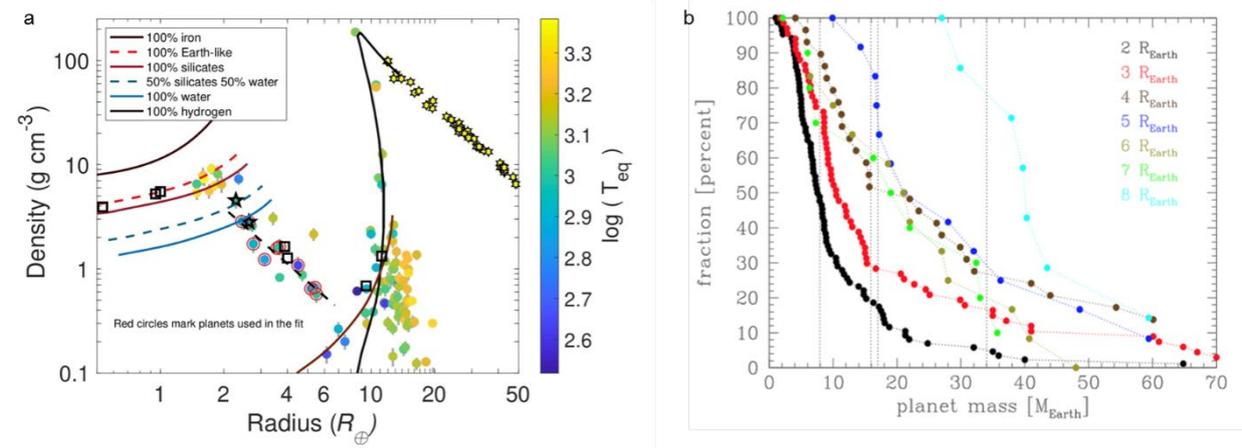

*Figure 16. Panel (a) Density–radius diagram of planets orbiting G-type host stars with masses determined within at least 15% and radii with an accuracy better than 5%. The two-star symbols represent TOI-763 b and c. The black squares are the Solar System planets, and the bright yellow star symbols at radii between 12 and 50 $R_{Earth}$ are red dwarf stars from Persson et al. (2018). The theoretical mass-radius curves are from Zeng et al. (2016) except the H-He model taken from Baraffe et al. (2003, 2008). The black dashed line represents a linear fit to the ice planets marked with red circles (Figure and description taken from Fridlund et al. 2020.). The colors correspond to the different equilibrium temperatures $T_{eq}$ of the planets. Panel (b) Cumulative frequency of planet masses for different planet radii between 2 and 8 $R_{Earth}$. For example, 50% of the planet with 2 $R_{Earth}$ have masses larger than 6 $M_{Earth}$. The dashed lines show the detection limits of 30 and 100 RV measurements with an instrument that delivers an accuracy of only 10 ms$^{-1}$ for planets with orbital periods of 1 and 10 days. From left to right: 100 RV-measurements and orbital period of one day, 30-RV measurements and orbital period of one day, 100 RV-measurements and orbital period of ten days and 30 RV-measurements and orbital period of ten days.*

2 $R_{Earth}$ have been discovered until 2020. Such exoplanets are called super-Earths or sometimes called Earth-like planets, although they could be super-Mercurys or sub-Neptunes, terrestrial planets that could not get rid of their primordial atmospheres after the disk dispersed (Lammer et al. 2020**b**).

These low mass planets are one of the most common group of exoplanets while the first discovered exoplanet 51 Pegasi b was a Jupiter-sized planet (Mayor and Queloz 1995). Fressin et al. (2013) estimated the occurrence rates of planets in different classes of planetary radii such as giant planets (6–22 $R_{Earth}$), large Neptunes (4–6 $R_{Earth}$), small Neptunes (2–4 $R_{Earth}$), super-Earths (1.25–2 $R_{Earth}$), and Earth-sized planets (0.8–1.25 $R_{Earth}$) using the data of planets and planet candidates measured by Kepler telescope. They showed that the occurrence rates of planets with orbital periods P ≤ 85 days per one star are 2 ± 0.22 % for giant planets, 1.97 ± 0.23 % for large Neptunes, 23.5 ± 1.6 % for small Neptunes, 23.0 ± 2.4 % for super-Earths and 18.4 ± 3.7 % for Earth-sized planets.

Also, Fulton et al. (2017, 2018) showed that the close-in small planet distribution splits into two classes consisting of planets with radii of < 1.5 $R$ and planets with radii of 2−3 $R_{Earth}$ in periods of < 100 days based on an occurrence rate analysis using the precise radius measurements from the California-Kepler Survey. Based on planetary formation and evolution models (e.g., Owen and Wu 2017, Jin and Mordasini 2018) which have reproduced the bimodal distribution of close-in exoplanets; most of the close-in small exoplanets which are included in the former class are likely bare rocky planets that have lost their primordial hydrogen-rich atmosphere due to photo-evaporation. Figure 17 shows the statistical mass-estimate of the known low

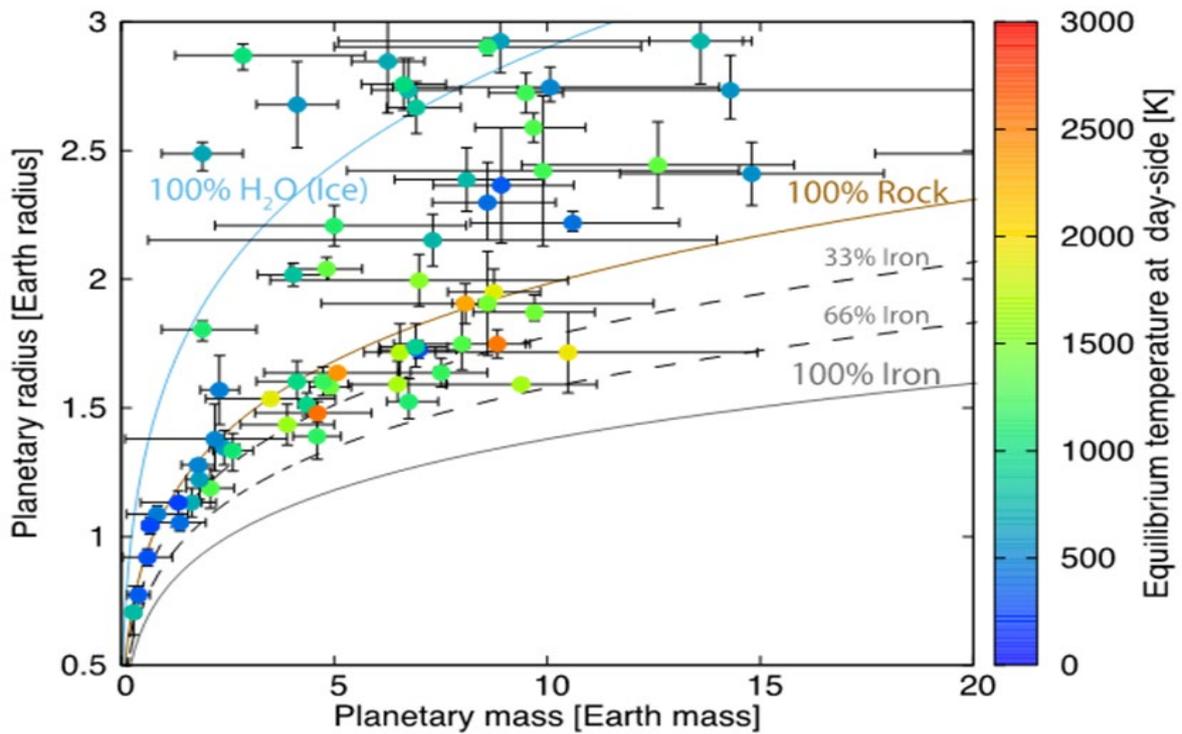

*Figure 17. The mass-radius diagram of exoplanets with radii of less than 3 $R_{Earth}$ and masses constrained not as only the upper or lower limit values. The bars show the error bar of measured values and the colors shows the different dayside equilibrium temperatures with zero planetary albedo. Solid lines show the mass and radius relationships of planets composed of only $H_2O$ (light blue), rock (brown) and iron (grey). The dashed lines also show the mass and radius relationships of rocky planets with iron fractions of 33% and 66%. The mass and radius relationships are calculated using the formula theoretically derived in Fortney et al. (2007). The data has been taken from an open exoplanet catalogue database (https://exoplanetarchive.ipac.caltech.edu, 27/7/2020).*

mass exoplanets.

**5.3. Stellar wind plasma interaction from Mercury to close-in rocky exoplanets**

If a rocky exoplanet orbits very close to its host star it looses all its volatiles and may form like airless bodies such as Mercury an exosphere or even atmosphere that consists of its minerals. The interaction of stellar wind plasma with the exospheres of such planets and possibly with their magnetospheric environment can be studied with numerical models that have been developed for Mercury or the Moon. Virtually any source of energy can give surface-bounded particles the necessary momentum to overcome the surface binding energy. At Mercury, and in general on any planetary surface, the energy source can be the planetary heat (TD, thermal desorption), the impact of solar wind or other plasma particles (IS, ion sputtering), the solar UV flux (PSD, photon-stimulated desorption), and micrometeorite impact vaporization (MMIV). Depending on the species (volatiles or refractories), local time, orbital position, and solar wind conditions, these different sources can have different relative importance (see Wurz et al. 2021, this issue). Because Na is the easiest element to be observed from Earth, and because of its intrinsic importance in the exosphere of Mercury, it is by far the most studied element but still the mechanisms of release are not completely understood and recent MESSENGER observations (Cassidy et al., 2016) seem to add complexity to previous established models (Mura et al, 2009, Leblanc et al., 2010). Once neutrals are released, they fill the exosphere travelling in ballistic orbits, and may be pushed in the anti-stellar direction by the radiation pressure of the star, forming a cometary-like tail or coma (Potter et al., 2002; see also Figure 18, Na population).

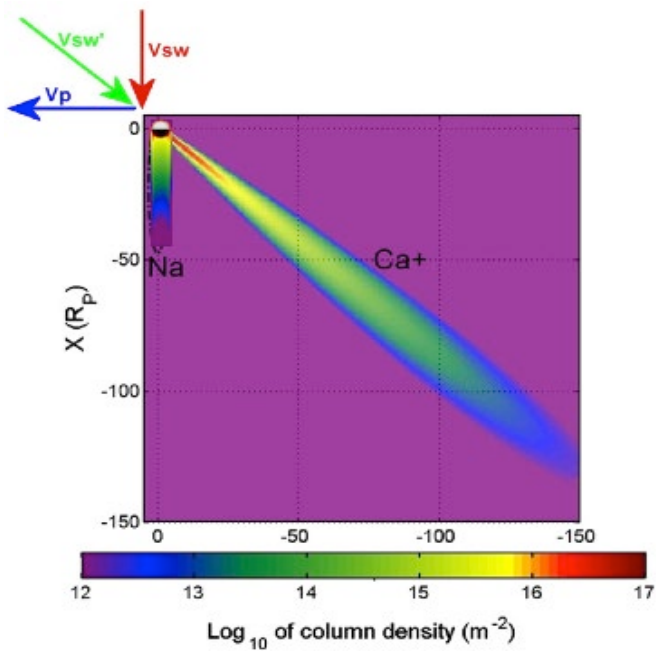

Figure 18. Simulated Na and Ca$^+$ tails for a close-in, super-Mercury exoplanet. The arrows indicate the velocity of the planet/reference frame ($V_p$), the velocity of the stellar wind in the inertial reference frame ($V_{sw}$) and in the non-inertial planetary reference frame ($V_{sw}^*$). The high Ca$^+$ tail inclination is exaggerated to match a case of high stellar wind aberration. The sodium tail is the feature on the left. (Adapted from From Mura et al, 2011).

Mercury's interaction with the solar wind is in some aspect similar to that of the Earth: it's magnetic field responses to the plasma flow resulting in a magnetosphere with reconnection events (Slavin et al., 2009); the solar wind can partially penetrate and charge-exchange with exospheric neutrals (Mura et al., 2003); exospheric particles are photoionized and can become pick-up ions in the solar wind (Sarantos et al., 2009). Other phenomena are more peculiar: Mercury's exosphere is partially generated by solar wind precipitating onto the polar cusps via ion sputtering (Sarantos et al., 2007); the solar wind also causes weathering of the surface (Domingue et al., 2014). Numerical models of the interaction of Mercury with the solar winds can be traced back as early as 2000 (Kabin et al., 2000; Ip and Kopp, 2002; Kallio and Janhunen, 2003a, 2003b, 2004; Massetti et al., 2003; Mura et al., 2003, 2005, 2006, Wurz et al., 2010, etc.).

Charge-exchange results in the formation of Energetic Neutral Atoms (ENA) that escape the planetary environment. While they are usually considered a diagnostic signal (Roelof et al., 1985; Orsini and Milillo, 1999) for orbiting instrumentation, in extreme conditions they may be detected over interplanetary distances and result in a substantial mass loss (Holmström et al., 2008; Kislyakova et al. 2014). In all respects, Mercury is the natural case study or paradigm for the investigation of the interaction of a close-in exoplanet with the stellar wind. We know that the interaction of the solar wind with a planet falls into 4 basic cases: planet with or without an atmosphere, and with or without an intrinsic magnetic field. The Earth, Mercury, Venus, and Mars can fill the 2×2 grid of all possibilities. In all 4 cases, the solar wind is known to impact to or to induce the formation of peculiar structures: a bow shock (a "fast shock" surface where the solar wind starts to be perturbed), a magnetopause (the tangential discontinuity surface where the passage of matter is not permitted, and the stellar wind flows around it), a magnetosheath (the region between these two surfaces), and a magnetosphere (the cavity inside the magnetopause). In the non-magnetized case, the conductivity of the ionosphere causes the formation of induced structures, even in extreme cases such as Mars where there is almost no atmosphere; the residual crustal magnetic field is capable of generating mini magnetospheres (Breus et al., 2005, see also Ness et al, 2000). The temporal variability of the solar wind parameters (pressure and IMF) is known to play a substantial role in changing the shape and size of the magnetosphere. In the simplest approach, one can calculate the stand-off point, where the pressures of the solar wind (left-hand side) and the magnetic field pressure (right-hand side) are balanced:

$$\rho v^2 = \frac{4}{2\mu_0}\left(\frac{M}{r_{so}^3}\right)^2,$$

where $\rho$ and $v$ are the stellar wind number density and bulk velocity, $r_{so}$ is the stand-off distance and $M$ is the planetary magnetic moment (Walker and Russel, 1985; Grießmeier et al., 2005).

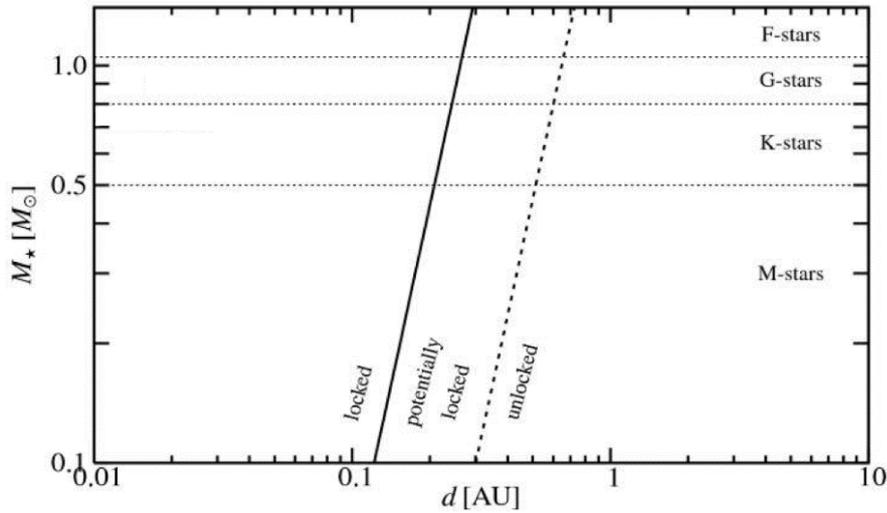

*Figure 19. Tidally locked (left) versus freely rotating (right) regime for "super-Earth" planets as a function of orbital distance d and mass M\* of the host star (Figure from Mura et al. 2011).*

The main problem is that, for almost all known exoplanets, none of these values are known, both on the stellar wind side and on the planetary side of the equation. The absence of an intrinsic magnetic field as that of Earth or stronger is sometimes associated with a tidally-locked planet (this has been recently questioned, see Reiners and Christensen 2010), so that, from the definition of the Love number k2 in Gladman et al. (1996) and their formula for the time of tidal locking, one can infer the preferred regime for exoplanets as a function of orbital distance and mass of the host star (Figure 19) which is, however, just an educated guess.

Mercury's magnetic field was measured for the first time by Mariner 10 (Connerney and Ness 1988). The updated value of the magnetic moment is $195 \pm 10$ nT $R_M^3$, which is sufficient to keep the magnetopause stagnation point at about 1.5 $R_M$ from planetary center (at the Earth, this is ~10 $R_E$). Because of the polarity of the intrinsic magnetic field, reconnection and plasma precipitation occur at the cusps, especially when the interplanetary magnetic field component $B_z$ is negative, while the $B_x$ component plays a role in determining north-south asymmetries (Sarantos et al., 2001). More recent MESSENGER observations (Anderson et al., 2011) refined our knowledge of the magnetic field and revealed an offset of the dipole of almost 500 kilometers northward of the geographic equator. Such shift could in principle lead to a larger cusp region in the south dayside hemisphere, in principle, even if the excess of induced sodium release has not been observed according to recent surveys (Milillo et al., 2021).

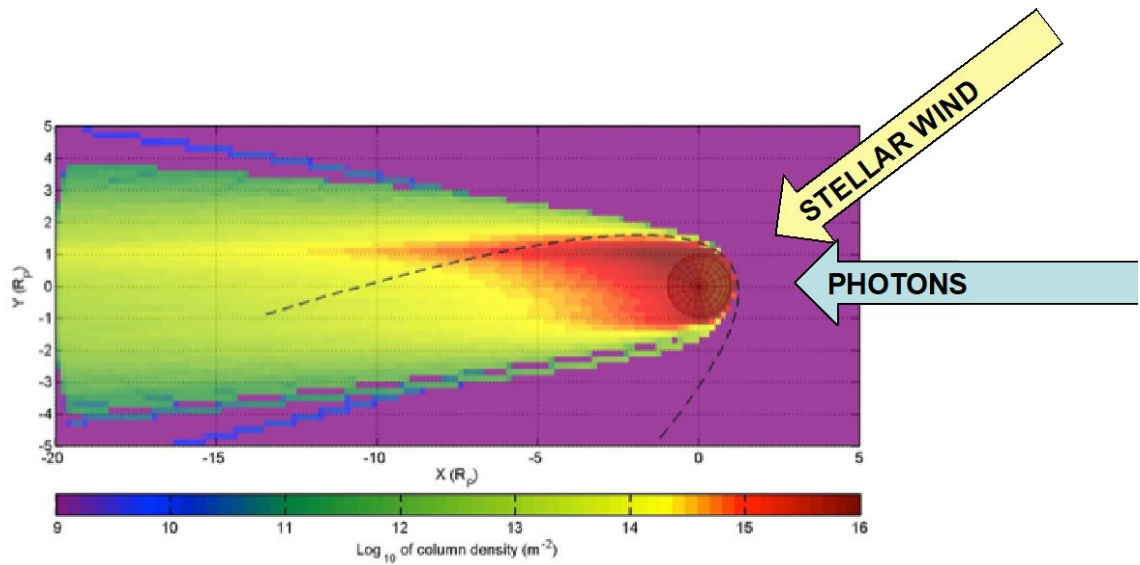

*Figure 20. Simulation of a Na population, with the formation of a tail, due to ion sputtering (or atmospheric sputtering), for a super-Mercury exoplanet with significant stellar wind aberration.*

In summary, since Mercury has a quite weak intrinsic dipole, it may be used, in some way, as an example for both exoplanetary cases (with/without an intrinsic magnetic field). In fact, in some extreme cases, the stand-off distance at Mercury can be so close to the planet that a large part of the daily surface is exposed to solar wind precipitation (Leblanc et al., 2003; Kallio et al., 2003). It is worth noting that, in case of a close-in exoplanet with an atmosphere, the situation is not dramatically different; protons are likely to impact the uppermost atmosphere layer. Such protons can produce energization of neutral particles, known as atmospheric sputtering (e.g., Johnson 1990; Lammer and Bauer, 1993) in a similar way as ion sputtering (Figure 20).

Part of the neutral particles of the exosphere are eventually photoionized by the solar radiation, leading to the formation of an 'exo-ionosphere'. At Mercury, the exo-ionosphere is confined inside the magnetosphere, but if the internal magnetic field is not strong enough the exo-ionosphere would be dragged by the stellar wind flow and form an ionized tail that is much larger than the Na tail (Figure 18). In this respect, Mercury is surely the most appropriate paradigm in the Solar System, as the long-studied sodium tail from both ground based observatories (e.g. Potter et al., 2003) and from space (McClintock et al., 2008, 2009) have revealed much on the evolution of this planet (Orsini et al., 2014). Exosphere transit observations (Schleicher et al., 2004, Mura et al., 2009) had greatly boosted the knowledge of the interaction between the solar wind and Mercury. Hence, it is possible to have an indication on the atmospheric stability by looking at the exospheric tails of exoplanets (Mura et al., 2011; Guenther et al. 2011), formed by those species that are subject to radiation pressure acceleration. This is a fashion similar to the well-studied Mercury case (Killen et al., 2007).

Such ion tails are much larger on close-in exo-Mercury's like CoRoT-7b. Figure 21 shows the colour-coded density of $Ca^+$ ions in the xy plane, integrated along the z direction, under the assumption that the close-in exoplanet has a negligible magnetic field (Mura et al. 2011). The

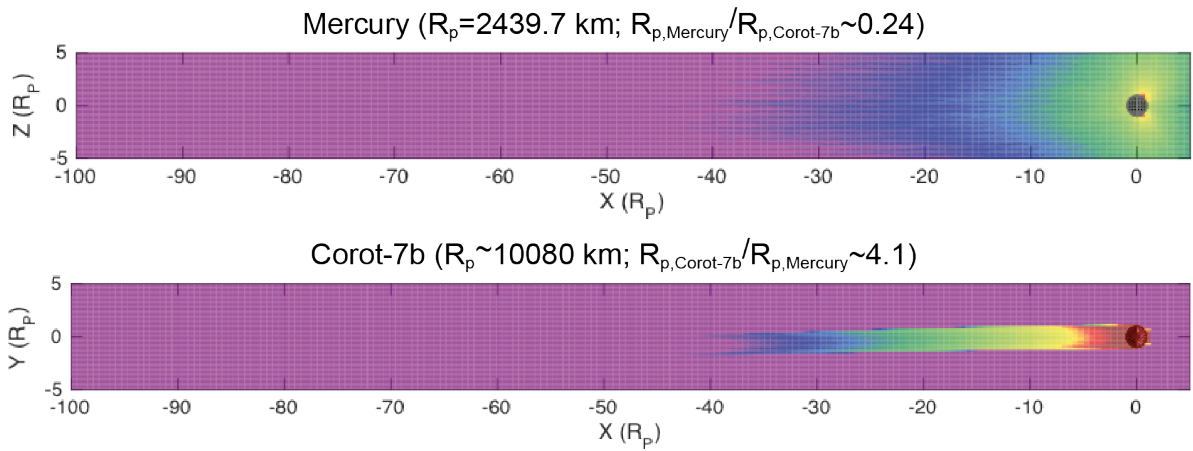

*Figure 21. Typical simulated case of Mercury's Na tail (top) in comparison with CoRoT-7b (bottom) with the stellar wind parameters as given in Mura et al. (2011). Please note that the radius of CoRoT-7b is about 4 times larger than Mercury's.*

$Ca^+$ tail has a scale length of ≈10 $r_{pl}$, when the $Ca^+$ ions become $Ca^{++}$. One should note that under such extreme conditions the $Ca^+$ density falls off exponentially but is still noticeable at a distance of ≈ 100 $r_{pl}$. The Na tail with its different inclination is also shown in Figure 18 as a reference. In this example, neutral particles in the exosphere of a close-in exoplanet are very rapidly ionized by stellar UV. The resulting ions are picked up by the stellar wind, dragged in the anti-star direction, and form a tail, in this case an ion tail.

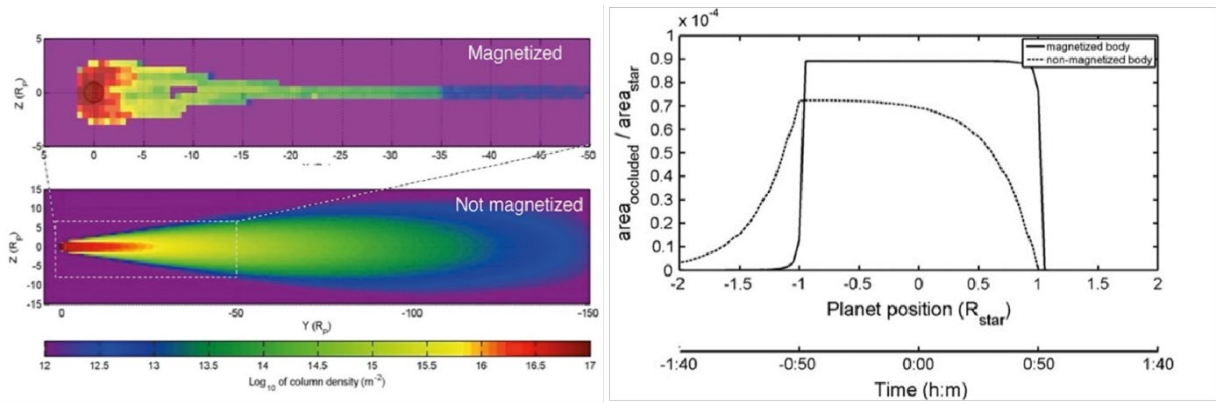

*Figure 22. Left: preliminary simulation of the Ca$^+$ tail, as column densities integrated along the line of sight, for a close in exoplanet, for two cases: with an intrinsic magnetic field (top-left) and without (bottom-left). Note that the scaling is different: in the case of a magnetized body, the Ca$^+$ population is confined in the magnetosphere, in the case of a non magnetized body, Ca$^+$ has a cometary shape. Because of the planet relative velocity w.r.t the stellar wind, the "magnetosphere" is not in the direction of the observer; for this reason, a rapidly orbiting close-in exoplanet has a favorable geometry condition, but, in general, the integrated column densities look different from almost any point of view. Right: simulation of an occultation experiment for the two cases. Fraction of occluded area, due to Ca$^+$ absorption, function of the planet position with respect to the host star (x = 0 is the center of the star, from Mura et al., 2011).*

As shown in Figure 22, depending on the presence of a significant dipole magnetic field of the planet, such an ion tail is either confined inside a magnetosphere (if the body has a magnetic field able to form one) or is dragged away by the stellar wind itself (Mura et al. 2011). While the direction of the neutral atom tail is regulated by the radiation pressure and hence is in the anti-stellar direction, the ionized tail is in the direction of the stellar wind that, in some case, may have a significant aberration.

Recently Vidotto et al. (2018) presented a 3D study of the formation of refractory-rich exospheres around the rocky planets HD219134b, a close-in exo-Mercury with a mass of ~4.7 $M_{Earth}$ and a radius of ~1.6 $R_{Earth}$. The planet orbits around an 11 Gyr old star of spectral class K3V at 0.0388 AU, which is about ten times closer than Mercury. Since the temperature (~1000 K) of this planet is too cool to host a magma ocean, it may have an exosphere formed by sputtering of surface particles due to the exposure to a dense stellar wind. If so, the sputtering will release refractory elements from the entire dayside of the planet. As seen Figure 23, sputtered elements such as O and Mg will create an extended neutral exosphere with densities ≥10 cm$^{-3}$ over several planetary radii. Vidotto et al. (2018) also investigated the detectability of such an exosphere with the current UV instruments and found that it is currently very unlikely to be observed.

In the next section we will discuss the different atmospheres that might evolve on exo-Mercurys.

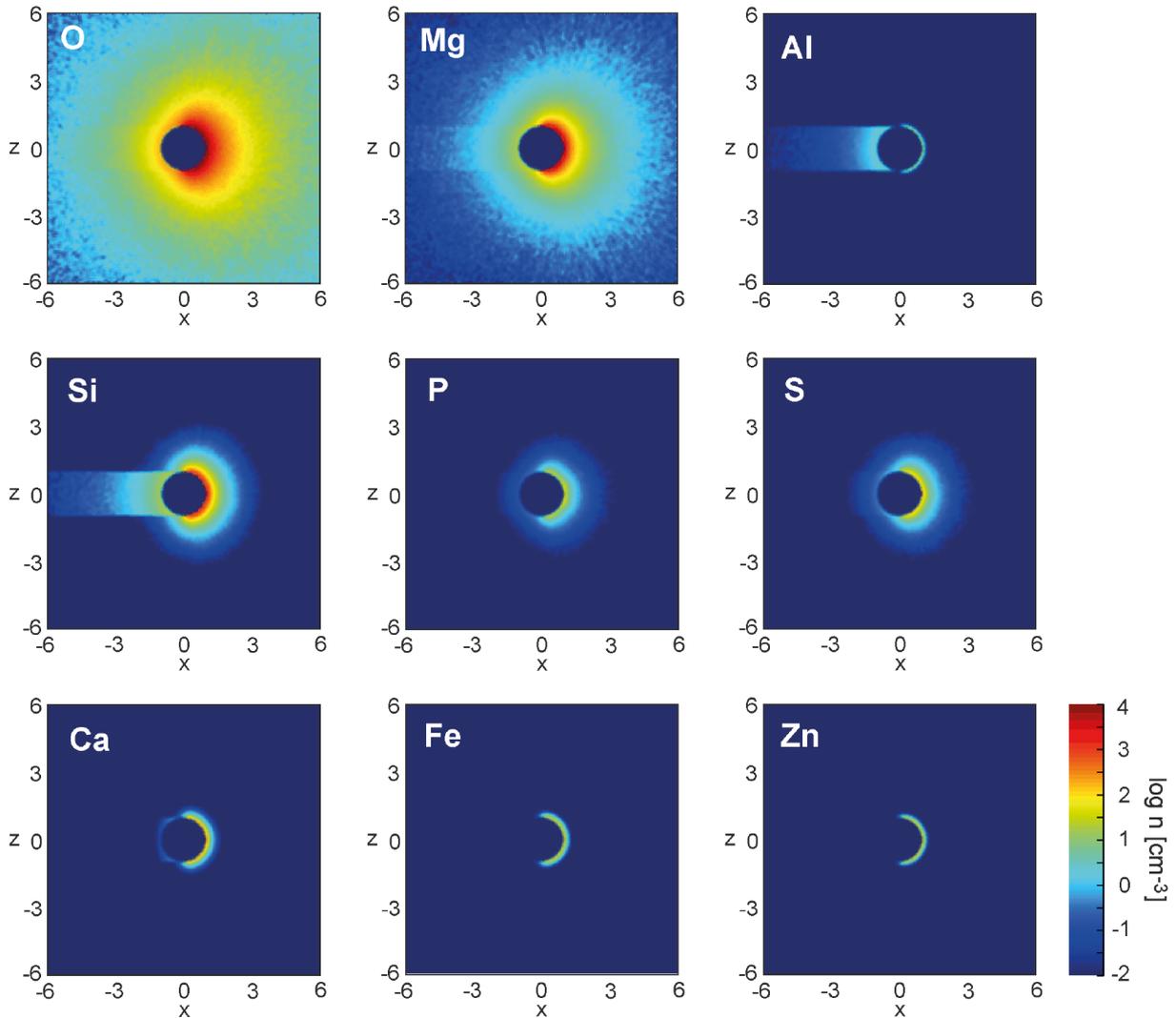

*Figure 23. The sputtering induced exosphere for different atomic species of HD219134b as simulated by Vidotto et al. (2018). Figure from Vidotto et al. (2018)*

### 5.3.1. What kind of atmospheres do we expect for close-in rocky exoplanets

The measured density (i.e., mass and radius) allows us to infer the bulk composition of the planet. Although one can expect a diverse iron fractionation of rocky exoplanets, the density may give an information about the fraction of the iron core for some of the close-in low mass exoplanets, although most of the masses of the small exoplanets have not been estimated yet, i.e., from 732 exoplanets with radii <1.5 $R_{Earth}$, only 74 have mass estimates (data from NASA

Exoplanet Archive as of 30 August 2021)[1]. Figure 17 shows the measured mass and radius of exoplanets with radii of < 3 $R_{Earth}$. Most of exoplanets with radii of < 2 $R_{Earth}$ distribute along the mass and radius relationships for rocky planets while the others should consist of large fractions of $H_2$-dominant, $H_2O$-vapor (see also, Zeng et al. 2019) or even He-dominant envelopes. If one considers rock and iron as the component of planetary bulk composition, there is a variety of rocky exoplanets' compositions such as pure rocky composition, Earth-like composition (Fe ~ 30%) and higher metal/silicate ratio composition (Fe ~ 70%). In addition, the radiative equilibrium temperatures of rocky exoplanets are ranging from that of Earth to high temperatures up to a few thousand K, as shown in Figure 16. Most of the rocky exoplanets with measured masses and radii have orbits less than 0.1 AU and, thus, they are likely tidally locked. Then, the day-side equilibrium temperatures of rocky exoplanets around G-type or K-type stars are high enough to melt and vaporize rock (1500 K is a typical melting point of rock) due to strong stellar irradiation while the planets with equilibrium temperatures of hundreds K are orbiting around M dwarfs.

*Table 3. List of parameters for HD219134b (values from Motalebi et al. 2015; Gillon et al. 2017).*

| Mass ($10^{24}$kg) | ≈28.3 | Rotation Period (hours) | 74.16 | Orbital Period (days) | 3.09 | Mean Temperature (C) | 1015 |
|---|---|---|---|---|---|---|---|
| Diameter (km) | ≈20435 | Length of Day (hours) | tidally locked | Orbital Velocity (km/s) | ≈136 | Surface Pressure (bars) | ? |
| Density (kg/m$^3$) | ≈6360 | Dist. from star ($10^6$ km) | 5.8 | Orbital Inclination (°) | 85.06 | Global Magnetic Field | ? |
| Gravity (m/s$^2$) | ≈18 | Perihelion ($10^6$ km) | 5.8 | Orbital Eccentricity | 0 | Bond albedo | ? |
| Escape Velocity (km/s) | ≈19.2 | Aphelion ($10^6$ km) | 5.8 | Obliquity to Orbit (°) | ? | Visual geom. albedo | ? |

Motivated by detections of exoplanets with higher metal/silicate ratio compositions, the evolution and formation scenarios explaining such a metal-rich composition have been also argued in an exoplanetary science field. The detection of the high-density rocky exoplanets, K2-229b and Keppler-107c, was reported by recent observations (Santerne et al. 2018, Bonomo et al. 2019). K2-229b has a radius of 1.16 (+0.06, -0.05) $R_{Earth}$, a mass of 2.59 ± 0.43 $M_{Earth}$ and a semi-major axis of 0.012 AU. Its bulk density and equilibrium temperature with zero planetary albedo at the dayside are 8.9 ± 2.1 g/cm$^3$ and 2300 K, respectively. Also, Kepler-107c has a radius of 1.60 ± 0.026 $R_{Earth}$, a mass of 9.39 ± 1.77 $M_{Earth}$ and a semi-major axis of 0.06 AU. Its bulk density and day-side equilibrium temperature are 12.65 ± 1.77 g/cm$^3$ and 1600 K, respectively. Their masses and radii are consistent with a metal-rich composition like Mercury (i.e., about 70 % metallic core and 30 % rocky mantle). As with the evolution and formation scenarios for Mercury, there are some possible origins for such a high-density exoplanet; The photoevaporation of a rocky mantle, giant impacts and formation in the metallic iron-rich region of a proto-planetary disk (e.g., Cameron 1985; Benz et al. 1988; Lewis 1972; see also Section 5.2). Especially, a giant impact is the likely origin of Kepler-107c (Bonomo et al. 2019). This is because the innermost planet, Kepler-107b, has similar density with pure rocky composition and, thus, is less dense than Kepler-107c. While the de-trend of the two planet's densities in orbits is inconsistent with the photoevaporation scenario and the iron-rich composition of Kepler-107c would not be primordial considering the Fe/Si and Mg/Si ratios derived from the host star abundance as a proxy of the protoplanetary disk composition, the mass and radius of Kepler-107c matches theoretical predictions from collisional mantle stripping through a giant impact (Marcus et al. 2010).

When comparing the mass and density measurements with models, we have to make an assumption about the relative abundance of the elements. It is a common habit to use simply the solar abundance. However, the abundances reflect the history of the material from which the star and planets have formed. That is, how many supernovae of type I and II, there were, whether the material was enriched by a Wolf-Rayet star, and so on. For example, the anomalous abundance in $^{26}$Al in the early solar-system was due the wind of a Wolf-Rayet star that had more than 20 $M_{Sun}$ (Portegies Zwart 2019). Thus, the relative abundances of all stars are different and so are their planets.

Furthermore, the abundances of elements within the proto-planetary disk differs from inside out. Metals are in the inner part, ices in the other part and silicates in the middle. As outlined, in Section 5.2, the high iron abundance of Mercury could also be due to the location where it formed, as a result of photophoresis which separated iron and silicates in the disk (Wurm et al. 2013). The temperature of the disk was 1300 and 1450 K at its current location (Gail 1998) which is too high for silicates. The fact that chemical abundances change with the distance from the host star also means that the compositions of rocky planets constrain where it formed (Kane et al. 2020). According to Plotnykov and Valencia (2020), iron enrichment and perhaps depletion must have happened before gas dispersal, if the chemical diversity of highly irradiated planets is the result of atmospheric evaporation. Studies of the composition of the atmospheres, or the material released from the surface, thus, promise to give us important insights in planet formation.

Unfortunately, at least observationally, there is a third class of low-mass planets. These are planets with intermediate densities. The prototype of this class is GJ1214b. The mass and radius and density of GJ 1214b are 6.16+/-0.91 $M_{Earth}$ and 2.71+/-0.24 $R_{Earth}$, and 1.6+/-0.6 g/cm$^3$, respectively (Anglada-Escudé et al. 2013). A model with a rocky core and a hazy atmosphere as well as a model with a mixture of rock and ice fits the data equally well (Rogers and Seager 2010). Observations of the atmospheres of such planets are possibly the best way to lift the ambiguity. Thus, atmospheric studies are of key importance for our understanding of these planets.

*Table 4. List of parameters for 55 Cnc e (values from Fischer et al. 2008; Dawson and Fabrycky 2010; Bourrier et al. 2018b).*

| Mass (10$^{24}$kg) | ≈48.3 | Rotation Period (hours) | 17.7 | Orbital Period (days) | 0.737 | Mean Temperature (C) | ≈2570 |
|---|---|---|---|---|---|---|---|
| Diameter (km) | ≈24340 | Length of Day (hours) | tidally locked | Orbital Velocity (km/s) | ≈234 | Surface Pressure (bars) | ? |
| Density (kg/m$^3$) | ≈6400 | Dist. from star (10$^6$ km) | 2.32 | Orbital Inclination (°) | 83.59 | Global Magnetic Field | ? |
| Gravity (m/s$^2$) | ≈22 | Perihelion (10$^6$ km) | 2.43 | Orbital Eccentricity | 0.05 | Bond albedo | ? |
| Escape Velocity (km/s) | ≈23 | Aphelion (10$^6$ km) | 2.20 | Obliquity to Orbit (°) | ? | Visual geom. albedo | ? |

So far, the best-studied close-in rocky exoplanet that is permanently molten on its tidally locked side is 55 Cnc e. Infrared-phase curve allows to determine the location of the hottest part on the surface of the planet and the phase-curve in the optical is a crude albedo map of the surface. Eclipse observations and phase curves have already been obtained for the rocky planet 55 Cnc e which has only 8.59 $M_{Earth}$-masses. The thermal emission phase curve shows that the hottest spot is not at the sub-stellar point but 41±12 degrees east of it (Demory et al. 2016a). This implies that the planet has an atmosphere that must even have a pressure of 1.4 bar (Angelo and Hu 2017). Repeated infrared observations have shown that the depth of the eclipse varies with time, which is further evidence for an atmosphere (Deming et al. 2015; Demory et al.

2016b; Tamburo et al. 2018). The best explanation for these results is that the substellar hemisphere is covered by material that has a high reflectivity, which varies in size. This material could be some kind of bright clouds, or haze in an atmosphere. Although 55 Cancri e is considered to be a rocky planet, the eclipse methods showed that it has an atmosphere. Combining the data from five transits, Ridden-Harper et al. (2016) presented hints of Na and the $Ca^+$ lines originating in the atmosphere of the planet. If real, this signal would correspond to an optically thick $Ca^+$ atmosphere that is five times larger than the Roche lobe. However, Jindal et al. (2020) recently showed that 55 Cnc e either has an atmosphere with a high mean molecular weight and/or clouds, or no atmosphere at all.

The atmospheres of close-in rocky planets are likely very different from the atmospheres of the Earth, or from that of the gas-giants that we have studied so far. Shortly after the discovery of CoRoT-7b (Léger et al. 2009), several authors came up with the idea of a lava or magma ocean (see also next section) at the substellar point (Briot and Schneider 2010, Barnes et al. 2010, and Rouan et al. 2011). The reason that such a planet should have a magma ocean is not only the intense radiation by the host star. If the host star is strongly magnetic and the planet along its close-in orbit crosses the stellar magnetic field lines, the sub-surface regions of ultra-short period planets can be subject to induction heating (Kislyakova et al. 2017; Kislyakova et al. 2018), which can also produce magmatic layers at the upper mantle or below the surface or can increase volcanic activity.

Table 5. List of parameters for CoRoT-7b (values from Leger et al. 2009, 2011; Hatzes et al. 2011).

| Mass ($10^{24}$kg) | ≈44.4 | Rotation Period (hours) | 1407 | Orbital Period (days) | 0.85 | Mean Temperature (C) | ≈1000 - 1500 |
|---|---|---|---|---|---|---|---|
| Diameter (km) | ≈20155 | Length of Day (hours) | tidally locked | Orbital Velocity (km/s) | ≈220 | Surface Pressure (bars) | ? |
| Density (kg/m$^3$) | ≈10400 | Dist. from star ($10^6$ km) | 2.57 | Orbital Inclination (°) | 80.1 | Global Magnetic Field | ? |
| Gravity (m/s$^2$) | ≈29 | Perihelion ($10^6$ km) | 2.57 | Orbital Eccentricity | 0 | Bond albedo | ? |
| Escape Velocity (km/s) | ≈24 | Aphelion ($10^6$ km) | 2.57 | Obliquity to Orbit (°) | ? | Visual geom. albedo | ? |

Mura et al. (2011) developed a model for a higher metal/silicate ratio atmosphere of CoRoT-7b. This model shows that the atmospheric loss rates of CoRoT-7b could be 2-3 orders of magnitude higher than that of Mercury, forming an ionized tail of escaping particles (see Figure 18 and Figure 21). This tail should be observable in the Na and Ca resonance lines. A first attempt to detect the atmosphere of CoRoT-7b using high-resolution spectroscopy, however, resulted only in upper limits (Guenther et al. 2011).

Luckily, we now know a number of ultra-short period planets orbiting stars that are closer and brighter than CoRoT-7b. We mentioned already 55 Cancri above. Another interesting target is HD 3167, which is 13 times brighter than CoRoT-7. Guenther and Kislyakova (2020) searched for the signatures of a higher metal/silicate ratio exosphere and volcanic activity at HD3167b and obtained upper limits of $I_p/I^* =10^{-4}$ to $10^{-3}$ for the most important tracers, where $I_p$ is the planetary, and $I^*$ the stellar flux. These upper limits are lower than claimed detections by Ridden-Harper et al. (2016) for 55 Cnc e. Another interesting object is Kepler-1520b, because it has a dusty tail containing material that is escaping from the planet (Schlawin et al. 2018). More observations of more objects are certainly required. Perhaps these phenomena are episodic and repeated observations will lead to a detection.

While on Earth-like planets inside the habitable zone $H_2O$ is the most abundant volcanic gas (Schmincke 2004), volcanically active exoplanets in close orbital distances are expected to eject mostly $SO_2$, which would then dissociate into oxygen and sulphur atoms (Kislyakova et al. 2018). While planets with a high surface pressure are expected to have carbon-rich and dry volcanic gases, low pressure atmospheres will contain mostly sulphur-rich gases (Gaillard and Scaillet 2014). Possible tracers for such an atmosphere are the same lines that have been detected in the plasma torus of Jupiter's Moon Io. These are [S III] 3722, [O II] 3726, [O II] 3729, [S II] 4069, [S II] 4076, [O III] 5008, [O I] 6300, [S III] 6312, [S II] 6716, [S II] 6731 Å lines, as well as the Na D-lines (Brown 1974; Brown et al. 1975; Brown and Yung 1976; Kupo 1976; Brown and Shemansky 1982; Morgan and Pilcher 1982; Thomas 1993,1996; Kueppers and Jockers 1995, 1997; Guenther and Kislyakova 2020).

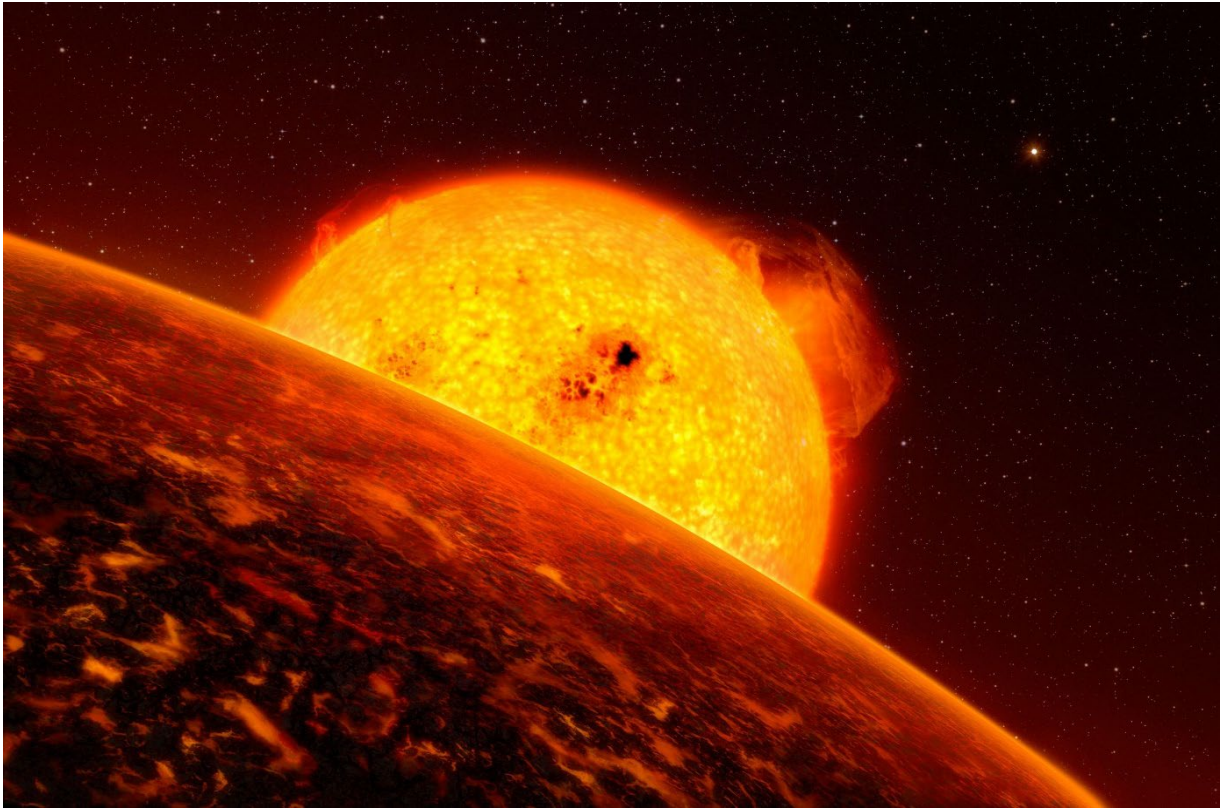

*Figure 24. Artist impression of a magma ocean planet, specifically of CoRoT-7b. Impression by ESO/L. Calcada.*

In the following Sect. 5.4 we will discuss in more detail magma ocean-related silicate atmospheres that should originate around very hot rocky close-in exoplanets.

**5.4. Magma ocean related silicate atmospheres**

Most of the potential rocky exoplanets are hot enough to melt and vaporize rock due to strong stellar irradiation, as shown in Figure 24. Also, about 500 discovered super-Earths or Earth-sized planets which are potentially rocky planets have substellar-point equilibrium temperatures high enough (>1500K) for rock to melt and vaporize itself (https://exoplanetarchive.ipac.caltech.edu, 27/7/2020). These include a first discovered high-density exoplanet, CoRoT-7 b (Leger et al. 2009, Queloz et al. 2009), whose substellar-point temperature are estimated to be about 2500 K with zero planetary albedo. If such planets are indeed rocky, they are thought to be entirely or partially covered with magma oceans and possibly have gases vaporized from the magma oceans as their secondary atmospheres composed of mainly sodium like Mercury (e.g., Schaefer and Fegley 2009, Léger et al., 2011). Such close-in rocky exoplanets are sometimes called lava planets or magma-ocean planets. One may think the category of lava planets or magma-ocean planets includes a young terrestrial planet just after its formation or a large collisional event. Thus, we refer to the rocky exoplanet as a close-in molten exoplanet just to clarify it.

### 5.4.1. Close-in molten exoplanets

Close-in molten exoplanets would have lost their primordial hydrogen-rich atmospheres due to hydrodynamic escape caused by X-ray and UV irradiation from their host stars (see Owen 2019, for a review). Escape of highly-irradiated hydrogen-rich atmospheres is known to occur in an energy-limited fashion (Sekiya et al. 1980, 1981; Watson et al. 1981) and the hydrodynamic escape process has been intensively studied (e.g., Lammer et al. 2003, Yelle 2004, García Muñoz 2007; Kubyshkina et al. 2018; Kubyshkina and Vidotto 2021). For example, in the case of CoRoT-7 b, Valencia et al. (2010) demonstrated that the cumulative escaped mass of a hydrogen-rich atmosphere could be a few Earth masses with an analysis based on energy-limited formula and thus, would have lost the primordial atmosphere. Therefore, they are expected to have secondary atmospheres. The compositions of the secondary atmospheres on rocky exoplanets are predicted to be diverse by theorists because the atmospheric components could be outgassed from their interiors and/or got material brought by impacts of smaller bodies (e.g., Elkins-Tanton and Seager, 2008).

The secondary atmospheres of close-in molten exoplanets are likely composed of materials directly vaporized from their magma ocean. Then, impacts of smaller bodies would not affect the main composition of the atmospheres because of the rapid vaporization/condensation (i.e., gas-melt equilibrium condition) unless the impacts are massive enough to change the magma composition. Some theoretical models have predicted the atmospheric composition based on gas-melt chemical equilibrium calculations. If close-in molten planets are dry, they likely have atmospheres composed of rocky materials such as Na, K, $O_2$ and SiO (Schaefer and Fegley 2009, Miguel et al. 2011, Ito et al. 2015, Ito and Ikoma 2021). On the other hand, if close-in molten planets have volatile elements such as H, C, N, S and Cl, they likely have atmospheres composed mainly of $H_2O$ and/or $CO_2$ with rocky vapors such as Na and SiO (Schaefer et al. 2012, Herbort et al. 2020). These volatile species should be efficiently outgassed from their magma ocean (e.g., Schaefer and Fegley 2007) but they could selectively disappear through massive escape due to strong stellar irradiation (Valencia et al. 2010, Léger et al., 2011), Then, the planets become dry having only low-volatility elements (i.e., refractory elements) and silicate atmospheres.

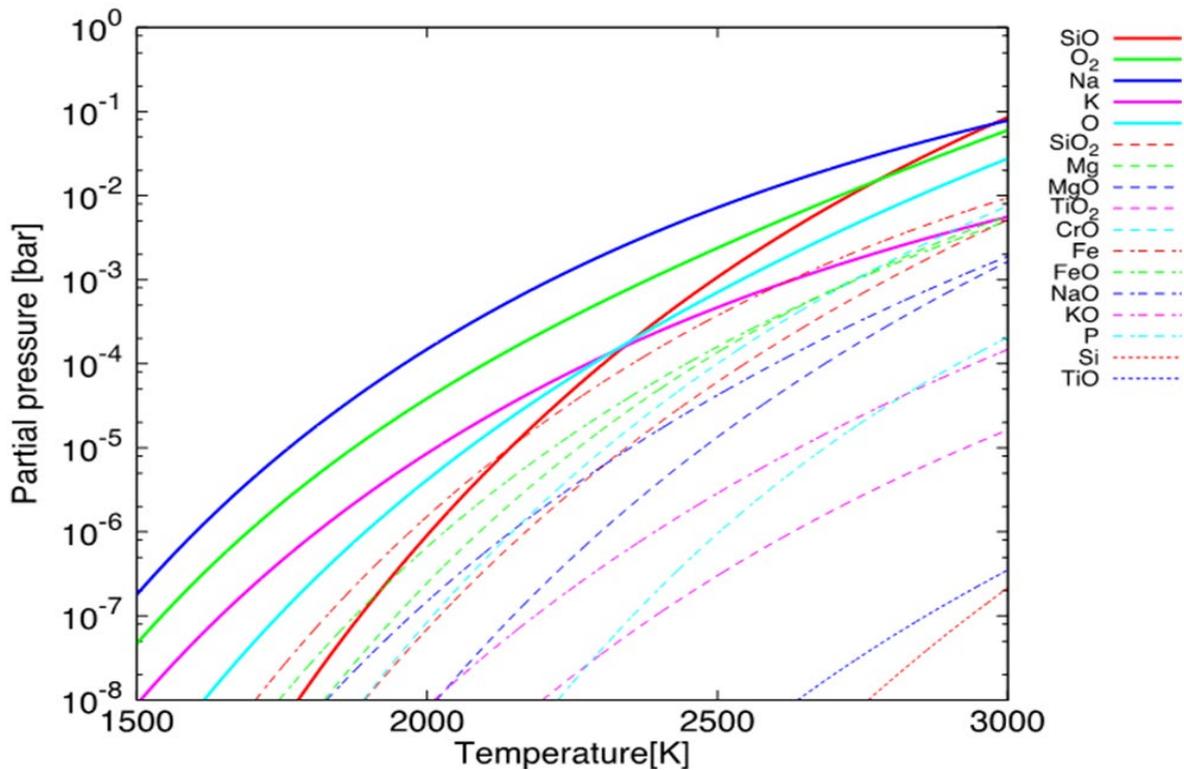

*Figure 25. Composition of gas in chemical equilibrium with molten silicate. Partial pressures of the gas species are shown as functions of magma temperature. These partial pressures are calculated by a chemical equilibrium model of Ito et al. (2015) assuming volatile-free bulk silicate Earth composition (see Ito et al. 2015, for the details).*

The total pressure and composition of a silicate atmosphere highly depend on the temperature of the magma ocean (Schaefer and Fegley 2009, Miguel et al. 2011, Ito et al. 2015). Figure 25 shows the partial pressures of gas species in a silicate atmosphere vaporized from magma ocean with volatile-free bulk silicate Earth composition. Na is the most abundant species in most of the temperature range, while SiO increases with temperature and becomes the most abundant one for T > 2800K. Also, the total vapor pressure is as small as about $10^{-7}$ bars at T = 1500K and 0.1 bar at T = 3000 K. These properties depend on magma composition, but they are almost same in typical rocky compositions of Earth; Earth's MORB, bulk crust and upper crust (Ito et al. 2015).

The strong dependence of the total pressure on temperature would cause strong winds from the planetary sub-stellar point to the terminator. If the magma composition are well mixed the atmospheric pressure remains close to their local saturation values because of the rapid vaporization/condensation of gases (Castan and Menou 2011) but, if not, the atmosphere and magma are compositionally variegated in latitudes/longitudes due to materials transport via the wind (Kite et al. 2016). In vertical, silicate atmospheres have thermal inversion structures for the substellar-point equilibrium temperature higher than 2300 K but have isothermal profiles for lower temperatures (Ito et al. 2015). The thermal inversion is caused by Far-UV absorption of SiO and visible absorption of Na and K.

Ito and Ikoma (2021) developed a 1-D hydrodynamic model for the study of UV-irradiated silicate atmospheres that contain Na, Mg, O, Si, their ions and electrons. Their model also includes the thermal and photochemistry, molecular diffusion, thermal conduction, X–ray and UV heating, and radiative line cooling. It was found that most of the host stars energy of these short wave lengths is lost by the radiative emission of Na, Mg, $Mg^+$, $Si_2^+$, $Na_3^+$ and $Si_3^+$. Ito and Ikoma (2021) found that a magmatic Earth-sized planet at an orbit of 0.02 AU around a young solar-type star, with the above given silicate atmosphere develops a low X-ray and UV heating efficiency which is in the order of $1 \times 10^{-3}$, which corresponds to a total thermal mass loss rate of $\approx 0.3\ M_{Earth}$/Gyr. It was found that efficient cooling in such a silicate atmosphere of a $1 M_{Earth}$-mass planet yields photo-evaporation rates that are not large enough for modifying the planetary mass and bulk composition largely. On the other hand, one should expect that no dense silicate atmosphere might accumulate because as discussed in Sect. 5.3 the dense stellar wind at 0.02 AU will most likely erode the atmosphere via ion pick up and other nonthermal loss processes.

Recent transit observations through photometric and spectroscopic methods have reported the signal of the atmospheric components or temperature of close-in exoplanets whose densities are consistent with rocky ones. Such characterization is important to know not only what exoplanets are like but also how they were evolved and formed. One of the most famous close-in super-Earths is 55 Cnc e (see Table 4 and Section 5.3.1). The exoplanet orbits a bright G8V (V = 5.95) star, which allows for measurements with a higher signal-to-noise ratio than that for other super-Earths. The 1 σ upper limit on the measured 55 Cnc e's density reaches the pure rocky regime without iron core. Thus, the planet is possibly a dense rocky planet with a relatively large atmosphere, or a planet made of lighter materials as water and carbon but with a small atmosphere.

The thermal phase curve and transmission spectra of 55 Cnc e are observed. Demory et al. (2016) found the day night temperature difference on 55 Cancri e are large by analyzing the thermal phase curve. The temperature averaged over the dayside is 2700 ± 270 K that is about twice higher than the night-side temperature, 1380 ± 400 K. The significant eastward hot-spot shift was also measured. These features seem imbalanced from the viewpoint of climate theory (Hammond and Pierrehumbert 2017). This is because, as the mean molecular weight of an atmosphere increases, the atmosphere tends to have a larger day–night temperature contrast but a smaller eastward phase shift (Zhang and Showman 2017). Also, the transmission spectra of 55 Cnc e have suggested the atmosphere contains a significant amount of light gases such as hydrogen (Tsiaras et al. 2016) but water vapor was not found (Esteves et al. 2017,, Jindal et al. 2020). These might suggest that 55 Cnc e has a volatile-rich atmosphere. On the other hand, Bourrier et al. (2018a) reported the detection of variability in 55 Cnc's FUV emission lines induced by O, $C^+$, $C_2^+$, $N_4^+$, $Si^+$, $Si_2^+$ and $Si_3^+$ before/during/after the transit of 55 Cnc e. They concluded the variations are unlikely to originate from purely the host star and purely the planet. Ridden-Harper et al. (2016) reported the possible detection of a sodium and calcium exosphere escaping from 55 Cnc e during transit while escaping hydrogen from 55 Cnc e was not detected (Ehrenreich et al. 2012). This might suggest that 55 Cnc e have a silicate atmosphere but, taken as a whole, these observations seem hard to reach a consensus with each other and atmospheric models for now.

**5.4.2. Evaporating close-in rocky exoplanets**

Rocky planets that may be currently evaporating have been detected around the three stars; KIC 12557548, K2-22 and KOI-2700 (e.g., Rappaport et al. 2012, 2014, Sanchis-Ojeda et al. 2015). Such planets are called evaporating planets. While planet transits are usually symmetric and periodic without any significant variations in their shape or depth over time, the variation in transit depth and the ingress-egress asymmetry of the transit light curve of KIC 12557548 were found (Rappaport et al. 2012). The variation and asymmetry of transit light curves have been interpreted as a piece of the evidence of catastrophic evaporation of small rocky planets with an evaporated dust tail (e.g., Rappaport et al. 2012, Perez-Becker and Chiang 2013, Kawahara et al. 2013, Croll et al. 2014). Also, since the variation period of transits for KIC 12257548 is consistent within 1-σ of the rotation period of its host star, the evaporation of the planet KIC 12257548 b may be correlated with stellar activity such as UV emission and magnetic field (Kawahara et al. 2013). The estimated mass loss rate for KIC 12257548 b is about one Earth mass per Gyr while an upper limit of the detected planet's radius is only about one Earth radius (e.g., Rappaport et al. 2012, Brogi et al. 2012, Kawahara et al. 2013, van Lieshout et al. 2014). As the equilibrium temperature of an evaporating planet is high (e.g., about 2000 K for KIC 12257548 b), a thermal ("Parker-type") wind from low-mass rocky planets is one of possible interspersions for such massive evaporation (Perez-Becker and Chiang 2013).

**5.5. What can we learn from future research on close-in rocky exoplanets?**

Low-mass rocky planets orbiting close to their host stars are very interesting to study, because they give us many new insights how such planets form and evolve. Precise measurements of the mass and radius indicate that some of them have not lost their primordial atmospheres that presumably contains only a few percent of the mass of the planet. These atmospheres are likely to be $H_2$/He-dominated but it is possible that they contain material released from the rocky cores underneath. The next generation of instruments will allow us to study these atmospheres. For the first time we might be able to find out what these planets are made of which puts strong constrains on how and where they formed.

ARIEL (Atmospheric Remote-sensing Infrared Exoplanet Large-survey) (Puig et al. 2018; Tinetti et al. 2016) is a satellite that is especial designed for the study of exoplanets atmospheres. The telescope has an elliptical primary of 1.1 × 0.7m, and the launch is scheduled for 2028. The VIS-channel has three photometric bands: 0.5-0.6 μm, 0.6-0.81 μm, and 0.81-0.1. μm. The NIR-channel obtains low-resolution spectra in the wavelength ranges 1.1-1.95 μm (with a resolving power $R$ of $R=\lambda/\Delta\lambda=20$ with $\Delta\lambda$ as the smallest difference that can be distinguished at wavelength $\lambda$), 1.95-3.9 μm ($R=100$), and 3.9-7.8 μm ($R=30$). Simulations show that a signal-to-nose ratio of 10 is already sufficient to detect the most prominent chemical species. The simulation of the 2.7 $R_{Earth}$ planet GJ1214b furthermore shows that it will be possible to detect features in the spectrum even if the ratio of the planetary radius $r_{pl}$ to the stellar radius $r_{star}$, i.e., $(r_{pl}/r_{star})2$ is only 1% larger in the lines than in the continuum if 100 transits are averaged (ARIEL mission proposal). This means that with ARIEL it is possible to study planets with less than 10 $M_{Earth}$ even if they have only slightly extended atmospheres.

Another important mission for exoplanet research will be JWST with its 6.5m aperture telescope. JWST offers the following observing modes: 0.7-5 μm ($R=100$; NIRspec prism), 0.7-2.5 μm ($R=700$; NIRISS grism+ prism), 2.5-5 μm ($R=1700$; NIRcam grisms), 5-12 μm ($R=70$; MIRI LRS prisms). Simulations by Greene et al. (2011) show that a single eclipse observation of GJ1214b in the 2.5-5 μm range will already constrain the most prominent molecular species in the atmosphere like $H_2O$, $CH_4$ and $NH_3$.

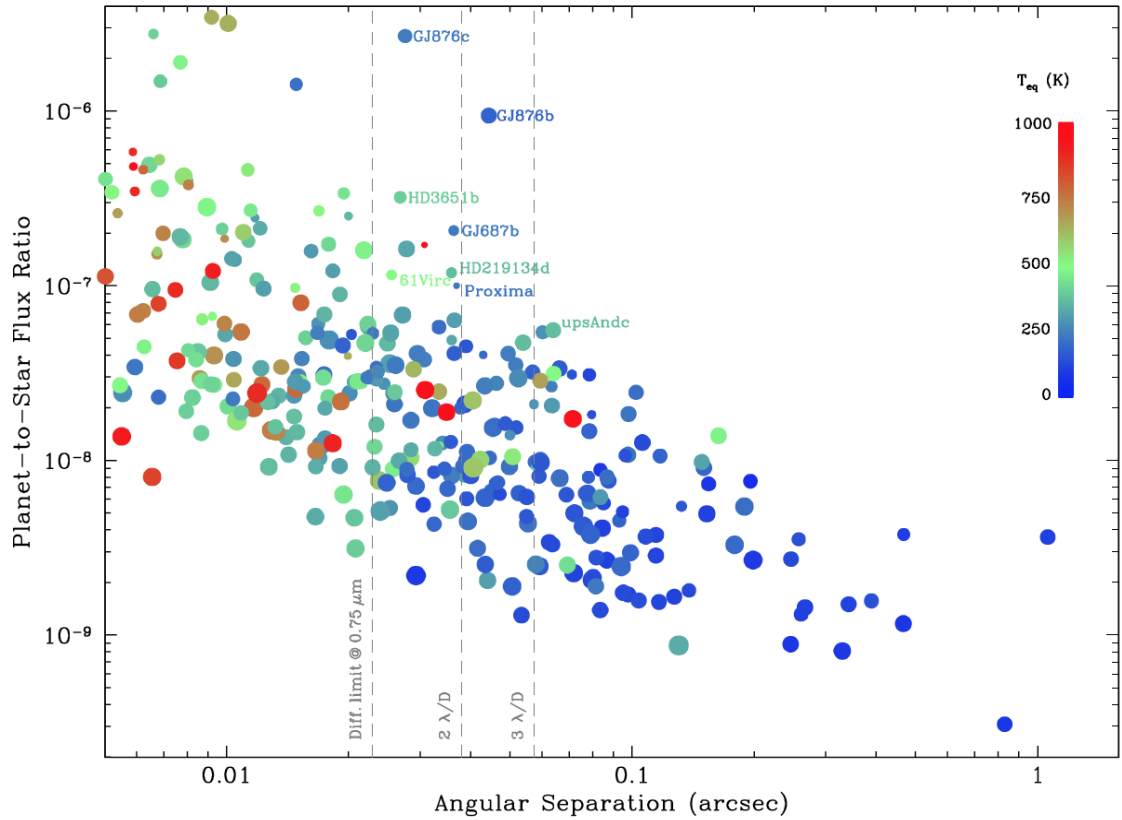

Figure 26. *Estimated planet-to-star contrast in reflected light for known exoplanets as a funtion of angular separation from their host star. Dot size is proportional to the logarithm of planet mass, while the color scale represents equilibrium temperature (assuming a Bond albedo of 0.3). Vertical dashed lines indicate the di fraction limit, 2 λ / D and 3 λ / D thresholds for the 8.2-m VLT at 750 nm (corresponding to the $O_2$ A-band). The combination of a $10^3$-$10^4$ contrast enhancement from SPHERE to the high spectral resolution of ESPRESSO can reveal the planetary spectral features and disentangle them from the stellar ones. The estimated planet-to-star contrast for Proxima b is $10^{-7}$ in reflected light. This is challenging, but a 5-σ detection would be possible by observing the object with SPHERE+ESPRESSO for 20-40 nights.*

Studies of magmatic rocky planets that lost their primordial atmospheres or never had some are also important. A big leap forward will be the high-resolution spectrographs on the next generation telescopes. One of the new instruments is the Visible Echelle Spectrograph – G-CLEF (Chicago-Large Earth Finder) of the Giant Magellan Telescope (GMT) is vacuum-enclosed and fiber-fed spectrograph that covers the wavelength region from 0.35 μm to 0.95 μm. The Precision Radial Velocity (PRV) mode provides a resolution $R$=108,000 (Szentgyorgyi 2014). High-resolution spectroscopy at infrared wavelength benefit from the higher brightness ratio between the planet and the star and they have the advantage that most molecular lines are at infrared wavelengths. METIS ('Mid-infrared Extremely Large Telescope (ELT) Imager and Spectrograph') is an L/M and N-band instrument that covers the wavelength region from 3 - 14 μm. In the L/M band (3-5.3 μm) it will provide a spectral resolution of R=100,000 (Brandl et al. 2010; Lenzen et a. 2010). High Resolution Echelle Spectrometer (HiRES) is an Initiative to realize a high-resolution spectrograph for the European-Extremely Large Telescope (E-ELT). The wavelength coverage will be unusually large, 0.36 to 1.8 μm, possibly even up to 2.48 μm. The spectral resolution will be of the order of R=100,000. HiRISE is the proposed fiber coupling between the direct imager SPHERE and the spectrograph CRIRES+ at the Very Large Telescope. Figure 26 shows the detection limits of 5 σ for HiRISE derived for a bright nearby young star (H = 3:5, 19 pc, 20 Myr), compared to the 20 % best SPHERE/SHINE detection limits. We overplot state-of-the-art population synthesis models based on the core accretion formation scenario (Mordasini 2018; Emsenhuber et al. 2020a,b).

The ELT instruments will not only outperform the existing ones because they are more modern, but also because this method is basically only limited by the signal-to-noise ratio that can be achieved in a given amount of time. Even high resolution than $R=100000$ could have its advantages (López-Morales et al. 2019).

The combination of high-contrast imaging and high-spectral-resolution can yield higher quality spectra than either of the two methods alone (Otten et al. 2021). This is even true if the planet is not seen directly in the high-contrast image, because the adaptive optics system still helps improving the light-ratio between the star and the planet. Combining the VLT instruments SPHERE and ESPRESSO might even allow to detect Earth-mass planet Proxima b despite the planet-to-star contrast ratio of 107. However, 20 to 40 observing nights on the VLT are needed for that (Lovis et al., 2017).

By studying such silicate atmospheres, we will obtain a deeper understanding of the mama ocean phase through which all rocky planets evolve. The detection of the outgassed silicate atmosphere would be a revelation. Perhaps many of these planets also have cometary-like tail structures, similar but more massive than those of Mercury. If we can detect them, we would also be able to put constrains on the composition of these planets. Perhaps the next generation of instruments, or just continued observations of known planets, or the discovery of such planets orbiting nearby stars may make such observations possible. After a long phase where we knew nothing about small exoplanets that they exist, we are now getting pieces of evidence what they really are and that they are not like any of the planets of our Solar System.

# 6. Conclusions

The origin and evolution of various airless planetary bodies from large planetesimals to planetary embryos to Earth's Moon and Mercury as well as recently discovered hot, rocky and most likely magmatic exoplanets were discussed. It is pointed out that by studying the origin and evolution of the Moon, Mercury and other airless planetary bodies, one has to consider that these bodies and hence their surfaces experienced a far more extreme bombardment by solar/stellar radiation and plasma during the first hundrets of million years compared to that of today. These extreme environments will have affected and modified together with frequent meteoritic impacts the uppermost surface layers of these bodies. We used the Moon as an example for that and addressed also the indications that the Lunar surface may contain "fingerprints" of escaped particles from early Earth's upper atmosphere that was heated and expanted due to the higher XUV-flux of the young Sun. We compared Mercury's solar wind interaction with that expected at known higher metal/silicate ratio planets that orbit very close to their host stars and speculate which kind of mineral-like atmospheric/exospheric environments these planets will have. In the near future one can expect that space-observatories such as the JWST or ARIEL and large telescopes like the VLT with its SPHERE and ESPRESSO instruments may allow us to detect atmospheres/exospheres that consist of rock-vapor or minerals by high-contrast-imaging and high-spectral-resolution. The detection of such atmospheres/exospheres around higher metal/silicate ratio planets may reveal their composition so that one can compare them with formation theories of Mercury in the Solar System.

**Acknowledgements**

The authors thank ISSI for supporting the fruitful ISSI-workshop "Surface Bounded Exospheres and Interactions in the Solar System". E. Guenther acknowledges that this work was generously supported by the Thüringer Ministerium für Wirtschaft, Wissenschaft und Digitale Gesellschaft. N.V. Erkaev acknowledges support by the Ministry of Science and Higher Education project No. 075-15-2020-780